# TITLE

## "The substructure of a  quantum oscillator of field "

### Author : Guido Giovanni

**Department of Mathematics and Physics; Senior High School " C.Cavalleri " Parabiago (Milan) - Italy**

## Abstract

| | |
|---|---|
| Un oscillatore (IQuO) più elementare di quello quantistico è formulato. | An oscillator (IQuO) more elementary than the quantum one is formulated. |
| Questo è espresso da operatori quantistici $(a, a^+)$ a due componenti ed è composto da sub-oscillatori, ciascuno con "semi-quanti" $(\frac{1}{2}\hbar\omega)$. | This is expressed by quantum operators $(a, a^+)$, with two-components and it is composed of sub-oscillators, each with "semi-quanta" $(\frac{1}{2}\hbar\omega)$. |
| La relazione di commutazione degli operatori di semi-quanto $(a, a^+)$ è calcolato e si definisce la struttura ad IQuO dei campi scalari con carica elettrica e senza. | The commutation relation of the $(a, a^+)$ "semi-quanta" operators is then calculated and the structure by IQuO of the scalar fields is defined with and without electric charge. |
| Si elabora un meccanismo di accoppiamento di IQuO elettricamente neutri che produce una coppia (materia –antimateria) con carica elettrica intera e non. | Processing a coupling mechanism of electrically neutral IQuO that produces a pair (matter-antimatter) with electric charge and not. |
| Tramite la struttura ad IQuO è possibile interpretare la carica elettrica e l'isospin (negli adroni) come manifestazioni di rotazioni in uno spazio interno ad una particella. | Through the IQuO-structure it is possible to interpret the electric charge and the isospin (in hadrons) as rotations in an inside space in the particle. |
| Infine, si congettura così una sub-struttura nei quarks e si definisce in termini probabilistici il valore non intero della carica elettrica. | Finally, a sub-structure of the  quarks is conjectured and one define in probabilistic terms the non-integer value of electrical charge |

## Introduzione

In questo studio rileveremo come l'oscillatore quantistico, rappresentativo di un campo quantistico, differisce da quello classico perché esprime la potenzialità di distinguere il verso di rotazione di fase associata con l'oscillazione.

Mostreremo che il segno della carica elettrica di una particella di campo è la manifestazione della esistenza di un grado di libertà "interno" all'oscillatore, collegato proprio al verso di rotazione della fase.

A tal proposito andiamo ad affrontare lo studio dell'oscillatore quantistico in modo originale.

All'inizio (nel paragrafo I) viene osservato come qualunque tipo di oscillatore, come quello forzato-smorzato o come quello elasticamente accoppiato ad altri oscillatori, mostri una soluzione delle equazioni del moto con due componenti oscillanti, definite componente elastica e componente inerziale.

Questo ci porta a costruire una rappresentazione a due componenti per l'oscillatore quantistico di campo.

Ne deriva che ciascuno dei due operatori di annichilazione (**a**) e di creazione (**a**$^+$) è espresso con due componenti: una componente elastica e l'altra inerziale:

[**a** ➜ (**a**$_{el}$, **a**$_{in}$) ; **a**$^+$ ➜ (**a**$^+$$_{el}$, **a**$^+$$_{in}$)]

Un così fatto oscillatore quantistico ci permette di definire una struttura più elementare, consistente in due sub-oscillatori accoppiati che contengono complessivamente due "semi quanti", ciascuno con energia di (½ℏω) ai quali corrispondono operatori definiti di "semi quanto".

Denotiamo con "**IQuO**" (acronimo di **I**ntrinsic **Qu**antum **O**scillator) l'oscillatore con una struttura a semi-quanti, precisando che esso può rappresentare un Φ-campo scalare quantistico (IQuO ⬤➜ oscillatore di campo).

## Introduction

In this study we'll note that the quantum oscillator, representative of a quantum field, differs by the classical one, because it expresses the possibility of distinguish the " direction" of phase rotation associated with the oscillation.

We'll show that the sign of electric charge of a particle-field is derived from the existence of a degree of freedom inside the oscillator, which is connected to the direction of rotation of the phase.

To such purpose we go to broach the study of the quantum oscillator in original way.

At beginning (in sect. I) it is remarked that any type of oscillator, as the driven-damped one or as the elastically coupled-to-other-oscillators one, shows a solution to the equations of motion with two oscillating components, defined as elastic and inertial component.

This leads us to build a two-component representation for the quantum oscillator of field.

It follows that each of the two operators of annihilation (**a**) and creation (**a**$^+$) is expressed by two components: an elastic component and the other inertial:

[**a** ➜ (**a**$_{el}$, **a**$_{in}$) ; **a**$^+$ ➜ (**a**$^+$$_{el}$, **a**$^+$$_{in}$)]

This original quantum oscillator allows us to define a more elementary structure of it, consisting in two coupled sub-oscillators which overall contain two "semi-quanta" ( with energy (½ℏω) ) described by "semi-quanta" operators.

The oscillator with a structure to semi-quanta is denoted by **"IQuO"** (the acronym of **I**ntrinsic **Qu**antum **O**scillator), stating that it can represent very well a Φ - scalar quantum field (IQuO- filed oscillator).

| | |
|---|---|
| Nel par. 2 una nuova forma esponenziale degli operatori ($\mathbf{a}$, $\mathbf{a}^+$) [1] si propone, che ci permette di determinare la forma matematica degli operatori di "semi-quanto", le loro nuove parentesi di commutazione e, infine, la distribuzione dei semi-quanti nei due sub-oscillatori. | In **Sect. 2** a new exponential form of the operators ($\mathbf{a}$, $\mathbf{a}^+$) [1] is proposed, which allows us to determine the mathematical form of the operators of "semi-quanta", as well as their new brackets of commutation and, finally, the distribution of semi-quanta in two sub-oscillators. |
| Tenendo conto sia della distribuzione dei semi-quanti all'interno dei due sub-oscillatori di un IQuO e sia del loro movimento in relazione alla fase ( espresso da un operatore di rotazione della fase ($\mathbf{r}$) ) , si può pensare che in un oscillatore quantistico a semi quanti sia possibile determinare un nuovo grado di libertà. | Taking into account both the distribution of semi-quanta inside the two sub-oscillators of an IQuO and both their movement in relation to the phase (expressed by an operator ($\mathbf{r}$) of phase rotation), one may think that a new degree of freedom it's possible to determine in a quantum oscillator at semi-quanta. |
| Questo grado di libertà all'interno dell'oscillatore quantistico di campo potrebbe consentire di rilevare il verso di rotazione della fase, che sarebbe poi interpretato come il segno della carica elettrica di una particella. | Inside the quantum oscillator of field this degree of freedom might allow to detect the direction of rotation of the phase, which would then be interpreted as the sign of the electric charge of a particle. |
| Successivamente, **nel par. 3**, formuliamo un meccanismo di accoppiamento tra due $\Phi$-IQuO di campo scalare reale [indicato con $\Phi_1 \oplus \Phi_2$], che conduce alla formazione finale di un sistema quantistico non separato costituito da due $\Psi$-IQuO di campo scalare complesso, noti in letteratura come soluzioni con carica elettrica dell'equazione d'onda di K-G (acronimo di Klein-Gordon). | Subsequently, in **Sect. 3**, we formulate a coupling mechanism between two real scalar fields $\Phi$-IQuO [indicated with $\Phi_1 \oplus \Phi_2$], which leads to the final formation of a quantum system consisting of two entangled complex scalar fields $\Psi$-IQuO, known in literature as an electrically charged solutions of the wave equation of KG (acronym for Klein-Gordon). |
| Nella sez. IV, utilizzando la struttura a semi-quanti dell'oscillatore quantistico, vengono calcolati i valori della carica elettrica Q, espressi dalla "coppia" di IQuO campi ($\Psi_+$, $\Psi_-$) adesso di un sistema fisico separato. | In sect. IV, using the structure by semi-quanta of quantum oscillator, the values of electric charge Q are calculated, expressed now by the "pair" of IQuO fields ($\Psi_+$, $\Psi_-$) which are not entangled. |
| Si mostra che un'eventuale struttura irregolare di semi-quanti di un $\Psi$-IQuO può essere "regolarizzata" da un ulteriore accoppiamento elastico con un neutro $\Phi$-IQuO. | It shows that any irregular structure of semi-quanta of $\Psi$-IQuO can be "regularized" by an additional elastic coupling with a neutral $\Phi$-IQuO. |
| Inoltre si dimostra che l' accoppiamento [$\Phi_1 \oplus \Phi_2$] di due $\Phi$-IQuO con autovalori opposti dell'operatore ($\mathbf{r}$), determina la "nascita" di una nuova coppia di campi-IQuO ($\Psi_1$ , $\Psi_2$) con carica elettrica opposta, diversa da quella iniziale, in cui ciascuno di essi ha gli operatori di semi-quanto con la stessa direzione della rotazione di fase. | Furthermore it is shown that the [$\Phi_1 \oplus \Phi_2$] coupling of two $\Phi$-IQuO with opposite eigenvalues of the operator ($\mathbf{r}$), gives rise to a new pair of IQuO-fields ($\Psi_1$, $\Psi_2$) with opposite electric charge, in which each has the operators of semi-quanta with the same direction of phase rotation. |

Si rileva che: $(\Psi_1 , \Psi_2) \equiv [\Psi_+ + \Psi_-]$

Il processo di accoppiamento $[\Phi_1 \oplus \Phi_2]$ ➔ $[\Psi_+ + \Psi_-]$ ci ricorda la ben nota creazione di coppie, descritta nella fisica dei campi quantistici.

Abbiamo il sospetto che il meccanismo di accoppiamento di due e più IQuO sia fondamentale per costruire gli stessi campi fisici e le loro possibili interazioni.

Nella sezione 5, riprendiamo la forma a matrici [1] del campo scalare complesso nelle componenti $(\Psi_+ , \Psi_-)$, rilevando la stretta associazione tra i coefficienti delle due matrici e la direzione della rotazione di fase della funzione d'onda.

Viene dimostrato che le soluzioni con energia negativa dell'equazione d'onda di Klein-Gordon sono collegate al verso orario della fase e che queste possono essere reinterpretate come soluzioni ad energia positiva con carica ed impulso opposti alla soluzione "partner" della Klein -Gordon.

Viene rilevato che questa reinterpretazione è possibile se un campo esploratore potesse rilevare e distinguere la "disposizione dei coefficienti", il senso di rotazione di fase della funzione d'onda e la direzione di movimento della particella nello spazio.

In tal caso un campo esploratore vedrebbe una particella proveniente "da destra", avente energia negativa e un verso orario di rotazione della fase, come se essa fosse una particella andante verso "sinistra", avente energia positiva e con un valore (-1) della variabile Q, diverso però da quello (+1) associato a una particella andante a "sinistra", originariamente però ad energia positiva!

Viene così evidenziato che quella caratteristica fisica, indicata come "carica elettrica (Q = ±1), non esprime altro che la relazione tra il verso di rotazione di fase e quegli aspetti di una particella espressi nei coefficienti della funzione d'onda.

One notes that: $(\Psi_1 , \Psi_2) \equiv [\Psi_+ + \Psi_-]$

The coupling process $[\Phi_1 \oplus \Phi_2]$ ➔ $[\Psi_+ + \Psi_-]$ reminds of the well-known creation of pairs, described in the quantum physics of fields.

We suspect that the coupling mechanism between two or more IQuO-particles is essential to build the different fields and their possible interactions.

In **Sect. 5**, we take the matrix form [1] of the complex scalar field in components $((\Psi_+, \Psi_-)$, noting the association between the coefficients of the two matrices and the direction of the rotation phase of the wave function.

It is demonstrated that solutions with negative energy of the wave equation of Klein-Gordon are connected with the clockwise direction of the phase and that these solutions can be reinterpreted as particles with positive energy and with momentum and charge opposite to those of the "partner" solution of Klein-Gordon ( that with positive energy).

It is noted that this interpretation is possible if a field explorer could detect and distinguish the "disposition of the coefficients" of the wave function, the direction of phase rotation and the direction of movement of the particle in space.

In this case, a explorer field would see a particle coming "right", with negative energy and a clockwise rotation of the phase, as if it is a particle going to the "left", with positive energy and a value (-1) of the variable Q, but different from the one (Q = +1) associated with a particle going to the "left", though originally positive energy!

Thus it is shown that the physical characteristic, designated as "electrical charge (Q = ±1), expresses the relationship between the direction of rotation of the phase and those aspects of a particle expressed in the coefficients of the wave function.

Si precisa che solamente attraverso la struttura a semi-quanti di un IQuO, è possibile individuare la direzione della rotazione di fase associata all'oscillazione: questa possibilità è proprio la "carica elettrica" (Q).

Sempre in questa sezione, si nota poi che la correlazione tra il segno della carica elettrica e il senso di rotazione di fase è in perfetta coerenza con l'invarianza di gauge delle interazioni elettromagnetiche.

La sezione viene conclusa facendo notare che un campo scalare reale, elettricamente neutro, è espressa da matrici a coefficienti "misti" nei due versi di rotazione della fase.

Pertanto si formula l'ipotesi dell'esistenza di un meccanismo che permette al campo esploratore di rilevare il verso di rotazione della fase delle funzioni d'onda e, attraverso alcuni indicatori relativi alla struttura a semi-quanti, "di interpretare" le particelle ad energia negativa (rotazione di fase in senso orario e coefficienti negativi ($\mathbf{\Phi}_-$, $\mathbf{X}_-$) come particelle ad energia positiva (antiparticella).

Nell'ultima sezione, riprendendo l'ipotesi della rotazione "interna" della fase vengono individuate ulteriori rotazioni "interne" ad una particella "composta" qual è un adrone.

Considerando difatti la struttura a multipletti di un adrone e la sua realtà di particella composta da quark, viene rilevato che la struttura a semi-quanto evidenzia uno spazio interno in ogni particella elementare (sia adrone sia quark) se alla fase, all'isospin e allo spin, vengono associate delle rotazioni "interne", di cui la carica elettrica (Q), l'isospin (T) e lo spin ne sono i generatori.

L'ipotesi di spazio interno diviene più calzante se si interpreta l'Ipercarica Y, formulata da Gell-Mann per descrivere la varietà di multipletti adronici ( Mesoni e Barioni), come indice di una struttura interna organizzata di IQuO-Quark (negli adroni) o di Sub-IQuO all'interno di un singolo quark!

---

Is thus made clear that only through the structure by semi-quanta of an IQuO, you can find the direction of the phase rotation associated with the oscillation: this possibility is precisely the "electric charge" (Q).

Always on this section, then you notice that the correlation between the sign of electric charge and the direction of phase rotation is in perfect coherence with the gauge invariance of electromagnetic interactions.

The section is concluded by noting that a real scalar field, electrically neutral, is expressed by matrices with coefficients "mixed" in the two directions of rotation of the phase.

Therefore, it formulates the hypothesis of a mechanism allowing for explorer fields to detect the direction of phase rotation of the wave functions and, through a number of indicators relating to the structure in a semi-quanta, allowing "to interpret" the particles with negative energy (or phase rotation in a clockwise direction and negative coefficients ($\mathbf{\Phi}_-$, $\mathbf{X}_-$) ) as the particles with positive energy (antiparticle).

In the last section, taking up the idea of "internal" rotation of the phase, further "internal" rotations into a "composed" particle (as is a hadron) are identified.

In fact, taking in consideration that hadrons are particles composed by quarks and that have a structure by multiplet , we find that the semi-quanta structure shows a space within each elementary particle ( even the quarks ) if we associate to the phase, to the isospin and to the spin, some internal rotations, of which the electrical charge (Q), the isospin (T) and the spin are the generators.

The hypothesis of internal space becomes more appropriate if one interprets the hypercharge Y, formulated by Gell-Mann to describe the variety of hadron multiplets (mesons and baryons), as an index of an internal structure organized of IQuO-Quark (in hadrons) or Sub-IQuO within a single quark!

| | |
|---|---|
| Rilevando poi le strette connessioni tra Q, T e Y si formula la matrice di calcolo della carica elettrica che applicata ad una matrice rappresentativa di un particolare IQuO, costruita attraverso il meccanismo di accoppiamento definito nella sezione 3, determina valori di carica elettrica non interi, corrispondenti a quelli possibili assegnati ad un quark.<br><br>Infine, si congettura una sub-struttura nei quark e si definisce in termini probabilistici il valore non intero della loro carica elettrica. | Noting after the close connections between Q, T and Y, we construct the matrix calculation of the electric charge. Using the coupling mechanism described in Section 3, we formulate the form of the representative matrix of a quark. Using the matrix calculation of Q the values are determined of electric charge of a IQuO-Quark. This non-integer values are identical to those assigned to precisely one quark.<br>Finally, we conjecture a sub-structure of the quarks and we define in probabilistic terms the non-integer value of their electrical charge. |

| **Par. 1) La forma a due componenti degli operatori (a,a⁺) di un oscillatore quantistico** | **Sect. 1) The form by two-component of the operators (a, a⁺) of quantum oscillator** |
|---|---|
| Consideriamo un oscillatore classico nelle coordinate (q, p) del piano delle fasi, con Hamiltoniana | We consider a classical oscillator in (q, p) coordinates of the phase plain, having Hamiltonian |

$$H = \left(\frac{p^2}{2m}\right) + \left(\frac{m\omega_0^2 q^2}{2}\right) \quad (1.1)$$

| ed equazioni del moto | and equations of motion |
|---|---|

$$\begin{cases} \dfrac{dp}{dt} = -m\omega_0^2 q \\ \dfrac{dq}{dt} = \left(\dfrac{p}{m}\right) \end{cases} \quad (2.1)$$

| Le soluzioni sono date da combinazioni lineari di due "oscillazioni" indipendenti: | The solutions are given by linear combinations of two independent oscillations |
|---|---|

$$\begin{cases} q(t) = q(0)\cos(\omega_0 t) + \left(\dfrac{p(0)}{m\omega_0}\right)\sin(\omega_0 t) \\ p(t) = p(0)\cos(\omega_0 t) - (m\omega_0)q(0)\sin(\omega_0 t) \end{cases} \quad (3.1)$$

| In presenza di smorzamento (Γ≠0) la soluzione generale può essere data da | in damped oscillator (Γ≠0) the general solution is |
|---|---|

$$\left(q(t)_{smorz}\right) = \left(e^{-(1/2)\Gamma t}\right)\left[A_s \sin(\omega_s t) + B_s \cos(\omega_s t)\right] \quad (4.1)$$

| con $\omega_s \neq \omega_0$ e q(t)$_{sm}$ una qualsiasi coordinata. Da notare che si evidenziano due "termini". L'energia dell'oscillatore sarà: | Where $\omega_s \neq \omega_0$ and q(t)$_{sm}$ is any coordinated. Note that there are two terms. The energy of the oscillator is: |
|---|---|

$$E(t) = E_0\left(e^{-\Gamma t}\right) = \left(e^{-\Gamma t}\right)\left[\frac{1}{2}M\left(\omega_s^2 + \omega_0^2\right)\left(\frac{1}{2}A_s^2 + \frac{1}{2}B_s^2\right)\right] \quad (5.1)$$

| Se (Γ≈ 0) il termine esponenziale di decadimento rimane approssimativamente costante durante un'oscillazione; in questo caso si ha: | If (Γ≈ 0) the exponential decay term is approximately constant during an oscillation, in which case we have: |
|---|---|

$$\left(q(t)_{smorz}\right) \approx (\cos t)\left[A_s \sin(\omega_s t) + B_s \cos(\omega_s t)\right] \quad (6.1)$$

| e l'energia diventa (con $\omega_s \approx \omega_0$) | The energy (with $\omega_s \approx \omega_0$) is |
|---|---|

$$E(t) \approx E_0 = \left[\frac{1}{2}M\left(2\omega_s^2\right)\left(\frac{1}{2}A_s^2 + \frac{1}{2}B_s^2\right)\right] = \frac{1}{2}M\omega_s^2\left(\frac{1}{2}A_s^2 + \frac{1}{2}B_s^2\right) \quad (7.1)$$

| Notare ancora la presenza di due componenti energetiche. | Note here again two energy components. |
|---|---|
| Una soluzione a due termini la ritroviamo anche nell'oscillatore forzato (con smorzamento). | A solution with two terms is given also in the driven oscillator (always with damping). |
| Considerando uno stato a regime, lontano dalla fase iniziale, l'oscillatore si è ormai adattato alla forza esterna oscillando con frequenza $\omega_a = \omega_{est} \neq \omega_0$ e manifesta una soluzione a due componenti: una assorbitiva ($A_{ass}$), l'altra elastica ($A_{el}$). | Away from the initial phase, the oscillator is adapted to the external force, it oscillates with frequency $\omega_a = \omega_{est} \neq \omega_0$ and it show a solution with two components: an absorptive ($A_{abs}$), the other elastic ($A_{el}$). |
| Si ha: | It's: |

$$(q(t)_{adattato}) = [A_{ass}\sin(\omega_a t) + B_{el}\cos(\omega_a t)] \quad (8.1)$$

| con un'energia media ( in un ciclo) data da: | with a mean energy (in a cycle ) given by: |
|---|---|

$$\langle E(t)\rangle_{T_a} = \left[\frac{1}{2}M\left(\omega_a^2 + \omega_0^2\right)\left(\frac{1}{2}A_{ass}^2 + \frac{1}{2}A_{el}^2\right)\right] \quad (9.1)$$

| espressa attraverso le due componenti assorbitiva ed elastica! | So energy is expressed through the two absorptive and elastic components! |
|---|---|
| Se ($\Gamma \approx 0$ e ($\omega_a = \omega_{est}$) $\approx \omega_0$) avremo lo stesso un modo adattato (risonanza) con energia ancora data da due componenti: | If $\Gamma \approx 0$ and ($\omega_a = \omega_{est}$) $\approx \omega_0$ the oscillator has an adapted oscillation (resonance) with energy to two components: |

$$\langle E(t)\rangle_{T_a} \approx \left[\frac{1}{2}M\left(\omega_a^2\right)\left(A_{ass}^2 + A_{el}^2\right)\right] \quad (10.1$$

| Con $T_a = 2\pi/\omega_a$. | Where $T_a = 2\pi/\omega_a$. |
|---|---|
| Consideriamo infine un oscillatore appartenente ad una catena di oscillatori accoppiati elasticamente. | Finally, consider an oscillator belonging to a chain of elastically coupled oscillators. |
| Pensiamo che ogni oscillatore della catena sia un oscillatore "forzato" periodicamente dagli oscillatori attigui con frequenza $\omega_{est} \neq \omega_0$ (frequenza propria del singolo oscillatore) imposta da una sorgente "lontana". | We think that each oscillator in the chain is as a driven oscillator by the neighboring oscillators with imposed frequency $\omega_{est} \neq \omega_0$ (where $\omega_0$ is frequency of the single oscillator) by a distant source. |
| Diremo che il sistema di oscillatori accoppiati si adatta così alla frequenza esterna ($\omega_{est} = \omega_a$). | We'll say that the system of coupled oscillators has adapted well to the external frequency ($\omega_{est} = \omega_a$) of the source. |
| Ne segue che il singolo oscillatore di campo diventa un oscillatore "adattato" a qualsiasi frequenza $\omega_a$ con un coefficiente di smorzamento $\Gamma \approx 0$. | It follows that the single-oscillator of field is like an adapted oscillator to any frequency $\omega_a$ having a damping coefficient $\Gamma \approx 0$. |
| In conclusione un oscillatore adattato può essere descritto sempre da due componenti oscillanti. | Finally the oscillator of field is an adapted oscillator which is described by two oscillating components. |
| Passando alle (**a**, **a**⁺) coordinate complesse, con ($\omega_a = \omega$) | Turning to (**a**, **a**⁺) complex coordinated with ($\omega_a = \omega$) |

$$a = \left[\left(\sqrt{\frac{m\omega}{2\hbar}}\right)\left(q + \frac{ip}{m\omega}\right)\right] \quad , \quad a^+ = \left[\left(\sqrt{\frac{m\omega}{2\hbar}}\right)\left(q - \frac{ip}{m\omega}\right)\right] \quad (11.1)$$

| | |
|---|---|
| avremo le seguenti equazione del moto: | The equations of motion are |

$$\begin{cases} \left(\dfrac{da_t}{dt}\right) = -i\omega a_t \\ \left(\dfrac{da_t^+}{dt}\right) = i\omega a_t^+ \end{cases} \qquad (12.1)$$

| | |
|---|---|
| con soluzioni: | With solutions |

$$(a_t = a_0 \exp(-i\omega t) \quad , \quad a_t^+ = a_0^+ \exp(i\omega t) \ ) \qquad (13.1)$$

| | |
|---|---|
| Dove $a_0 = a(0)$ , $a_0^+ = a^+(0)$. | Where [$a_0 = a(0)$ , $a_0^+ = a^+(0)$]. |
| Nel piano delle coordinate (q, p) le soluzioni (**a$_t$**, **a$_t$**⁺) potrebbero essere correlate ad una coppia di vettori (**Oa**, **Oa**⁺) con rotazione di fase opposta. | In the coordinate plane (q, p) the solutions (**a$_t$**, **a$_t$**⁺) could be related to a pair of vectors (**Oa**, **Oa**⁺) with opposite phase rotation. |

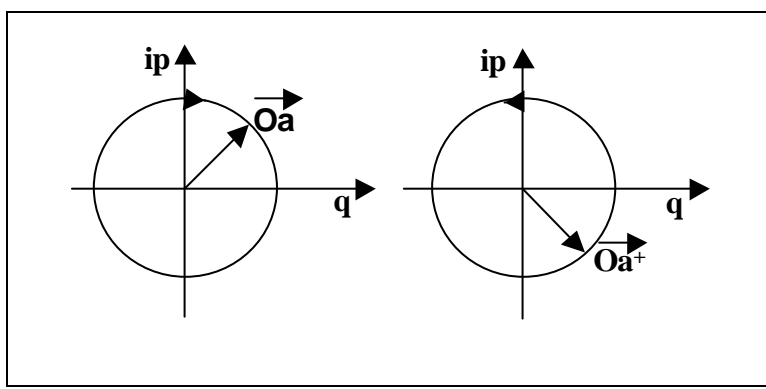

Fig. 1

Le due soluzioni possono essere connesse con due opposte rotazioni della fase.

The two solutions can be connected with the two opposite rotations of the phase.

Nella meccanica classica queste due rotazioni non sono evidenziate.

In classical mechanics this two "rotations" aren't underlined.

Le due possibilità sono nascoste: un pendolo che oscilla a ritroso nel tempo (rotazione oraria della curva rappresentativa in fig. 1) è indistinguibile da uno che oscilla avanti nel tempo (rotazione antioraria ).

The two possibilities are hidden: an oscillating pendulum going back in time (clockwise rotation of the representative curve in Fig.1) is indistinguishable by the one going ahead in time (anticlockwise rotation).

Noi crediamo che "un'originale fisica" potrebbe esistere in cui queste due differenti possibilità (stati) potrebbero essere evidenziate in qualche fenomeno fisico sia "simultaneamente" sia separatamente.

We believe that "an original physics" could exists in which these two different possibilities (states) could be revealed in some physical phenomenon.

In questa nuova e originale fisica le due differenti possibilità "in simultanea" dovranno essere espresse attraverso una "differenza" tra i vettori ($\mathbf{Oa}$, $\mathbf{Oa}^+$) della rotazione di fase:

In this new "original physics" the two different possibilities are expressed by difference between vectors ($\mathbf{Oa}$, $\mathbf{Oa}^+$) of phase rotation:

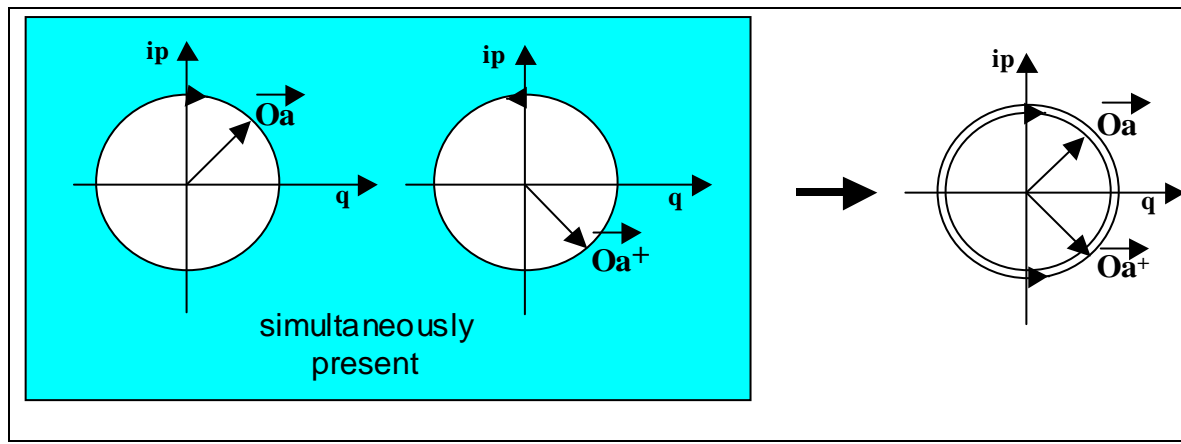

simultaneously present

Fig. 2

Questa nuova fisica è la fisica quantistica, dove i singoli stati (possibilità o località) di un sistema possono esistere singolarmente oppure, in termini complementari, insieme nello stato di sovrapposizione (non località).

The new physics is the quantum mechanics, where the possible local states of a system are individually existing, but, in complementary terms, they can exist also in the superposition state (non-locality).

In questa "nuova" fisica può accadere anche che una particolare non località (quella degli stati degeneri) in particolari condizioni possa manifestarsi nelle sue componenti individuali (vedi livelli energetici degeneri dell'atomo H ).

In this new physics can happen that a particular non local state (such as that of degenerate states) in some conditions may show itself in its individual components (see degenerate energy levels of H atom).

Potremmo ritenere che, nell'oscillatore quantistico [1] la differenza ($\mathbf{Oa} \neq \mathbf{Oa}^+$) sia in relazione con la parentesi di commutazione (a valore non zero) degli operatori non hermitiani corrispondenti ($\mathbf{\hat{a}}$, $\mathbf{\hat{a}}^+$):

About quantum oscillator we think that [1] the ($\mathbf{Oa} \neq \mathbf{Oa}^+$) difference is related to the bracket of commutations (non-zero value) of the corresponding Hermitian operators ($\mathbf{\hat{a}}$, $\mathbf{\hat{a}}^+$):

$$\boxed{[\hat{a}, \hat{a}^+] = 1} \qquad (14.1)$$

| | |
|---|---|
| Riteniamo inoltre che anche le due diverse possibilità di rotazione della fase associate ad un oscillatore quantistico (ricordiamo che l'energia è degenere per questi due autostati) possono invece trovare, in particolari condizioni, una manifestazione osservabile.<br>Per capire quali possono essere queste condizioni rivediamo lo studio dell'oscillatore quantistico.<br>Le soluzioni [1] dell'eq. del moto nel piano delle fasi quantistico degli operatori hermitiani $(\hat{q}, \hat{p})$ sono: | We also believe that the two different directions of rotation of the phase assigned to a quantum oscillator (note that the energy is degenerate in these two eigenstates) may instead find, under certain conditions, an observable event.<br>To discover these special conditions we must review the study the quantum oscillator.<br><br>The solutions [1] of the motion equations in the phase plane of the quantum Hermitian operators $(\hat{q}, \hat{p})$ are: |

$$\boxed{\begin{cases} \hat{q}(t) = \hat{q}(0)\cos\omega t + \left(\dfrac{\hat{p}(0)}{m\omega}\right)\sin\omega t \\ \hat{p}(t) = \hat{p}(0)\cos\omega t - (m\omega)\hat{q}(0)\sin\omega t \end{cases}} \qquad (15.1)$$

| | |
|---|---|
| Se consideriamo un oscillatore "adattato" ad un agente esterno, le soluzioni ( vedi la (8.1)) dell'oscillatore classico sono: | In an adapted oscillator to a external agent, the solutions (see eq. (8.1)) of classical oscillator are: |

$$\boxed{q(t)_{adattato} = q_{el}(0)\cos(\omega t) + q_{ass}(0)\sin(\omega t)} \qquad (16.1)$$

| | |
|---|---|
| per il corrispondente oscillatore quantistico, espresso tramite operatori, avremo: | In quantum oscillator, expressed by operators, we well have: |

$$\boxed{\hat{q}(t) = \hat{q}_{el}(0)\cos(\omega t) + \hat{q}_{ass}(0)\sin(\omega t)} \qquad (17.1)$$

| | |
|---|---|
| dove $q_{el}$ è la componente elastica in fase con la forza esterna, mentre $q_{ass}$ è in ritardo di fase di $\pi/2$ con la forza esterna.<br>Noi ricordiamo tuttavia che qualunque oscillatore ha due componenti: elastica e inerziale; pertanto porremo la seguente equivalenza: $\mathbf{q_{ass} \equiv q_{inertial}}$<br>Segue che | where $q_{el}$ is the elastic component in-phase with the external force, while $q_{abs}$ is in phase delay of $\pi/2$ with the external force.<br>We remember that any oscillator has two components: elastic and inertial one; therefore we admit the following equivalence:<br>$\mathbf{q_{abs} \equiv q_{inertial}}$<br>It follows |

$$\boxed{\hat{q}(t) = \hat{q}_{el}(0)\cos(\omega t) + \hat{q}_{in}(0)\sin(\omega t)} \qquad (18.1)$$

| | |
|---|---|
| Questa relazione può anche descrivere, come abbiamo già detto, un oscillatore appartenente ad una catena di oscillatori accoppiati (campo). | This equation can also describe, as we have already said, an oscillator belonging to a chain of coupled oscillators (field). |
| Diremo allora che un **oscillatore quantistico di campo** dovrà essere sempre rappresentato a due componenti. | We say that an **oscillator quantum of field** should always be represented with two components. |
| Graficamente avremo: | The graphic representation of the quantum oscillator with two components is: |

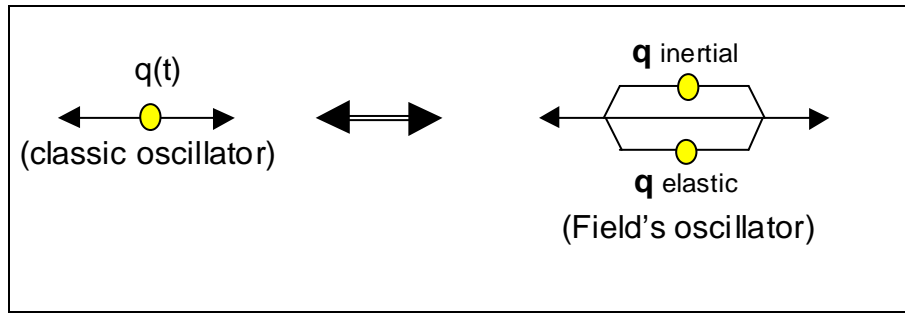

Fig.3

| | |
|---|---|
| Considerando la (11.1) espressa con operatori, le relazioni (13.1) e ricorrendo alle relazioni di Eulero si trova facilmente che: | Now we use the eq. (11.1), but by operators, and the eq. (13.1) resorting to the Euler's relations so we easily find that: |

$$\begin{cases} \hat{a}_t = \left(\sqrt{\frac{m\omega}{2\hbar}}\right)\left[\hat{q}_0 e^{-i\alpha} + i\left(\frac{\hat{p}_0}{m\omega}\right)e^{-i\alpha}\right] = \hat{k}_q e^{-i\alpha} + i\hat{k}_p e^{-i\alpha} = \hat{k}_q e^{-i\alpha} + \hat{k}_p e^{-i(\alpha - \pi/2)} \\ \hat{a}_t^+ = \left(\sqrt{\frac{m\omega}{2\hbar}}\right)\left[\hat{q}_0 e^{i\alpha} - i\left(\frac{\hat{p}_0}{m\omega}\right)e^{i\alpha}\right] = \hat{k}_q e^{i\alpha} - i\hat{k}_p e^{i\alpha} = \hat{k}_q e^{i\alpha} + \hat{k}_p e^{i(\alpha - \pi/2)} \end{cases} \quad (19.1)$$

| | |
|---|---|
| dove [q$_0$= q(0), p$_0$= p(0)] e che | where it's [q$_0$= q(0), p$_0$= p(0)] and |

$$\begin{cases} \left(\hat{k}_q = \left(\sqrt{\frac{m\omega}{2\hbar}}\right)\hat{q}_0, \quad \hat{k}_p = \left(\sqrt{\frac{1}{2m\hbar\omega}}\right)\hat{p}_0\right) \\ \left[\hat{k}_q\hat{k}_p - \hat{k}_p\hat{k}_q\right] = \frac{i}{2} \\ \left[\hat{a}_0, \hat{a}_0^+\right] = 1 \end{cases} \quad (20.1)$$

| | |
|---|---|
| Notiamo così che anche gli operatori (**â, â⁺**) possono essere espresse da due componenti (vedi anche l'eq. 18.1): una elastica l'altra inerziale. | Thus, we note that the (**â, â⁺**) operators can be expressed by two components (see also Eq. 18.1): an elastic component and an inertial another. |
| Pertanto l'eq. (19.1) può essere espressa con queste due nuove componenti: | We could say then that the operators ($\hat{\mathbf{a}}$, $\hat{\mathbf{a}}^+$) will be expressed by the following equation: |

$$\begin{cases} \hat{a}_t = \hat{k}_{(elastic)} e^{-i\omega t} + \hat{k}_{(inertial)} e^{-i(\omega t - \pi/2)} \\ \hat{a}_t^+ = \hat{k}_{(elastic)} e^{i\omega t} + \hat{k}_{(inertial)} e^{i(\omega t - \pi/2)} \end{cases} \qquad (21.1)$$

| | |
|---|---|
| Facilmente si dimostra (considerando le relazioni presenti nell'eq. (20.1) che $[\hat{\mathbf{a}}_t, \hat{\mathbf{a}}_t^+] = 1$.<br><br>Potremmo affermare che gli operatori ($\hat{\mathbf{a}}$, $\hat{\mathbf{a}}^+$) siano espressi dalla seguente generalizzazione dell'eq. 21.1): | It's easy to show (taking into account the present relations in Eq. (20.1) that $[\hat{\mathbf{a}}_t, \hat{\mathbf{a}}_t^+] = 1$.<br><br>We could say then that the operators ($\hat{\mathbf{a}}$, $\hat{\mathbf{a}}^+$) will be expressed by the following equation (you see eq. 21.1): |

$$\begin{cases} \hat{a}_t = \hat{a}(t)_{elastic} + \hat{a}(t)_{inertial} \equiv O\vec{a} \\ \hat{a}_t^+ = \hat{a}^+(t)_{elastic} + \hat{a}^+(t)_{inertial} \equiv O\vec{a}^+ \end{cases} \Leftrightarrow \begin{cases} \hat{a}_t = \hat{a}_{(el)} e^{-i\omega t} + \hat{a}_{(in)} e^{-i(\omega t - \pi/2)} \\ \hat{a}_t^+ = \hat{a}^+_{(el)} e^{i\omega t} + \hat{a}^+_{(in)} e^{i(\omega t - \pi/2)} \end{cases} \qquad (22.1)$$

| | |
|---|---|
| Che è la generalizzazione della (21.1)<br>Osserviamo bene che la componente elastica è in ritardo di fase di $\pi/2$, in confronto con quella inerziale.<br>D'ora in poi ricorreremo ad un oscillatore a due componenti per descrivere un oscillatore di campo associato ad una particella.<br>Sospettiamo che proprio in un oscillatore di campo si potrebbero verificare, per la sua struttura a due componenti, quelle condizioni "particolari" che permetterebbero la rilevazione del verso di rotazione della fase.<br>Se consideriamo la fig. 2, una successiva configurazione nel tempo sarà ( vedi eq. 22.1 per $\omega t = (3/4)\pi$): | This is the generalization of Eq. (21.1)<br>Let's observe well that the elastic component is in phase delay of $\pi/2$, in comparison to inertial one!<br>We will resort to an oscillator with two components to describe an oscillator of field associated with a particle.<br>We conjecture that the oscillator field, with two components, could allow to us, , to detect the phase rotation during the interactions between particles.<br><br>We give the following representation of quantum oscillator with two components (you see fig. 2 and eq. 22.1) if we set that $\omega t = (3/4)\pi$): |

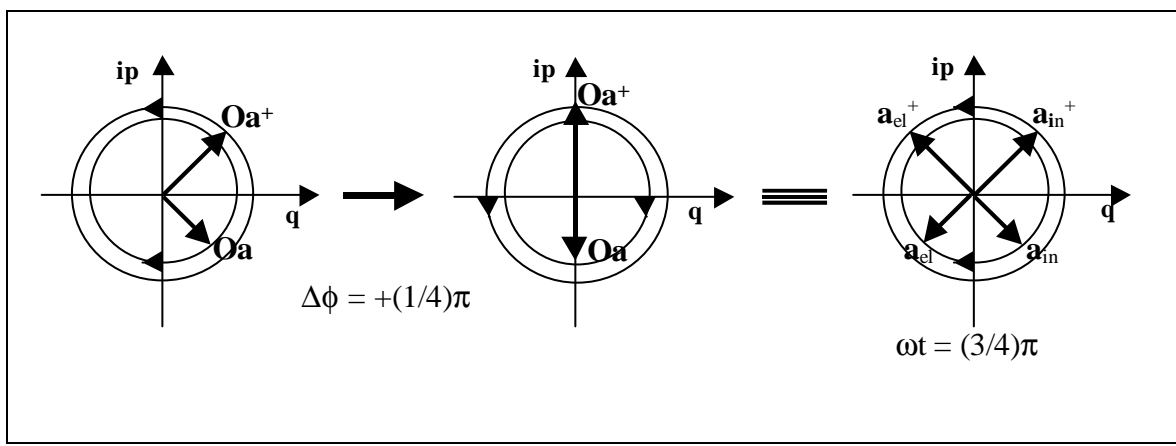

Fig. 4

| | |
|---|---|
| Noi pensiamo che un oscillatore quantistico di campo sia a due componenti anche nello stato di vuoto.<br><br>Pertanto lo stato di vuoto (con energia $\varepsilon = 1/2\ \hbar\omega$) sarà a due componenti ( una elastica e l'altra inerziale ) ciascuna con energia $\varepsilon = 1/4\ \hbar\omega$.<br><br>Definiamo " **semi-quanto pieno (●)**" il quanto d'oscillazione con $\varepsilon = ½\ \hbar\omega$, mentre definiamo " semi-quanto vuoto (**o**)" il quanto con $\varepsilon = 1/4\ \hbar\omega$.<br><br>Noi ricordiamo che l'operatore **â** è un operatore di annichilazione mentre **â⁺** è di creazione.<br><br>L'operatore Hamiltoniano è: | We think that a quantum oscillator is with two oscillating components also in the empty fundamental state.<br><br>Therefore the vacuum state (with energy $\varepsilon = 1/2\ \hbar\omega$) is at two components, everyone with energy $\varepsilon = 1/4\ \hbar\omega$<br><br>We define **"full semi-quanta (●)"** the quanta of oscillation with energy $\varepsilon = 1/2\ \hbar\omega$ while we define **"empty semi-quanta (o)" "** the quanta with $\varepsilon = 1/4\ \hbar\omega$<br><br>We remember that the (**â**) operator is an annihilation operator while (**â⁺**) is a creation operator.<br><br>The Hamiltonian operator is: |

$$\hat{H} = \left( \hat{a}^{+}\hat{a} + \frac{1}{2} \right)\hbar\omega = \left( \hat{n} + \frac{1}{2} \right)\hbar\omega \qquad (23.1)$$

| | |
|---|---|
| dove **n** è l'operatore numero di quanti.<br><br>Avendo autovettori simultaneamente definibili sia (**H**) che (**n**), gli operatori (**H**) e (**n**) commutano.<br><br>Infatti, gli autovalori dell'op. hamiltoniano sono: | where (**n**) is the operator of quanta number.<br><br>The operators (**H**) and (**n**), having the simultaneously definable eigenvectors, are commuting.<br><br>In fact the eigenvalues of Hamiltonian operator are: |

$$E_n = \left( n + \frac{1}{2} \right)\hbar\omega \qquad (24.1)$$

| | |
|---|---|
| Dove n (1,2….) sono gli autovalori dell'op. (**n**).<br><br>Se al verso di rotazione della fase associamo, in qualunque tipo d'oscillazione, un operatore (**r**) ( con autovalori r' = ±1 ), dovremmo ammettere che esso deve commutare con l'op. (**H**) e di conseguenza con l'op. (**n**).<br><br>L'energia dell'oscillatore quantistico "isolato", degenere negli autovalori di (**r**) ovvero nel verso della rotazione di fase [2], sarà: | where n (1,2,…) are the eigenvalues of (**n**) operator.<br><br>In any type of oscillation, (**r**) is an operator assigned to the direction of phase rotation: this operator has the eigenvalues r' = ±1 and commutes with the op. (**H**) and, accordingly, with the op. (**n**).<br><br>In the quantum isolated oscillator the energy can be degenerated in eigenvalues of (**r**) as it is degenerated in the direction of phase rotation. We will have [2]: |

$$\left[\left[ E_n = \left( n + \frac{1}{2} \right)\hbar\omega \right]\right]_{(r'=\pm 1)} \qquad (25.1)$$

| | |
|---|---|
| Si può cosi definire uno spazio vettoriale **S(H,n)** $_{(r'=\pm1)}$ (degenere negli autovalori dell'op. **r**) con autovettori di base relativi agli op. **(H, n)**$_{(r'=\pm1)}$ dove sarà [$(\|\Psi\rangle)_{(r'=\pm1)} \equiv (\|\Psi_H\rangle \|\Psi_n\rangle)_{(r'=\pm1)}$ ]. | Therefore we define a **S(H,n)** $_{(r'=\pm1)}$ vectorial space with base eigenvectors of **(H, n)**$_{(r'=\pm1)}$ operators where is [$(\|\Psi\rangle)_{(r'=\pm1)} \equiv (\|\Psi_H\rangle\|\Psi_n\rangle)_{(r'=\pm1)}$ ]. |
| Chiameremo **"IQuO"** l'oscillatore di campo a semi quanti dove IQuO è l'acronimo di "**I**ntrinsec **Qu**antum **O**scillator". | We'll call **"IQuO"** the field oscillator defined by semi-quanta, where "IQuO" is the acronym of "**I**ntrinsec **Qu**antum **O**scillator". |
| **Par. 2 ) The Intrinsec Quantum Oscillator (IQuO)** | **Sect. 2) The Intrinsec Quantum Oscillator (IQuO)** |
| Dirac ([2] pp 156-157) ha tentato di proporre le seguenti relazioni per gli operatori (**a**, **a**$^+$): | Dirac ([ 2] pp 156-157) tried to propose the following relations for (**a**, **a**$^+$) operators: |

$$\begin{cases} \hat{a} = \big(\exp(i\hat{\varphi})\big)\big(\sqrt{\hat{n}}\,\big) \\ \hat{a}^+ = \big(\sqrt{\hat{n}}\,\big)\big(\exp(-i\hat{\varphi})\big) \end{cases} \qquad (1.2)$$

| | |
|---|---|
| dove viene introdotto l'operatore angolo di fase (**φ**). Tuttavia risulta evidente che questo operatore (**φ**) non commuta con gli operatori **H**, **n** ed **r**: l' angolo di fase (φ) è indeterminato se l'autovalore (n) è determinato. | Where it is introduced the **φ**-operator of the phase angle. However, it's evident that this operator (**φ**) does not commute with the operators (**H**), (**n**) and (**r**): the phase angle [θ = ωt] is uncertain if instead the (n)-eigenvalue of (**n**) operator is certain. |
| Segue che se l'energia è determinata, la fase deve essere indeterminata. | Note that if the energy of oscillator (degenerated in eigenvalue of (**r**)) is certain then the phase is uncertain. |
| Come è ben dimostrato (vedi Glover [2]) queste relazioni conducono ad alcune contraddizioni; ciò induce a considerare la fase non un operatore ma una variabile correlata all'op. **n**. | However, the eq. (1.2) (see Glover [2]) leads to some contradictions; this push us to see the phase as a physical variable and no operator. |
| In [2] si definisce l'operatore **Φ** funzionale: | In [2] is defined the functional operator (**Φ**) |

$$\hat{\Phi}(\varphi) = \exp(i\varphi) \quad , \quad \hat{\Phi}^+(\varphi) = \exp(-i\varphi) \qquad (2.2)$$

| | |
|---|---|
| tale che al posto delle (1.2) si abbia: | Which transforms the eq. (1.2) in |

$$\begin{cases} \hat{a} = \big(\exp(i\varphi)\big)\big(\sqrt{\hat{n}}\,\big) \\ \hat{a}^+ = \big(\sqrt{\hat{n}}\,\big)\big(\exp(-i\varphi)\big) \end{cases} \qquad (3.2)$$

| | |
|---|---|
| che conduce ad avere le variabili (n), (φ) correlate da una relazione di indeterminazione. | Where the variables (n) and (φ) are related by an uncertainty relation. |
| Tuttavia in questa trattazione con le variabili (n), (φ) ( vedi il lavoro di Fain citato nella bibl. **[2]** ) si rileva ancora una volta che l'energia risulta degenere al segno di (φ). | However, (see the work of Fain in bibl. [2]) it is not clear why the energy is degenerate in the sign of (φ). |
| Se invece consideriamo un oscillatore di campo a due componenti possiamo pensare che la sua "duplice" struttura possa permetterci di rilevare il verso di rotazione della fase relativa all'operatore (**r**). | Instead we think that the oscillator of field with two components allows to understand the degenerate energy in sign of (φ) why it allows of detect the direction of rotation of the phase through the introduction of the operator (**r**). |
| A tal fine introduciamo un'originale forma dell'operatore (**Φ**) in cui figuri la presenza dell'operatore (**r**) accanto alla variabile di fase (φ) . | For this aim an original form of the operator (**Φ**) is introduced where the operator (**n**) and (**r**), next to the variable phase (φ), are defined. |
| Si propone la forma: | It is proposed the form: |

$$\begin{cases} \hat{a} = (\hat{\Phi}(\hat{r}, \varphi))\left(\sqrt{\hat{n}}\ \right) \\ \hat{a}^+ = \left(\sqrt{\hat{n}}\ \right)(\hat{\Phi}^+(\hat{r}, \varphi)) \end{cases} \qquad (4.2)$$

| | |
|---|---|
| In cui la funzione operatore (**Φ**) consiste in | In which the Φ-operator function consists in |

$$\hat{\Phi}(\hat{r}, \varphi) = \exp(i\hat{r}\varphi) \quad , \quad \hat{\Phi}^+(\hat{r}, \varphi) = \exp(-i\hat{r}\varphi) \qquad (5.2)$$

| | |
|---|---|
| Ritornando all'oscillatore con due componenti, seguirebbe (vedi eq. 21.1) | In oscillator with two components, it follows (see eq 21.1). |

$$\begin{cases} \hat{a} = (\Phi(\hat{r}, \varphi))(\sqrt{\hat{n}}) = \left[\left(e^{-i\hat{r}\alpha t}\right)\left(\hat{k}_{el} + i\hat{k}_{in}\right)\right] \\ \hat{a}^+ = (\sqrt{\hat{n}})\Phi*(\hat{r}, \varphi) = \left[\left(\hat{k}_{el} - i\hat{k}_{in}\right)\left(e^{i\hat{r}\alpha t}\right)\right] \end{cases} \qquad (6.2)$$

| | |
|---|---|
| Ma | But |

$$\begin{cases} \hat{a}^+\hat{a} = \hat{n} = (\sqrt{\hat{n}})(\sqrt{\hat{n}}) \\ \hat{a}^+\hat{a} = \hat{n} = \left[\left(\hat{k}_{el} - i\hat{k}_{in}\right)\left(e^{i\hat{r}\alpha t}\right)\right] \cdot \left[\left(e^{-i\hat{r}\alpha t}\right)\left(\hat{k}_{el} + i\hat{k}_{in}\right)\right] = \left(\hat{k}_{el} - i\hat{k}_{in}\right)\left(\hat{k}_{el} + i\hat{k}_{in}\right) \end{cases} \qquad (7.2)$$

| | |
|---|---|
| dove osserviamo che l'ultima uguaglianza è in contraddizione con la prima perché $(k_{el}-ik_{in}) \neq (k_{el}+ik_{in})$. | Note that the last equality implicates a contradiction with the first because it's $(k_{el}-ik_{in}) \neq (k_{el}+ik_{in})$. |
| Per evitare questa contraddizione ricordiamo quanto ipotizzato in precedenza: la differenza tra i vettori (**Oa** ≠ **Oa**⁺) trova espressione nella parentesi di commutazione [**â, â⁺**] a valore non nullo. | To avoid this contradiction we remember than assumed previously: the difference (**Oa** ≠ **Oa**⁺) between vectors (**Oa , Oa**⁺) is contained implicitly in the bracket of commutations [**â, â⁺**] (non-zero value). |
| Poiché la non commutazione [**â, â⁺**] è una realtà fisica allora la differenza tra i vettori (**Oa** ≠ **Oa**⁺) deve poter essere ammessa. | Because the no commutations [**â, â⁺**] is a physical reality then the difference (**Oa** ≠ **Oa**⁺) should be admitted. |
| Ciò significa che possiamo distinguere i moduli di questi operatori ovvero possiamo distinguere l'operatore $\left(\sqrt{\hat{n}}\right)$ contenuto in (**â**) da quello contenuto in (**â⁺**); congetturiamo allora la seguente forma: | This means that we can distinguish the norm of these operators or we can distinguish the $\left(\sqrt{\hat{n}}\right)$ operator contained in (**â**) from that contained in (**â⁺**); then we conjecture the following form: |

$$\begin{cases} \hat{a} = \hat{\Phi}(\hat{r},\varphi)\left(\sqrt{\hat{n}_+}\right) \\ \hat{a}^+ = \left(\sqrt{\hat{n}_-}\right)\left(\hat{\Phi}^+(\hat{r},\varphi)\right) \end{cases} \qquad (8.2)$$

| | |
|---|---|
| Dove | Where |

$$\sqrt{\hat{n}_+} = (\hat{k}_{el} + i\hat{k}_{in}) \;\; ; \sqrt{\hat{n}_-} = (\hat{k}_{el} - i\hat{k}_{in}) \qquad (9.2)$$

| | |
|---|---|
| Con | With |

$$\begin{cases} \sqrt{\hat{n}_+} \neq \sqrt{\hat{n}_-} \\ \left(\sqrt{\hat{n}_+}\right)^+ = \sqrt{\hat{n}_-} \Leftrightarrow \left(\sqrt{\hat{n}_-}\right)^+ = \sqrt{\hat{n}_+} \end{cases} \qquad (10.2)$$

| | |
|---|---|
| Avremo che: | We'll have |

$$\left(\sqrt{\hat{n}_+}\right)\left(\sqrt{\hat{n}_-}\right) = (\hat{k}_{el} + i\hat{k}_{in}) \;\; (\hat{k}_{el} - i\hat{k}_{in}) = \left(\hat{k}_{el}^2 + \hat{k}_{in}^2\right) - \left(\frac{1}{2}\right) \qquad (11.2)$$

| | |
|---|---|
| Così troviamo dalle (6.2) e (8.2) che | So we find ( by eq. 6. 2 and eq. 8.2) that |

$$\begin{cases} \hat{a}^+\hat{a} = \left[\left(\hat{k}_{el}\right)\left(e^{i\hat{r}\alpha}\right) - \left(i\hat{k}_{in}\right)\left(e^{i\hat{r}\alpha}\right)\right]\left[\left(e^{-i\hat{r}\alpha}\right)\hat{k}_{el} + \left(ie^{-i\hat{r}\alpha}\right)\hat{k}_{in}\right] = \left[\left(\sqrt{\hat{n}_+}\right)\left(\sqrt{\hat{n}_-}\right)\right] \\ \hat{a}^+\hat{a} = \hat{n} \end{cases} \qquad (12.2)$$

| | |
|---|---|
| Confermando che | Confirm so that |

$$\hat{n} = \left[\left(\sqrt{\hat{n}_+}\right)\left(\sqrt{\hat{n}_-}\right)\right] \qquad (13.2)$$

| In questo modo le (8.2) diventano: | In this way the eq. (8.2) become: |

$$\begin{cases} \hat{a} = (\hat{\Phi}(\hat{r}, \varphi))(\sqrt{\hat{n}_+}) = \left[ \left( e^{-i\hat{r}\alpha t} \right)\left( \hat{k}_{el} + i\hat{k}_{in} \right) \right] \\ \hat{a}^+ = (\sqrt{\hat{n}_-})\hat{\Phi}^+(\hat{r}, \varphi) = \left[ \left( \hat{k}_{el} - i\hat{k}_{in} \right)\left( e^{i\hat{r}\alpha t} \right) \right] \end{cases} \qquad (14.2)$$

| La nuova forma degli operatori di Dirac (**a**, **a**+) sarà: | A new Dirac's form of the (**a**, **a**+) operators will be |

$$\begin{cases} \hat{a} = (\hat{\Phi}(\hat{r}, \varphi))\left(\sqrt{\hat{n}_+}\right) = \left[ \left( \hat{\Phi}(\hat{r}, \varphi)\right)\left(\hat{k}_{el}\right) + \left(\hat{\Phi}(\hat{r}, \varphi)\right)\left(i\hat{k}_{in}\right) \right] = \left(\hat{a}_{el} + \hat{a}_{in}\right) \\ \hat{a}^+ = \left(\sqrt{\hat{n}_-}\right)\left(\hat{\Phi}^+(\hat{r}, \varphi)\right) = \left[ \left(\hat{k}_{el}\right)\left(\hat{\Phi}^+(\hat{r}, \varphi)\right) - \left(i\hat{k}_{in}\right)\left(\hat{\Phi}^+(\hat{r}, \varphi)\right) \right] = \left(a_{el}^+ + \hat{a}_{in}^+\right) \end{cases} \qquad (15.2)$$

| Si osserva che gli operatori a due componenti (**a**, **a**+) individuano operatori prodotto di operatori "complessi" quali l'operatore (**Φ**) (con modulo reale ρ = 1) e l'operatore $\left(\sqrt{\hat{n}_\pm}\right)$.<br><br>Questa forma inoltre è coerente con le azioni degli operatori di occupazione (**â**, **â**+) sugli autovettori \|n> dell'operatore (**n**): | Note that the (**a**, **a**+) become complex operators composed by the operator (Φ) (with real norm ρ = 1) and the $\left(\sqrt{\hat{n}_\pm}\right)$ operator.<br><br>This form is also consistent with the actions of the operators (**a**,**a**+) on the eigenvectors \| **n** > of the operator (**n**): |

$$\begin{cases} \hat{a}\vert n\rangle = \left(\sqrt{n}\right)\vert n\text{-}1\rangle \\ \hat{a}^+\vert n\rangle = \left(\sqrt{n+1}\right)\vert n+1\rangle \end{cases} \qquad (16.2)$$

| Osserviamo che questa nuova forma degli operatori (**a**, **a**+) contiene un'informazione in più ($\left(\sqrt{\hat{n}_\pm}\right)$) rispetto alla "vecchia" forma.<br>**Ciò mostra che l'oscillatore a due componenti possiede uno spazio interno che in qualche fenomeno deve poter essere rilevato.**<br>Riteniamo così che il passaggio dall'oscillatore a quello quantistico ( espresso con due componenti) potrebbe permetterci di rilevare (indirettamente) i due diversi versi della rotazione di fase (autovalori dell'op. **r**).<br>In un autostato del numero di quanti e del verso di rotazione della fase (**H**, **n**, **r**) è possibile descrivere un oscillatore in termini di semiquanti pieni ( elastico e inerziale ) e vuoti! | Note that this new form of the operators (**a**,**a**+) contains more information (see the $\left(\sqrt{\hat{n}_\pm}\right)$) than the old form.<br>**This shows that the oscillator of field with two components has an internal space that some phenomenon must be detected.**<br><br>We believe so that the transition from quantum oscillator to one of field (but expressed with two components) could allow to detect (indirectly) two different directions of phase rotation (eigenvalues of op. **r** ).<br><br>Note that in an eigenstate of (**H**, **n**, **r**) is possible to describe an oscillator with of full semi-quanta (elastic and inertial) and empty ones! |

| Sempre nello spazio degli autostati S(**H,n,r**) si dimostra [1] facilmente dalle proprietà degli operatori di quanto che | Always in the eigenstate of (**H, n, r**) we easily demonstrate [1], by properties of the quantum operators, that |
|---|---|

$$\hat{U}|n\rangle = \left(\frac{1}{2}m\omega^2\hat{x}^2\right)|n\rangle = \left(\frac{1}{4}\hbar\omega\right)_U \left[\hat{a}^2 + (\hat{a}^+)^2 + \hat{a}^+\hat{a} + \hat{a}\,\hat{a}^+\right]|n\rangle$$

$$\hat{K}|n\rangle = \left(\frac{\hat{p}^2}{2m}\right)|n\rangle = \left(\frac{1}{4}\hbar\omega\right)_K \left[-\hat{a}^2 - (\hat{a}^+)^2 + \hat{a}^+\hat{a} + \hat{a}\,\hat{a}^+\right]|n\rangle$$

(17.2)

| Così calcoliamo il valore dell'energia potenziale e cinetica per l'autostato \| 0 > | So we can calculate the value of the potential and kinetics energy in \| 0 > eigenstate |
|---|---|

$$\hat{U}|0\rangle = \left(\frac{1}{2}m\omega^2\hat{x}^2\right)|0\rangle = \left(\frac{1}{4}\hbar\omega\right)_U \left[(\hat{a}^+)^2 + 1\right]|0\rangle$$

$$\hat{K}|0\rangle = \left(\frac{\hat{p}^2}{2m}\right)|0\rangle = \left(\frac{1}{4}\hbar\omega\right)_K \left[-(\hat{a}^+)^2 + 1\right]|0\rangle$$

(18.2)

| Otteniamo la forma dell'operatore hamiltoniano nello stato di vuoto | Thus the form of Hamiltonian operator is found |
|---|---|

$$\hat{H}|0\rangle = \hat{U}|0\rangle + \hat{K}|0\rangle = \left(\frac{1}{4}\hbar\omega\right)_U |0\rangle + \left(\frac{1}{4}\hbar\omega\right)_K |0\rangle$$

(19.2)

| Per gli autostati \| n > avremo | While in \|n> eigenstate it's |
|---|---|

$$\hat{U}|n\rangle + \hat{K}|n\rangle = \left(\frac{1}{4}\hbar\omega\right)_U \left[\hat{a}^+\hat{a} + \hat{a}\,\hat{a}^+\right]_U |n\rangle + \left(\frac{1}{4}\hbar\omega\right)_K \left[\hat{a}^+\hat{a} + \hat{a}\,\hat{a}^+\right]_K |n\rangle$$

(20.2)

| Dimostrando così che vale: | We demonstrate also that |
|---|---|

$$\hat{H}|n\rangle = \hat{U}|n\rangle + \hat{K}|n\rangle = \left(2\hat{n}+\hat{1}\right)\left(\frac{1}{4}\hbar\omega\right)_U |n\rangle + \left(2\hat{n}+\hat{1}\right)\left(\frac{1}{4}\hbar\omega\right)_K |n\rangle$$

(21.2)

| con autovalori  (r') in S(**H,n**) | with eigenvalues in the eigenstate S(**H,n**) such |
|---|---|

$$\left[H_{(n)}\right]_r = \left[U_{(n)} + K_{(n)}\right]_r = \left[\left(2n+1\right)\left(\frac{1}{4}\hbar\omega\right)_U + \left(2n+1\right)\left(\frac{1}{4}\hbar\omega\right)_K\right]_{r'}$$

(22.2)

| oppure | or |

$$\left(\mathrm{H}_{(n)}\right)_{r'} = \left(\mathrm{U}_{(n)} + \mathrm{K}_{(n)}\right)_{r'} = \left[\mathrm{n}\left(\left(\frac{1}{2}\hbar\omega\right)_{U}\right) + \left(\frac{1}{4}\hbar\omega\right)_{U}\right]_{r'} + \left[\mathrm{n}\left(\left(\frac{1}{2}\hbar\omega\right)_{K}\right) + \left(\frac{1}{4}\hbar\omega\right)_{K}\right]_{r'} \quad (23.2)$$

| Abbiamo qui l'ulteriore conferma della rappresentazione di un oscillatore quantistico mediante semiquanti pieni e vuoti.<br>Inoltre guardando in [3] le forme delle funzioni d'onda $\Psi(x)$ e $|\Psi(x)|^2$ associate all'oscillatore quantistico nell'autostato |1>, si può ammettere (data la presenza del nodo centrale) che esso sia composto da due unità elementari d'oscillazioni, che noi definiamo "**sub-oscillatori**" (**sub-osc.**) | Here it is further confirmation that the quantum oscillator has some full semi-quanta and some empty semi-quanta.<br>Besides, [3] looking at the forms of the wave functions $\Psi(x)$ and $|\Psi_1(x)|^2$ of the quantum oscillator in |1> eigenstate, we note that the probability function $P_1(x) = |\Psi_1(x)|^2$ can be interpreted (note the central node) as a function describing an oscillator with two basic units of oscillations (**sub-oscillators**). |

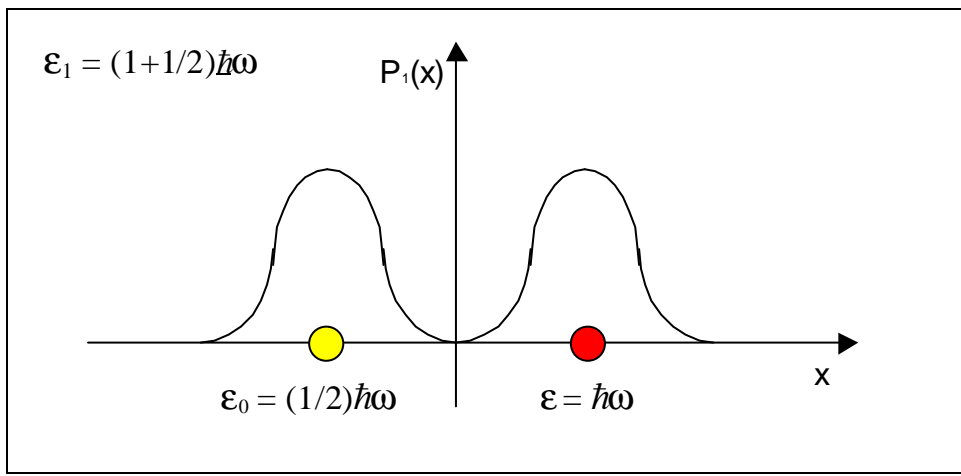

Fig. 5

| L'energia [$\varepsilon_1 = (1)\hbar\omega + (1/2)\hbar\omega$] è distribuita in due sub-osc., di cui uno (sub-quantum) è con energia [$\varepsilon = (1/2)\hbar\omega$] e l'altro ha energia [$\varepsilon = (1)\hbar\omega$]. In |0> eigenstate abbiamo [3] un solo sub-osc: | The [$\varepsilon_1 = (1)\hbar\omega + (1/2)\hbar\omega$] energy is distribute in two sub-osc., of which one (sub-quanta) with energy [$\varepsilon = (1/2)\hbar\omega$] and the other with energy [$\varepsilon = (1)\hbar\omega$]. In |0> eigenstate [3] there is only one sub-oscillator: |

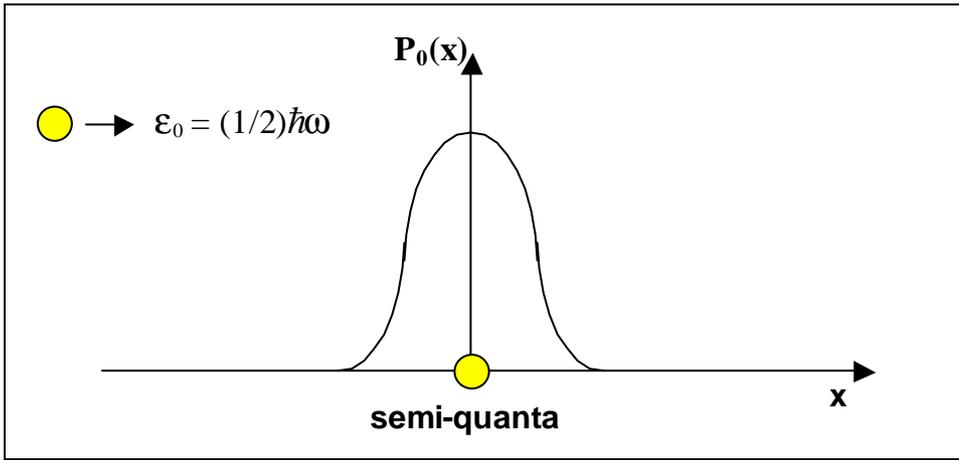

Fig. 6

| Tuttavia guardando l'eq. (21.2) e l'eq. (22.2) abbiamo: | Looking the eq. (21.2) and (22.2) one has: |
|---|---|

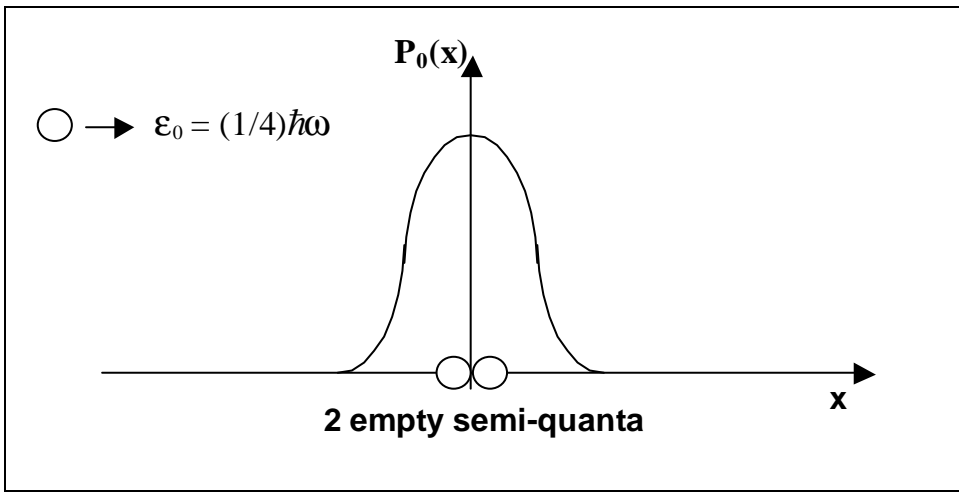

Fig. 7

| come | As |
|---|---|

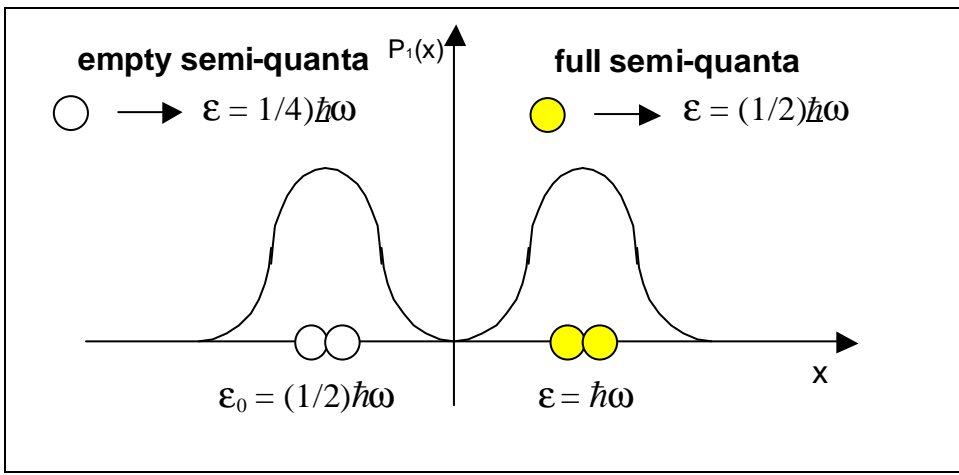

Fig. 8

| | |
|---|---|
| Invece nello stato \|n > il numero di sub-osc. accoppiati (dai nodi presenti in \|Ψ(x)\|² ) è dato da (n+1), dove n è il numero di quanti.<br><br>Riassumendo, asseriamo che un oscillatore in uno stato eccitato (n-livello) è composto da (n+1) sub-osc., con due semi-quantum vuoti ( o ) ad energia **ε = 1/4 ℏω** e n semi-quanti pieni (●) ad energia **ε = 1/2 ℏω** ciascuno (2 semi-quantum pieni sono uguali ad un quanto **ε = ℏω** ).<br><br>Anche se l'energia potenziale e cinetica nella rappresentazione degli stati d'occupazione assumono sempre un valore indipendente dal tempo, i semiquanti si muovono dentro il sub-osc. avanti e indietro!<br><br>Adesso possiamo costruire le seguenti corrispondenze: | In \|n > state the numero of coupled oscillators is (n+1), where n is the numero di quanta.<br><br>In summary we can affirm that an oscillator going to n-level it will consists of (n +1) sub-osc., one of whom has two ( o ) empty semi-quanta at energy **ε = 1/4 ℏω** while the remaining n other have each of them two full semi-quanta (●) with energy **ε = 1/2 ℏω** each of them (2 full semi-quanta are equal to one quanta or [(●) + (●) = 1 quanta]).<br><br>Even if the potential energy maximum and the maximum kinetic always have a constant value over time, the semi-quanta goes forward and back from a sub-osc. to another.<br><br>Now we can build the following correspondences: |

$$\left\{ \begin{matrix} \mathcal{E}_{\|>}= \\ \end{matrix} \begin{matrix} U \equiv (\bullet + o) \\ K \equiv (\bullet + o) \end{matrix} \right\} <=> \left\{ \vec{\Psi}_{osc} \equiv \begin{pmatrix} \vec{O}a = \vec{O}a_{el} + \vec{O}a_{in} \\ \vec{O}a^+ = \vec{O}a_{el}^+ + \vec{O}a_{in}^+ \end{pmatrix} \right\} <=> \left\{ \begin{matrix} a_{el} \equiv U_{el}(\bullet) & a_{in} \equiv K_{in}(o) \\ a_{el}^+ \equiv U_{el}(o) & a_{in}^+ \equiv K_{in}(\bullet) \end{matrix} \right\} \quad (24.2)$$

| | |
|---|---|
| Avremo anche la seguente equivalenza tra matrici (invarianza per scambio (o)◀▶ (●) ) | There is also the following equivalence (invariance for (o) ◀▶ (●) change) between matrices |

$$\left\{ \begin{pmatrix} a_{el} \equiv U_{el}(\bullet) & a_{in} \equiv K_{in}(o) \\ a_{el}^+ \equiv U_{el}(o) & a_{in}^+ \equiv K_{in}(\bullet) \end{pmatrix} \right\} <=> \left\{ \begin{pmatrix} a_{el} \equiv U_{el}(o) & a_{in} \equiv K_{in}(\bullet) \\ a_{el}^+ \equiv U_{el}(\bullet) & a_{in}^+ \equiv K_{in}(o) \end{pmatrix} \right\} \quad (25.2)$$

| | |
|---|---|
| Guardando la fig. 4, avremo la seguente rappresentazione (vettoriale) | Looking the fig. 4, it is the following (vectorial) representation |

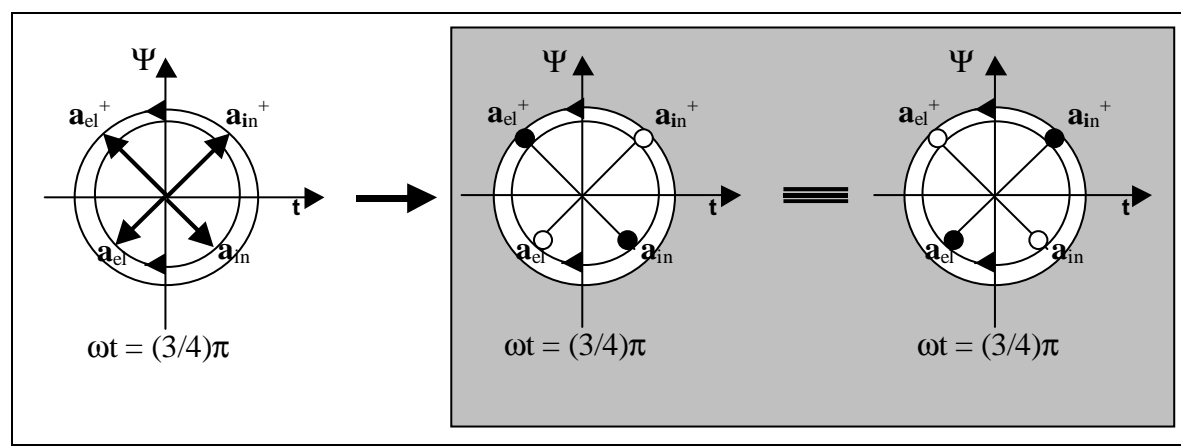

Fig. 9

| Per gli operatori (**a**, **a⁺**) possiamo stabilire i seguenti simboli: | About (**a**, **a⁺**) operators we can establish the following symbols: |
|---|---|

$$\left\{\begin{pmatrix} \hat{a}_{el} \equiv (\hat{\bullet})_{el} & \hat{a}_{in} \equiv (\hat{o})_{in} \\ \hat{a}_{el}^+ \equiv (\hat{o}^+)_{el} & \hat{a}_{in}^+ \equiv (\hat{\bullet}^+)_{in} \end{pmatrix}\right\} \Longleftrightarrow \left\{\begin{pmatrix} \hat{a}_{el} \equiv (\hat{o})_{el} & \hat{a}_{in} \equiv (\hat{\bullet})_{in} \\ \hat{a}_{el}^+ \equiv (\hat{\bullet}^+)_{el} & \hat{a}_{in}^+ \equiv (\hat{o}^+)_{in} \end{pmatrix}\right\} \quad (26.2)$$

$B_1$ - Matrix $\qquad\qquad$ $B_2$ - Matrix

| Dove ((**o**⁺)ₑₗ, (**o**⁺)ᵢₙ) sono gli operatori di creazione di semi-quanti vuoti ($\varepsilon = 1/4\,\hbar\omega$), mentre ((**●**⁺)ₑₗ, (**●**⁺)ᵢₙ) sono gli operatori di creazione di semi-quanti pieni ($\varepsilon = 1/2\,\hbar\omega$). Invece ((**o**)ₑₗ, (**o**)ᵢₙ) sono gli operatori di annichilazione di semi-quanti vuoti, mentre ((**●**)ₑₗ,(**●**)ᵢₙ) sono gli operatori di annichilazione di semi-quanti pieni. In un autostato degli operatori **r** ed **n** avremo (si vedano l'eq. 21.1 e 22.1) che | Where ((**o**⁺)el, (**o**⁺)in) are the creating operators of empty semi-quanta ( with $\varepsilon = 1/4\,\hbar\omega$ ), while ((**●**⁺)el, (**●**⁺)in) are the creating operators of full semi-quanta ( $\varepsilon = 1/2\,\hbar\omega$ ). Instead ((**o**)el, (**o**)in) are annihilation operator of empty semi-quanta, while ((**●**)el, (**●**)in) are the annihilation operator of full semi-quanta. In an eigenstate of (**r**) and (**n**) operators will be ( You see the eq. 21.1 and eq. 22.1) |
|---|---|

$$\begin{cases} \overrightarrow{O\hat{a}}_{n,r'}^+(t) \equiv \hat{a}_{n,r'}^+(t) = (\hat{\bullet}^+)_{el}^{(n)}[\exp(ir'\omega t)] + (\hat{o}^+)_{in}^{(n)}[\exp(i(r'\omega t - \pi/2))] \\ \overrightarrow{O\hat{a}}_{n,r'}(t) \equiv \hat{a}_{n,r'}(t) = (\hat{o})_{el}^{(n)}[\exp(-ir'\omega t)] + (\hat{\bullet})_{in}^{(n)}[\exp(-i(r'\omega t - \pi/2))] \end{cases} \quad (27.2)$$

| O equivalentemente | or equivalent |
|---|---|

$$\begin{cases} \overrightarrow{O\hat{a}}_{n,r'}^+(t) \equiv \hat{a}_{n,r'}^+(t) = (\hat{o}^+)_{el}^{(n)}[\exp(ir'\omega t)] + (\hat{\bullet}^+)_{in}^{(n)}[\exp(i(r'\omega t - \pi/2))] \\ \overrightarrow{O\hat{a}}_{n,r'}(t) \equiv \hat{a}_{n,r'}(t) = (\hat{\bullet})_{el}^{(n)}[\exp(-ir'\omega t)] + (\hat{o})_{in}^{(n)}[\exp(-i(r'\omega t - \pi/2))] \end{cases} \quad (28.2)$$

| Queste relazioni ci danno la forma matematica dell'oscillatore di campo definito da noi con l'acronimo **"IQuO"** Per un IQuO sarà: | These mathematical relationships give to us the mathematical form of the field oscillator defined by the acronym **"IQuO"** For an IQuO will be |
|---|---|

$$\begin{cases} \vec{\Psi}_{IQuo}(t)_{n,r'} = \overrightarrow{O\hat{a}}_{n,r'}(t) + \overrightarrow{O\hat{a}}_{n,r'}^+(t) & \text{eigenstate } |n,r'> \\ \vec{\Psi}_{IQuo}(t) = \vec{O}\hat{a}(t) + \vec{O}\hat{a}^+(t) & \text{superposition state} \end{cases} \quad (29.2)$$

| Un IQuO al primo livello è | In an IQuO at first level it is |
|---|---|

$$\begin{cases} \hat{a}_{1,r'}^+(t) = (\hat{\bullet}^+)_{el}[\exp(ir'\omega t)] + (\hat{o}^+)_{in}[\exp(i(r'\omega t - \pi/2))] \\ \hat{a}_{1,r'}(t) = (\hat{o})_{el}[\exp(-ir'\omega t)] + (\hat{\bullet})_{in}[\exp(-i(r'\omega t - \pi/2))] \end{cases} \quad (30.2)$$

| Avremo | We'll have |
|---|---|

$$\left( \varepsilon_1 \right)_{r'} = \left\{ \begin{matrix} U \equiv (\bullet + o) \\ K \equiv (\bullet + o) \end{matrix} \right\}_{r'} <=> \left\{ \begin{matrix} \hat{a}_{el} \equiv U_{el}(\hat{o}) & \hat{a}_{in} \equiv K_{in}(\hat{\bullet}) \\ a^+{}_{el} \equiv U_{el}(\hat{\bullet}) & a^+{}_{in} \equiv K_{in}(\hat{o}) \end{matrix} \right\}_{r'} \qquad (31.2)$$

| con | with |
|---|---|

$$(\hat{r})' = +1 => \begin{bmatrix} [\hat{a}_{el} = \hat{o}_{el} (\text{clockwise direction})] \\ [\hat{a}_{in} = \hat{\bullet}_{in} (\text{clockwise direction})] \end{bmatrix}$$
$$(\hat{r})' = +1 => \begin{bmatrix} [\hat{a}^+{}_{el} = \hat{\bullet}^+{}_{el} (\text{anti - clockwise direction})] \\ [\hat{a}^+{}_{in} = \hat{o}^+{}_{in} (\text{anti - clockwise direction})] \end{bmatrix} \qquad (32.2)$$

$$(\hat{r})' = -1 => \begin{bmatrix} [\hat{a}_{el} = \hat{o}_{el} (\text{anti - clockwise direction})] \\ [\hat{a}_{in} = \hat{\bullet}_{in} (\text{anti - clockwise direction})] \end{bmatrix}$$
$$(\hat{r})' = -1 => \begin{bmatrix} [\hat{a}^+{}_{el} = \hat{\bullet}^+{}_{el} (\text{clockwise direction})] \\ [\hat{a}^+{}_{in} = \hat{o}^+{}_{in} (\text{clockwise direction})] \end{bmatrix} \qquad (33.2)$$

| Possiamo usare la seguente rappresentazione (vedi eq. 33.2) per un IQuO con (r' = -1): | We could use the following representation (you see eq. 33.2) about an IQuO with (r' = -1): |
|---|---|

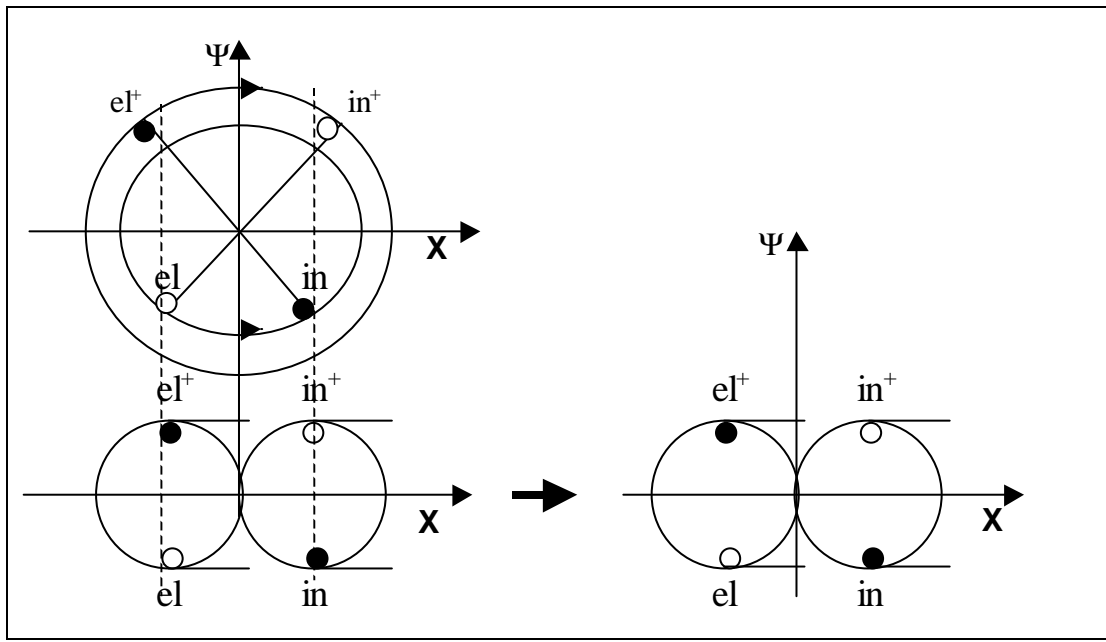

FIg. 10

| Questa rappresentazione sarà utile nel rappresentare una particella in movimento lungo l'asse X, intendendo questa come un quanto in movimento lungo una linea di oscillatori accoppiati di campo. | This representation will be useful to represent a particle in motion along the X axis, seeing this like a quanta in motion along a line of coupled oscillators of field. |
|---|---|

| Guardando la fig. 3, fig. 5, fig. 6, Fig. 7, Fig. 8 e Fig. 10 diamo la seguente rappresentazione (rappresentazione a semi-quanti): | Looking the Fig. 3, Fig. 5, Fig. 6, Fig. 7, Fig. 8 and Fig. 10 we give the follows representation ( representation at semi-quanta): |
|---|---|

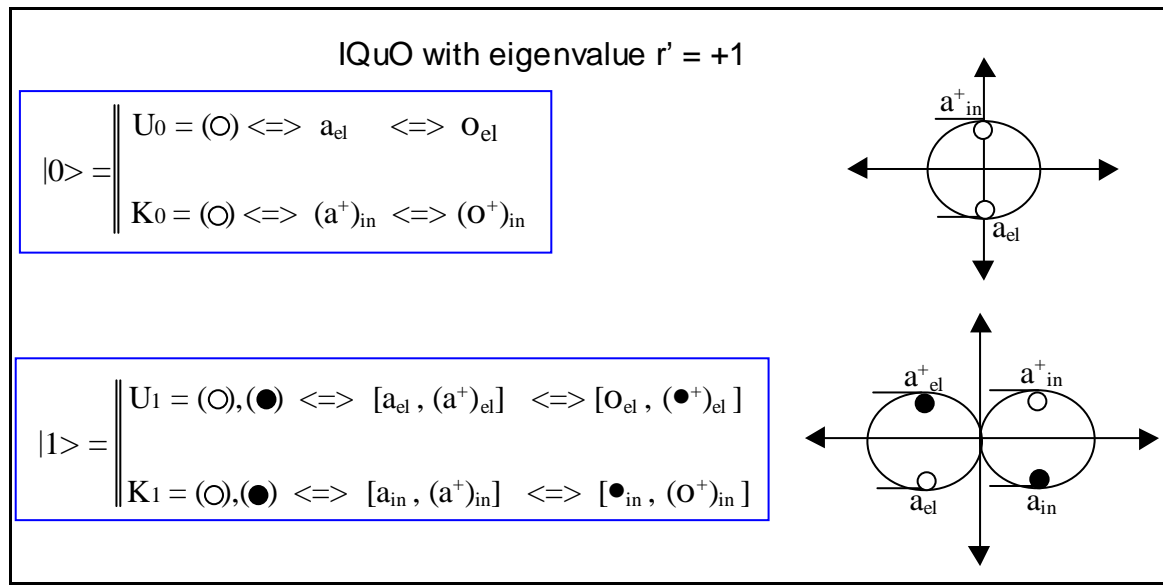

Fig. 11

| Rileviamo ancora una volta che l'oscillatore con due componenti (oscillatore forzato) evidenzia la presenza di un grado di libertà interno. | The oscillator with two components (forced oscillator) establishes the presence of an inside degree of liberty. |
|---|---|
| Si dirà che l'oscillazione è 1-dim. lungo l'asse x ma avente un grado di libertà interno che potrebbe evidenziare il verso di rotazione della fase. | It will be said that the oscillation is 1-dim., along x-axis, but with an inside degree of liberty which they could underline the verse of phase rotation. |
| I due (r') autovalori originano due IQuO differenti nel verso di rotazione della fase: r' = (+1, -1). | Two different IQuO in rotation verse of phase: r' = (+1, -1) are associated to two (r') eigenvalues. |
| E' anche evidente che un IQuO non potrebbe cambiare spontaneamente l'autovalore, di r' passando da r' = 1 a r'= - 1. | It's even evident that an IQuO couldn't change spontaneously the r' eigenvalue, passing from r' = +1 to r' = - 1. |
| Mostriamo (per esempio) le prime tre possibili configurazioni dell'IQuO durante la variazione di fase, nel caso (r'= 1): | Let's show (for example) the first three possible configurations of the IQuO during the phase variation, in the case (r' = 1): |

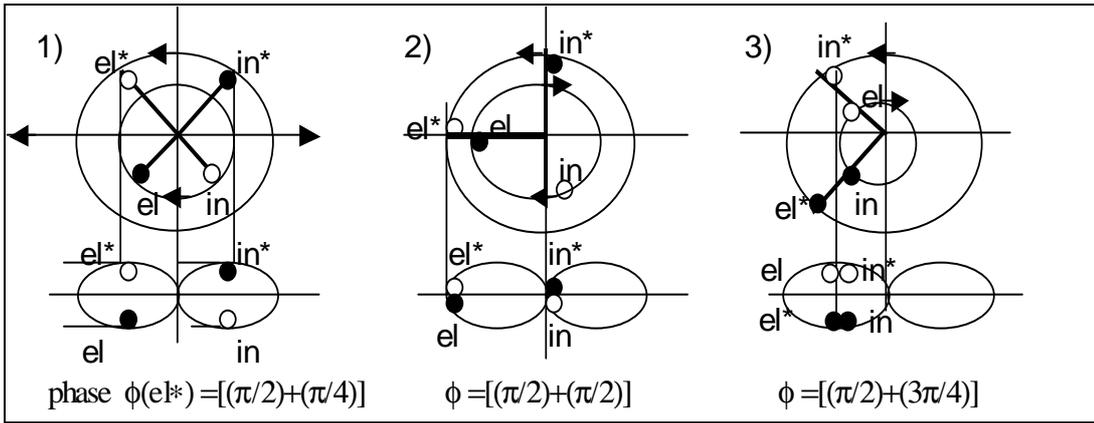

Fig. 12

| | |
|---|---|
| Proiettando sull'asse x avremo ( in analogia con quello classico ) il movimento del quanto rappresentativo dell'oscillatore. | In analogy with the classical oscillator, we project on the x axis the movement of the quanta which represents the oscillator; we'll have: |

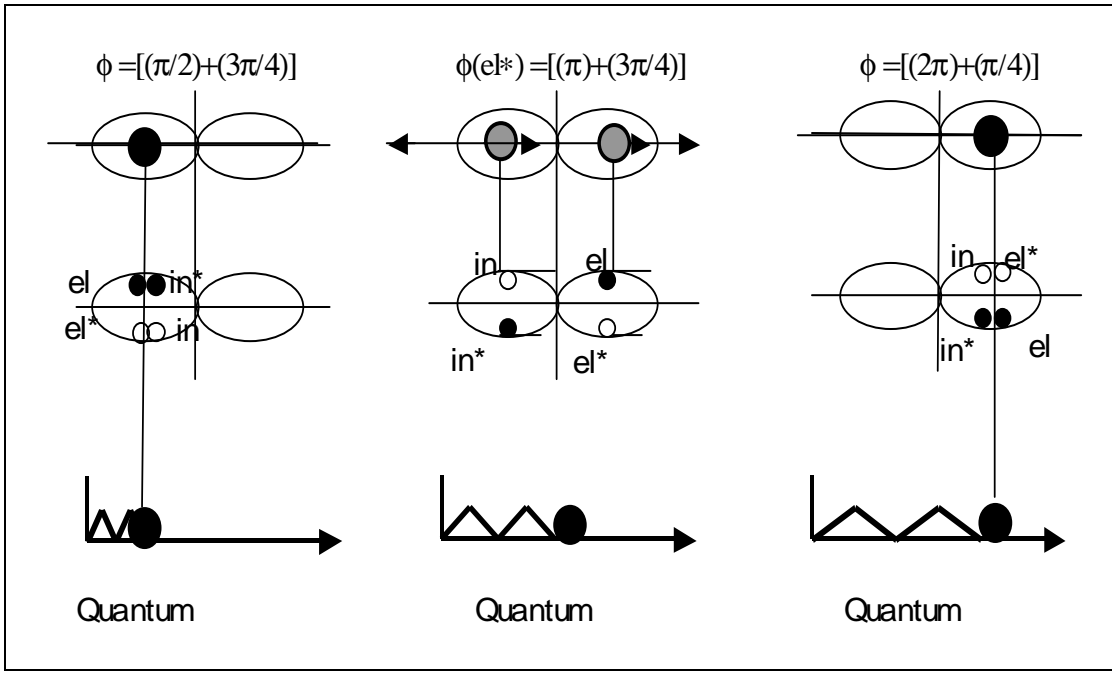

Fig. 13

| | |
|---|---|
| Dove la posizione del quanto segue la legge probabilistica di $\|\Psi_1(x)\|^2$.<br>Sottolineamo che nell'autostato $\| r >$ (stato locale) la fase è indeterminata (stato non locale delle configurazioni possibili).<br>Lo stato non locale della fase comporta che l'istante di transizione del semi-quanto da un sub-osc. all'altro è indeterminato, anche se le configurazioni, durante l'oscillazione, si susseguono ordinatamente. | Where the quanta position follow the probabilistic law of the function $\|\Psi_1(x)\|^2$.<br>We underline that in $\|r>$ eigenstate (local state) the phase is uncertain (no local state of the possible configurations).<br>The no local state of the phase involves that the transition instant of semi-quanta from sub-osc. to other is uncertain, even if the configurations, during the oscillation follow one another in an orderly. |

| Mostriamo la forma dell'IQuO con autovalore (r') = - 1 | We show the IQuO's form with [r'] = - 1 eigenvalue |
|---|---|

$$\begin{cases} \hat{a} = e^{-i\bar{r}\omega t}\hat{k}_{el} + ie^{-i\bar{r}\omega t}\hat{k}_{in}\big|_{r'=-1} = (\hat{k}_{el} + i\hat{k}_{in})e^{i\omega t} = (\hat{k}_{el} + i\hat{k}_{in}e^{i\pi/2})e^{i\omega t} \\ \hat{a}^+ = \hat{k}_{el}e^{i\bar{r}\omega t} - i\hat{k}_{in}e^{i\bar{r}\omega t}\big|_{r'=-1} = (\hat{k}_{el} - i\hat{k}_{in})e^{-i\omega t} = (\hat{k}_{el} + \hat{k}_{in}e^{-i\pi/2})e^{-i\omega t} \end{cases}$$  (34.2)

| Avremo allora per ( r' = -1 ) | Now we show the IQuO's configurations for: |
|---|---|

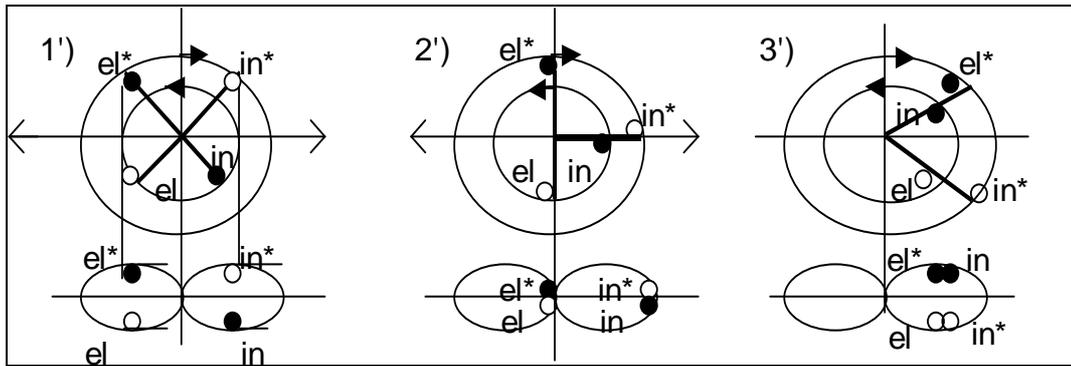

Fig. 14

| Definiamo Φ(B)-matrice quella associata ad ogni configurazione ( appartenente a r' =± 1 ) | We define the Φ(B)-matrix associated at each configuration: |
|---|---|

$$[\Phi(B)]_{(r'=+1)} \equiv \left\{ \begin{pmatrix} (a_{el})_{(cl)} & (a_{in})_{(cl)} \\ (a_{el}^+)_{(c\bar{l})} & (a_{in}^+)_{(c\bar{l})} \end{pmatrix} <=> \left[ \begin{pmatrix} (o_{el})_{(cl)} & (\bullet_{in})_{(cl)} \\ (\bullet_{el}^+)_{(c\bar{l})} & (o_{in}^+)_{(c\bar{l})} \end{pmatrix} \equiv \begin{pmatrix} (\bullet_{el})_{(cl)} & (o_{in})_{(cl)} \\ (o_{el}^+)_{(c\bar{l})} & (\bullet_{in}^+)_{(c\bar{l})} \end{pmatrix} \right] \right\}$$  (35.2)

$$[\Phi(B)]_{(r'=-1)} \equiv \left\{ \begin{pmatrix} (a_{el})_{(c\bar{l})} & (a_{in})_{(c\bar{l})} \\ (a_{el}^+)_{(cl)} & (a_{in}^+)_{(cl)} \end{pmatrix} <=> \left[ \begin{pmatrix} (o_{el})_{(c\bar{l})} & (\bullet_{in})_{(c\bar{l})} \\ (\bullet_{el}^+)_{(cl)} & (o_{in}^+)_{(cl)} \end{pmatrix} \equiv \begin{pmatrix} (\bullet_{el})_{(c\bar{l})} & (o_{in})_{(c\bar{l})} \\ (o_{el}^+)_{(cl)} & (\bullet_{in}^+)_{(cl)} \end{pmatrix} \right] \right\}$$  (36.2)

| dove l'indice (cl) indica una rotazione della fase oraria e $\overline{cl}$ una rotazione antioraria. Usando le proprietà degli operatori quantistici ( **a**, **a⁺** ) (come le parentesi di commutazione) e le proprietà del **H** operatore del [hamiltonian], otteniamo le parentesi di commutazione degli (**●, o**) operatori semi-quantistici, come: | where cl index ≡ clockwise direction and $\overline{cl}$ index ≡ anti-clockwise direction . Using the properties of the ( **a**, **a⁺** ) quantum operators (as commutation brackets) and the properties of the (**H**) hamiltonian operator, we obtain the commutation brackets of the (**●, o**) semi-quanta operators, as: |
|---|---|

| I° case : $[(\bullet_{el}^+)_{cl}, (o_{el})_{\overline{cl}}] = 0$ ; $[(\bullet_{in}^+)_{cl}, (o_{in})_{\overline{cl}}] = 0$ <br> I° case : $[(o_{el}^+)_{cl}, (\bullet_{el})_{\overline{cl}}] = 0$ ; $[(o_{in}^+)_{cl}, (\bullet_{in})_{\overline{cl}}] = 0$ |  (37.2) |
|---|---|

| Le altre relazioni degli operatori di semi-quanto e delle loro proprietà di simmetria possono essere sintetizzate nella seguente parentesi di commutazione: | The other relations about the semi-quanta operators and about their symmetry properties can be synthesized in the following commutation brackets: |

$$I°)\ case\ :\ \left[\bullet^+_{el}, o_{in}\right] = \frac{i}{2} \quad ; \quad \left[o^+_{el}, \bullet_{in}\right] = \frac{i}{2}$$

$$II°)\ case\ :\ \left[\bullet^+_{in}, o_{el}\right] = \frac{i}{2} \quad ; \quad \left[o^+_{in}, \bullet_{el}\right] = \frac{i}{2}$$

(38.2)

| Gli altri casi (vedi appendice) derivano dalla proprietà che un eventuale scambio globale ($\bullet \longleftrightarrow o$) non cambia il risultato delle parentesi di commutazione degli operatori ($\mathbf{a}, \mathbf{a}^+$). Troviamo allora che vale : | The other cases (you see the appendix) derive by property that a possible global exchange ($\bullet \longleftrightarrow o$ ) doesn't change the commutation brackets of the ( $\mathbf{a}, \mathbf{a}^+$ ) operators. . Then it is: |

$$I°)\ case\ :\ \left[\bullet^+_{el}, \bullet_{in}\right] = \frac{i}{2} \quad ; \quad \left[o_{el}, o^+_{in}\right] = \frac{i}{2}$$

$$II°)\ case\ :\ \left[\bullet^+_{in}, \bullet_{el}\right] = \frac{i}{2} \quad ; \quad \left[o_{in}, o^+_{el}\right] = \frac{i}{2}$$

(39.2)

| Il calcolo delle parentesi di commutazione è riportato in appendice. | The calculus of the commutation relations is reported in appendix. |

## Par. 3) Meccanismo di accoppiamento tra due IQuO rappresentativi di due campi

Consideriamo due campi scalari reali. Nella rappresentazione degli operatori di quanto ($\mathbf{a}, \mathbf{a}^+$) abbiamo:

## Sect. 3) Coupling Mechanism between two representative IQuO of two fields

Let's take two real scalar fields in the representation of quantum operators ($\hat{\mathbf{a}}, \hat{\mathbf{a}}^+$):

$$\hat{\Phi}_{1R} = \sum_k \varpi_k \left[\left(\hat{a}_{1k}\left(e^{-i\hat{r}\omega_k t + \alpha}\right) + \hat{a}^+_{1(-k)}\left(e^{i\hat{r}\omega_k t + \alpha}\right)\right)\left(e^{ikx}\right)\right]$$

$$\hat{\Phi}_{2R} = \sum_k \varpi_k \left[\left(\hat{a}_{2k}\left(e^{-i\hat{r}\omega_k t + \beta}\right) + \hat{a}^+_{2(-k)}\left(e^{-i\hat{r}\omega_k t + \beta}\right)\right)\left(e^{ikx}\right)\right]$$

(1.3)

| L'accoppiamento elastico tra i due campi si realizza mediante un accoppiamento locale ( spazialmente ) tra gli IQuO rappresentativi che denotiamo con: $\mathbf{\Phi_1 \oplus \Phi_2}$ | The elastic coupling between the two fields ($\mathbf{\Phi_1, \Phi_2}$) is achieved in the space through a local coupling between the two respective IQuO ($I_1, I_2$). Denote this coupling with $\mathbf{\Phi_1 \oplus \Phi_2}$ . |
| Questo accoppiamento è espresso da una particolare **"azione reciproca"** che : | A particular " **reciprocal action**" between the two IQuO ($I_1$, $I_2$) expresses this coupling: |
| **1)** opera un reciproco sfasamento: $\varphi_1 = -\varphi_2$ | 1) the action operates a reciprocal variation of phase in two IQuO: $\Delta\varphi_1 = -\Delta\varphi_2$ |
| **2)** può effettuare uno scambio di energia tra i due oscillatori, realizzato mediante i semiquanti, ma con la condizione di indeterminazione del tempo correlata all'indeterminazione della fase. | 2) can happening an energy exchange (through couples of the semi-quanta) between the two oscillators, but the time of exchange is uncertain as well as the phase |
| Lo schema degli autostati dell'operatore **r** relativo al sistema dei due IQuO è : | The eigenstates scheme of the **r** operator relative to the physical system of the two IQuO is: |

$$|\Phi\rangle_{(IQuO1+IQuO2)} = |\Phi\rangle_{(1)} \oplus |\Phi\rangle_{(2)} => \begin{cases} |\Phi(r'=\pm1)\rangle_{(1)} \oplus |\Phi(r'=\pm1)\rangle_{(2)} \\ |\Phi(r'=\pm1)\rangle_{(1)} \oplus |\Phi(r'=\mp1)\rangle_{(2)} \end{cases} \qquad (2.3)$$

| | |
|---|---|
| La prima combinazione la diremo ad autovalore non nullo dell'operatore globale **r** = (**r₁+r₂**), mentre la seconda la diremo ad autovalore nullo. Esaminiamo le due possibili combinazioni relative all'autovalore non nullo di **r**: (|**Φ₁**( ±1)>) ⊕ (|**Φ₂**( ±1)>) **Primo caso:** (|**Φ₁**( +1)>) ⊕ (|**Φ₂**( +1)>) Raggiunto l'adattamento (for ex. (**r₁**)'= (**r₂**)'= +1), avremo (per semplicità poniamo $\alpha = \beta$ nell'eq. 1.3) | The first combination is at "not zero eigenvalue" of operator (**r**), while the second is "at zero eigenvalue". We examine the two possible combinations of the "not zero eigenvalue" of the (**r**): (|**Φ₁**( ±1)>) ⊕ (|**Φ₂**( ±1)>) **First case:** (|**Φ₁**( +1)>) ⊕ (|**Φ₂**( +1)>) Once reaching the adaptation (for ex. (**r₁**)'= (**r₂**)'= +1), we'll have (for simplicity we set $\alpha = \beta$ in the eq. 1.3 ) |

$$\hat{\Phi}_{1R} = \sum_k \varpi_k \left[ \left( \hat{a}_{1k} e^{-i\omega_K t} + \hat{a}^+_{1(-k)} e^{i\omega_K t} \right) \left( e^{ikx} \right) \right]$$

$$\hat{\Phi}_{2R} = \sum_k \varpi_k \left[ \left( \hat{a}_{2k} e^{-i\omega_K t} + \hat{a}^+_{2(-k)} e^{i\omega_K t} \right) \left( e^{ikx} \right) \right] \qquad (3.3)$$

| | |
|---|---|
| Il particolare " accoppiamento ", indicato con un nuovo elemento algebrico ⊕, esprime una particolare combinazione di somma tra due elementi matematici, in questo caso gli operatori di campo, e che opera nel seguente modo : | This particular " coupling", denoted with a new algebraic element ⊕, consist of a particular sum between two mathematical elements, in this case two field operators, which operates in the following way: |

$$\hat{\Phi}_{1R} \oplus \hat{\Phi}_{2R} =$$
$$= \left\{ \sum_k \varpi_k \left[ \left( \hat{a}_{1k} \left( e^{-i\hat{r}\omega_k t+\alpha} \right) + \hat{a}_{1(-k)}^+ \left( e^{i\hat{r}\omega_k t+\alpha} \right) \right) \left( e^{ikx} \right) \right] \right\} \oplus \left\{ \sum_k \varpi_k \left[ \left( \hat{a}_{2k} \left( e^{-i\hat{r}\omega_k t+\beta} \right) + \hat{a}_{2(-k)}^+ \left( e^{-i\hat{r}\omega_k t+\beta} \right) \right) \left( e^{ikx} \right) \right] \right\} \quad (4.3)$$

| | |
|---|---|
| seguendo | following |

$$\left( \hat{\Phi}_{1R} \oplus \hat{\Phi}_{2R} \right)_{|r\rangle} =$$
$$= \left[ \sum_k \varpi_k \left( \hat{a}_{1k} e^{-i\omega_k t} + \hat{a}_{1(-k)}^+ e^{i\omega_k t} \right) \left( e^{ikx} \right) \right] \oplus \left[ \sum_k \varpi_k \left( \hat{a}_{2k} e^{-i\omega_k t} + \hat{a}_{2(-k)}^+ e^{i\omega_k t} \right) \left( e^{ikx} \right) \right] =$$
$$= \left[ \sum_k \varpi_k \left( \hat{a}_{1k} e^{-i\omega_k t} + \hat{a}_{1(-k)}^+ e^{i\omega_k t} \right) \left( e^{ikx} \right) \right] \left( e^{i\varphi_1} \right) + \left( e^{i\varphi_2} \right) \left[ \sum_k \varpi_k \left( \hat{a}_{2k} e^{-i\omega_k t} + \hat{a}_{2(-k)}^+ e^{i\omega_k t} \right) \left( e^{ikx} \right) \right] \quad (5.3)$$

| | |
|---|---|
| ponendo φ₁ = -φ₂ = -φ  e  φ = (π/4) segue: | setting  φ₁ = -φ₂ = -φ  and φ = (π/4) it follows: |

$$\hat{\Phi}_{1R} \oplus \hat{\Phi}_{2R} = \left\{ \sum_k \varpi_k \left[ \left( \hat{a}_{1k} + \hat{a}_{2k} e^{i\pi/2} \right) \left( e^{-i\pi/4} \right) \left( e^{-i\omega_k t} \right) + \left( \hat{a}_{1(-k)}^+ + \hat{a}_{2(-k)}^+ e^{i\pi/2} \right) \left( e^{-i\pi/4} \right) \left( e^{i\omega_k t} \right) \right] \left( e^{ikx} \right) \right\}$$
$$= e^{-i\pi/4} \cdot \left\{ \sum_k \varpi_k \left[ \left( \hat{a}_k \right) \left( e^{-i\omega_k t} \right) + \left( \hat{b}_{(-k)}^+ \right) \left( e^{i\omega_k t} \right) \right] \left( e^{ikx} \right) \right\} \quad (6.3)$$

| | |
|---|---|
| dove | Where |

$$\left( \hat{a}_k \right) = \left( \hat{a}_{1k} + i \hat{a}_{2k} \right) \; ; \; \left( \hat{b}_{(-k)}^+ \right) = \left( \hat{a}_{1(-k)}^+ + i \hat{a}_{2(-k)}^+ \right) \qquad (7.3)$$

| | |
|---|---|
| Dalla forma degli operatori (**a**, **b**⁺) ricaviamo anche che | From the ( **a**, **b**⁺) form of the operators we also obtain that |

$$\left\{ \left( \hat{a}_k^+ \right) = \left( \hat{a}_{1k}^+ - i \hat{a}_{2k}^+ \right) ; \left( \hat{b}_{(-k)} \right) = \left( \hat{a}_{1(-k)} - i \hat{a}_{2(-k)} \right) \right\} \qquad (8.3)$$

| | |
|---|---|
| Il campo [**Φ** = (**Φ₁** ⊕ **Φ₂**)] che abbiamo qui ottenuto è perfettamente coincidente con il campo complesso di  K-G (eq. (1)).<br>Rileviamo subito (vedi le 6.3) che [**Φ** = (**Φ₁** ⊕ **Φ₂**)]  risulta a carattere non locale ovvero  non  separato  nelle  componenti | Note that the field [**Φ** = (**Φ₁** ⊕ **Φ₂**)] is perfectly coincident with the Klein-Gordon's complex field (K-G).<br>Believe (you see the eq. 6.3) that  the field [**Φ** = (**Φ₁** ⊕ **Φ₂**)]  is a not local aspect or not separate  of  two  states |
| $\boldsymbol{\Psi_+}$ **(a, a⁺)** e $\boldsymbol{\Psi_-}$ **(b, b⁺)** correlate ai due campi carichi di K-G. Sintetizzando si ha: | $\boldsymbol{\Psi_+}$ **(a, a⁺)** and $\boldsymbol{\Psi_-}$ **(b, b⁺)** in correlation with the two (K-G) fields having electric charge. Summarising we have : |

$$\hat{\Phi}_{1R} \oplus \hat{\Phi}_{2R} \equiv \hat{\Psi}_C(\hat{a},\hat{b}^+) = (\hat{\Psi}_+ \cap \hat{\Psi}_-) \qquad (9.3a)$$

| | |
|---|---|
| o | Or |

$$\hat{\Phi}_{1R} \oplus \hat{\Phi}_{2R} \equiv \hat{\Psi}_C(\hat{a}^+,\hat{b}) = (\hat{\Psi}_+ \cap \hat{\Psi}_-) \qquad (9.3b)$$

| | |
|---|---|
| dove l'ultima relazione indica lo stato di non separazione. Si rileva che l'operazione descritta da (⊕) coincide con quella di prodotto tra matrici; esattamente avremo: | where the last relation denotes the not separation state. Note that the **(⊕)-operation** coincides with a multiplication of matrices; exactly we'll have: |

$$\hat{\Phi}_{1R} \oplus \hat{\Phi}_{2R} => ((\hat{\Phi}_{1R})) \oplus ((\hat{\Phi}_{2R})) = \begin{pmatrix} \hat{\Phi}_{1R} \\ 1 \end{pmatrix} \cdot \left[ \begin{pmatrix} e^{-i\varphi} & 0 \\ 0 & e^{i\varphi} \end{pmatrix} \cdot \begin{pmatrix} 1 \\ \hat{\Phi}_{2R} \end{pmatrix} \right] \qquad (10.3)$$

| | |
|---|---|
| Per avere una manifestazione di $\boldsymbol{\Psi_+}$ e $\boldsymbol{\Psi_-}$ indipendente e separata occorrerebbe un eventuale disaccoppiamento, causato da un campo-agente esterno. Tuttavia questa possibilità potrebbe essere realizzata tramite l'accoppiamento $[\boldsymbol{\Phi} = (\boldsymbol{\Phi_1} \oplus \boldsymbol{\Phi_2})]$ dei due campi appartenenti all' "autovalore zero" dell'op. **(r)** globale, cioè: $(\|\boldsymbol{\Psi_1}(\mathbf{r'} = \pm 1)>) \oplus (\|\boldsymbol{\Psi_2}(\mathbf{r'} = \mp 1)>)$ **Second case:** $(\|\boldsymbol{\Psi_1}(\pm 1)>)\oplus(\|\boldsymbol{\Psi_2}(\mp 1)>)$ Matematicamente avremo : | To have an independent and separate manifestation of $\boldsymbol{\Psi_+}$ and $\boldsymbol{\Psi_-}$, we must have an eventual uncoupling, caused by an external agent-field. However, this possibility can be implemented via the coupling $[\boldsymbol{\Phi} = (\boldsymbol{\Phi_1} \oplus \boldsymbol{\Phi_2})]$ of the two fields belonging to '"eigenvalue zero" of the global **(r)** global, that is: $(\|\boldsymbol{\Psi_1}(\mathbf{r'} = \pm 1)>) \oplus (\|\boldsymbol{\Psi_2}(\mathbf{r'} = \mp 1)>)$ **Second case:** $(\|\boldsymbol{\Psi_1}(\pm 1)>)\oplus(\|\boldsymbol{\Psi_2}(\mp 1)>)$ Mathematically we have : |

$$\hat{\Phi}_{1R} \oplus \hat{\Phi}_{2R} = \left[ \sum_k \varpi_k \left( \hat{a}_{1k}\left(e^{i\omega_k t}\right) + \hat{a}^+_{1(-k)}\left(e^{-i\omega_k t}\right) \right)\left(e^{ikx}\right) \right] \oplus \left[ \sum_k \varpi_k \left( \hat{a}_{2k}\left(e^{-i\omega_k t}\right) + \hat{a}^+_{2(-k)}\left(e^{i\omega_k t}\right) \right)\left(e^{ikx}\right) \right] =$$

$$= \left[ \sum_k \varpi_k \left( \hat{a}_{1k}\left(e^{i\omega_k t}\right) + \hat{a}^+_{1(-k)}\left(e^{-i\omega_k t}\right) \right)\left(e^{ikx}\right) \right] e^{i\varphi_1} + e^{i\varphi_2} \left[ \sum_k \varpi_k \left( \hat{a}_{2k}\left(e^{-i\omega_k t}\right) + \hat{a}^+_{2(-k)}\left(e^{i\omega_k t}\right) \right)\left(e^{ikx}\right) \right] =$$

$$= \left[ \sum_k \varpi_k \left( \hat{a}_{1k}\left(e^{i\omega_k t+\varphi_1}\right) + \hat{a}^+_{1(-k)}\left(e^{-i\omega_k t+\varphi_1}\right) \right)\left(e^{ikx}\right) + \sum_k \varpi_k \left( \hat{a}_{2k}\left(e^{-i\omega_k t+i\varphi_2}\right) + \hat{a}^+_{2(-k)}\left(e^{i\omega_k t+i\varphi_2}\right) \right)\left(e^{ikx}\right) \right]$$

(11.3)

| Ponendo $\varphi_1 = -\varphi_2 = -\varphi$ e $\varphi = (\pi/4)$ segue: | setting $\varphi_1 = -\varphi_2 = -\varphi$ it follows : |
|---|---|

$$\hat{\Phi}_{1R} \oplus \hat{\Phi}_{2R} = \left\{ \sum_k \varpi_k \left[ \left( \hat{a}^+_{1(-k)} \left( e^{-i\varphi} \right) + \hat{a}_{2k} \left( e^{i\varphi} \right) \right) \left( e^{-i\omega_k t} \right) + \left( \hat{a}_{1k} \left( e^{-i\varphi} \right) + \hat{a}^+_{2(-k)} \left( e^{i\varphi} \right) \right) \left( e^{i\omega_k t} \right) \right] \left( e^{ikx} \right) \right\}$$

$$= \left\{ \sum_k \varpi_k \left[ \left( \hat{a}^+_{1(-k)} + \hat{a}_{2k} \left( e^{2i\varphi} \right) \right) \left( e^{-i\varphi} \right) \left( e^{-i\omega_k t} \right) + \left( \hat{a}_{1(k)} + \hat{a}^+_{2(-k)} \left( e^{2i\varphi} \right) e^{2i\varphi} \right) \left( e^{-i\varphi} \right) \left( e^{i\omega_k t} \right) \right] \left( e^{ikx} \right) \right\}$$

(12.3)

| Ponendo $\varphi = (\pi/4)$ segue: | setting $\varphi = (\pi/4)$ it follows : |
|---|---|

$$\hat{\Phi}_{1R} \oplus \hat{\Phi}_{2R} = \left\{ \sum_k \varpi_k \left[ \left( \hat{a}^+_{1(-k)} + \hat{a}_{2k} \left( e^{i\pi/2} \right) \right) \left( e^{-i\pi/4} \right) \left( e^{-i\omega_k t} \right) + \left( \hat{a}_{1(k)} + \hat{a}^+_{2(-k)} \left( e^{i\pi/} \right) e^{i\pi/2} \right) \left( e^{-i\pi/4} \right) \left( e^{i\omega_k t} \right) \right] \left( e^{ikx} \right) \right\}$$

$$= \left( e^{-i\pi/4} \right) \cdot \left\{ \sum_k \varpi_k \left[ \left( \hat{a}^+_{1(-k)} + i\hat{a}_{2k} \right) \left( e^{-i\omega_k t} \right) + \left( \hat{a}_{1(k)} + i\hat{a}^+_{2(-k)} \right) \left( e^{i\omega_k t} \right) \right] \left( e^{ikx} \right) \right\}$$

$$= \left( e^{-i\pi/4} \right) \cdot \left\{ \sum_k \varpi_k \left[ \left( \hat{\alpha}_{(k,-k)} \right) \left( e^{-i\omega_k t} \right) + \left( \hat{\beta}_{(k,-k)} \right) \left( e^{i\omega_k t} \right) \right] \left( e^{ikx} \right) \right\}$$

$$= \hat{\Psi}_1 \left( \hat{\alpha} \right) + \hat{\Psi}_2 \left( \hat{\beta} \right)$$

(13.3)

| Abbiamo ottenuto in questo caso un risultato molto importante : <br> **l'accoppiamento ha prodotto IQuO ( $\Psi(\alpha)$, $\Psi(\beta)$ ) i cui rispettivi operatori di annichilazione e di creazione sono equiversi nella rotazione della fase!** <br> L'operatore Hamiltoniano **H** del sistema si determina, al solito modo, ricorrendo all'operatore campo e al suo coniugato (ponendo $\beta = \gamma^+$) | A very important result is here: <br> **the fields ( $\Psi(\alpha)$, $\Psi(\beta)$ ) are expressed respectively by IQuO with operators of annihilation and creation having the same direction of phase rotation.** <br> The **H** Hamiltonian operator of the physical system is established, as usual, by the field operator and his conjugated (setting $\beta = \gamma^+$) |
|---|---|

$$\begin{cases} \hat{\Psi} = \left\{ \sum_k \varpi_k \left[ \left( \hat{\alpha}_{(k,-k)} \right) e^{-i\omega_k t} + \left( \gamma^+_{(k,-k)} \right) e^{i\omega_k t} \right] e^{ikx} \right\} \\ \hat{\Psi}^+ = \left\{ \sum_k \varpi_k \left[ \left( \hat{\gamma}_{(k,-k)} \right) e^{-i\omega_k t} + \left( \hat{\alpha}^+_{(k,-k)} \right) e^{i\omega_k t} \right] e^{-ikx} \right\} \end{cases}$$

(14.3)

| Ottenendo così | So obtaining |
|---|---|

$$\hat{H} = \sum_{-\infty}^{\infty} \hbar\omega_k \left( \hat{\alpha}^+_k \hat{\alpha}_k + \gamma^+_k \gamma_k + 1 \right)$$

(15.3)

| Prendendo la struttura a semi-quanti (eq. 30.2 ) segue | Using the semi–quanta structure ( you see the eq. 30.2) it follows: |
|---|---|

$$\hat{\Phi}_{1R} \oplus \hat{\Phi}_{2R} = e^{-i\pi/4} \cdot \left\{ \sum_k \varpi_k \left[ \left( \hat{a}^+_{1(-k)} + i\hat{a}_{2k} \right) e^{-i\omega_k t} + \left( \hat{a}_{1(k)} + i\hat{a}^+_{2(-k)} \right) e^{i\omega_k t} \right] e^{ikx} \right\}$$

$$= e^{-i\pi/4} \cdot \left\{ \left[ \left( \hat{\bullet}^+_{1el} - i\hat{o}^+_{1in} \right)_{-k} + i\left( \hat{\bullet}_{2el} + i\hat{o}_{2in} \right)_k \right) e^{-i\omega_k t} + \left( \left( \hat{o}_{1el} + i\,\hat{\bullet}_{1in} \right)_k + i\left( \hat{o}^+_{2el} - i\,\hat{\bullet}^+_{2in} \right)_{-k} \right) e^{i\omega_k t} \right] \right\}$$

$$= e^{-i\pi/4} \cdot \left\{ \sum_k \varpi_k \left[ \left( \hat{\alpha}_{(k,-k)} \right) e^{-i\omega_k t} + \left( \gamma^+_{(k,-k)} \right) e^{i\omega_k t} \right] e^{ikx} \right\}$$

(16.3)

| Con | with |
|---|---|

$$\begin{cases} \left( \hat{\alpha}_{(k,-k)} \right) e^{-i\omega_k t} e^{-i\pi/4} = \left( \hat{\bullet}^+_{1el} - i\hat{o}^+_{1in} \right)_{-k} + i\left( \hat{\bullet}_{2el} + i\hat{o}_{2in} \right)_k \right) e^{-i\omega_k t} e^{-i\pi/4} \\ \left( \gamma^+_{(k,-k)} \right) e^{i\omega_k t} e^{-i\pi/4} = \left( \left( \hat{o}_{1el} + i\,\hat{\bullet}_{1in} \right)_k + i\left( \hat{o}^+_{2el} - i\,\hat{\bullet}^+_{2in} \right)_{-k} \right) e^{i\omega_k t} e^{-i\pi/4} \end{cases}$$

(17.3)

| Possiamo sintetizzare l'eq. 3.15 usando | We can summarise, using the eq. 3.15 |
|---|---|

$$\begin{cases} \hat{\alpha} = \hat{c}^+_1 + i\hat{c}_2 \\ \hat{\gamma}^+ = \hat{d}_1 + i\hat{d}^+_2 \end{cases}$$

(18.3)

| Dove gli operatori (c) e (d) soddisfano alle usuali parentesi di commutazione.<br>Utilizzando la iniziale configurazione dell'IQuO con fase $\omega t = \pm(\pi/2 + \pi/4)$, si ottiene: | Where the (c) operators and (d) satisfy the usual commutation brackets.<br>Using the initial configuration of the IQuO with phase $\omega t = \pm(\pi/2 + \pi/4)$, one obtains : |
|---|---|

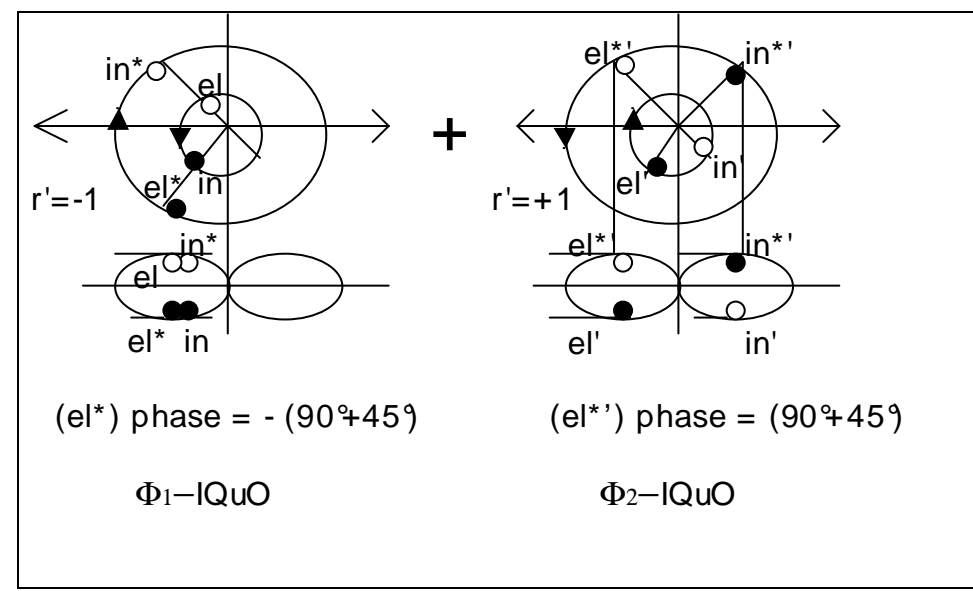

(el*) phase = - (90°+45°)          (el*') phase = (90°+45°)

$\Phi_1$–IQuO          $\Phi_2$–IQuO

Fig. 10

| Avremo infine | Finally we'll have |
|---|---|

$$\begin{cases} \left(\hat{\alpha}_{(k,-k)}\right)e^{-i\omega_K t}e^{-i\pi/4} = \left((\hat{\bullet}_{1el}^{+}e^{-i\pi} + \hat{o}_{1in}^{+}e^{-i3\pi/2})_{(cl,-k)} + (\hat{\bullet}_{2el}e^{-i\pi/2} + \hat{o}_{2in})_{(cl,k)})\right) = (\hat{\alpha}(t*))_{(k,-k)}^{cl} \\ \left(\gamma_{(k,-k)}^{+}\right)e^{i\omega_K t}e^{-i\pi/4} = \left((\hat{o}_{1el}e^{i\pi/2} + \hat{\bullet}_{1in}e^{i\pi})_{(c\bar{l},k)} + (\hat{o}_{2el}^{+}e^{i\pi} + \hat{\bullet}_{2in}^{+}e^{i\pi/2})_{(c\bar{l},-k)})\right) = (\gamma^{+}(t*))_{(k,-k)}^{c\bar{l}} \end{cases} \qquad (19.3)$$

| **Par.4) IQuO con carica elettrica** | **Sect. 4)  IQuO electrically charged** |
|---|---|
| Consideriamo lo stato di sovrapposizione di tutte le possibili configurazioni relative alla combinazione dei due autostati $\|\alpha_r\rangle$ e $\|\gamma^{+}_r\rangle$ | Consider the state of superposition of the all possible configurations of the combination of two eigenstates $\|\alpha_r\rangle$ and $\|\gamma^{+}_r\rangle$ : |
| $\|\Psi\rangle \equiv [ (\|\alpha_1\rangle + \|\gamma^{+}_1\rangle) + (\|\alpha_2\rangle + \|\gamma^{+}_2\rangle) + ... (\|\alpha_n\rangle + \|\gamma^{+}_n\rangle)].$ | $\| \Psi \rangle \equiv [ (\|\alpha_1\rangle + \|\gamma^{+}_1\rangle) + (\|\alpha_2\rangle + \|\gamma^{+}_2\rangle) + ... (\|\alpha_n\rangle + \|\gamma^{+}_n\rangle)]$ |
| Dove gli autostati sono differenti in fase e nella distribuzione degli operatori di semi-quanto ($\bullet$, o ) (vedi eq. 26.2). | Where the eigenstates are different in distribution of the ($\bullet$, o ) semi-quanta ( you see eq. 26.2 ) and in phase. |
| Infatti considerando l'altra forma equivalente della eq. (30.2) avremo: | In fact if one consider the other equivalent form of the eq. (30.2)  will be |

$$\begin{cases} \left(\hat{\alpha}'_{(k,-k)}\right)e^{-i\omega_K t}e^{-i\pi/4} = \left((\hat{o}_{1el}^{+}e^{-i\pi} + \hat{\bullet}_{1in}^{+}e^{-i\pi})_{(cl,-k)} + (\hat{o}_{2el}e^{-i\pi/2} + \hat{\bullet}_{2in})_{(cl,k)})\right) = (\hat{\alpha}'(t*))_{(k,-k)}^{cl} \\ \left(\hat{\beta}'_{(k,-k)}\right)e^{i\omega_K t}e^{-i\pi/4} = \left((\hat{\bullet}_{1el}e^{i\pi/2} + \hat{o}_{1in}e^{i\pi})_{(\bar{cl},k)} + (\hat{\bullet}_{2el}^{+}e^{i\pi} + \hat{o}_{2in}^{+}e^{i\pi/2})_{(\bar{cl},-k)})\right) = (\hat{\beta}'(t*))_{(k,-k)}^{c\bar{l}} \end{cases} \qquad (1.4)$$

| Avuta dopo uno scambio di energia o <=> $\bullet$ ). Combinando le due configurazioni equivalenti | got after an exchange ($\bullet$ ←→ o ). Combining the two equivalent configurations |
|---|---|

$$(\alpha + \alpha') + (\beta + \beta') =$$
$$\left((\hat{\bullet}_{1el}^{+}e^{-i\pi} + \hat{o}_{1in}^{+}e^{-i3\pi/2})_{(cl,-k)} + (\hat{\bullet}_{2el}e^{-i\pi/2} + \hat{o}_{2in})_{(cl,k)})\right) + \left((\hat{o}_{1el}^{+}e^{-i\pi} + \hat{\bullet}_{1in}^{+}e^{-i3\pi/2})_{(cl,-k)} + (\hat{o}_{2el}e^{-i\pi/2} + \hat{\bullet}_{2in})_{(cl,k)})\right) +$$
$$\left((\hat{o}_{1el}e^{i\pi/2} + \hat{\bullet}_{1in}e^{i\pi})_{(\bar{cl},k)} + (\hat{o}_{2el}^{+}e^{i\pi} + \hat{\bullet}_{2in}^{+}e^{i\pi/2})_{(\bar{cl},-k)})\right) + \left((\hat{\bullet}_{1el}e^{i\pi/2} + \hat{o}_{1in}e^{i\pi})_{(\bar{cl},k)} + (\hat{\bullet}_{2el}^{+}e^{i\pi} + \hat{o}_{2in}^{+}e^{i\pi/2})_{(\bar{cl},-k)})\right)$$
$$(2.4)$$

| si ottengono tutte le varie possibilità di ricombinazione tra i due IQuO iniziali che determinano , alla fine, due nuovi IQuO uscenti: | We obtain the different possibilities of recombination between the two initial IQuO which determine, at the end, two new "outgoing" IQuO: |
|---|---|
| **1) Primo doppietto di coppie di IQuO** | **1) First pair of IQuO couples** |

$$I°) \begin{cases} 1) \begin{cases} (\hat{\Psi}_1)_{cl} = [(\hat{\bullet}_{1el}^{+}e^{-i\pi} + \hat{o}_{1in}^{+}e^{-i3\pi/2})_{(cl,-k)} + (\hat{o}_{2el}e^{-i\pi/2} + \hat{\bullet}_{2in})_{(cl,k)}] \\ (\hat{\Psi}_1)_{\overline{cl}} = [(\hat{o}_{1el}e^{i\pi/2} + \hat{\bullet}_{1in}e^{i\pi})_{(\overline{cl},k)} + (\hat{\bullet}_{2el}^{+}e^{i\pi} + \hat{o}_{2in}^{+}e^{i\pi/2})_{(\overline{cl},-k)}] \end{cases} \\ 2) \begin{cases} (\hat{\Psi}_2)_{cl} = [(\hat{o}_{1el}^{+}e^{-i\pi} + \hat{\bullet}_{1in}^{+}e^{-i3\pi/2})_{(cl,-k)} + (\hat{\bullet}_{2el}e^{-i\pi/2} + \hat{o}_{2in})_{(cl,k)}] \\ (\hat{\Psi}_2)_{\overline{cl}} = [(\hat{\bullet}_{1el}e^{i\pi/2} + \hat{o}_{1in}e^{i\pi})_{(\overline{cl},k)} + (\hat{o}_{2el}^{+}e^{i\pi} + \hat{\bullet}_{2in}^{+}e^{i\pi/2})_{(\overline{cl},-k)}] \end{cases} \end{cases} \qquad (3.4)$$

| 2) Secondo doppietto di coppie di IQuO | 2) Second pair of IQuO couples |

$$II°) \quad \begin{cases} 3) \begin{cases} (\hat{\Psi}_3)_{cl} = \left[(\hat{\bullet}_{1el}^+ \, e^{-i\pi} + \hat{\bullet}_{1in}^+ \, e^{-i3\pi/2})_{(cl,-k)} + (\hat{o}_{2el} e^{-i\pi/2} + \hat{o}_{2in})_{(cl,k)}\right] \\ (\hat{\Psi}_3)_{\overline{cl}} = \left[(\hat{\bullet}_{1el}^+ e^{i\pi/2} + \hat{\bullet}_{1in} e^{i\pi})_{(\overline{cl},k)} + (\hat{o}_{2el}^+ e^{i\pi} + \hat{o}_{2in}^+ e^{i\pi/2})_{\overline{cl},-k)}\right] \end{cases} \\ 4) \begin{cases} (\hat{\Psi}_4)_{cl} = \left[(\hat{o}_{1el}^+ e^{-i\pi} + \hat{o}_{1in}^+ e^{-i3\pi/2})_{(cl,-k)} + (\hat{\bullet}_{2el} e^{-i\pi/2} + \hat{\bullet}_{2in})_{(cl,k)}\right] \\ (\hat{\Psi}_4)_{\overline{cl}} = \left[(\hat{o}_{2el} e^{i\pi/2} + \hat{o}_{1in} e^{i\pi})_{(\overline{cl},k)} + (\hat{\bullet}_{2el}^+ e^{i\pi} + \hat{\bullet}_{2in}^+ e^{i\pi/2})_{\overline{(cl,-k)}}\right] \end{cases} \end{cases} \quad (4.4)$$

| Guardando queste relazioni possiamo asserire che: 1) il primo doppietto corrisponde a due forme equivalenti di una sola coppia di due IQuO ovvero | Looking this relations we can affirm that 1) the first pair corresponds to two equivalent forms of one only couple of two IQuO or |

$$I°) \quad \begin{cases} IQuO(cl) \equiv \begin{cases} (\hat{\Psi}_1)_{cl} \equiv \left[(\hat{\bullet}_{1el}^+ e^{-i\pi} + \hat{o}_{1in}^+ e^{-i3\pi/2})_{(cl,-k)} + (\hat{o}_{2el} e^{-i\pi/2} + \hat{\bullet}_{2in})_{(cl,k)})\right] \\ (\hat{\Psi}_2)_{cl} \equiv \left[(\hat{o}_{1el}^+ e^{-i\pi} + \hat{\bullet}_{1in}^+ e^{-i3\pi/2})_{(cl,-k)} + (\hat{\bullet}_{2el} e^{-i\pi/2} + \hat{o}_{2in})_{(cl,k)}\right] \end{cases} \\ IQuO(\overline{cl}) \equiv \begin{cases} (\hat{\Psi}_1)_{\overline{cl}} \equiv \left[(\hat{o}_{1el} e^{i\pi/2} + \hat{\bullet}_{1in} e^{i\pi})_{(\overline{cl},k)} + (\hat{\bullet}_{2el}^+ e^{i\pi} + \hat{o}_{2in}^+ e^{i\pi/2})_{(\overline{cl},-k)}\right] \\ (\hat{\Psi}_2)_{\overline{cl}} \equiv \left[(\hat{\bullet}_{1el} e^{i\pi/2} + \hat{o}_{1in} e^{i\pi})_{(\overline{cl},k)} + (\hat{o}_{2el}^+ e^{i\pi} + \hat{\bullet}_{2in}^+ e^{i\pi/2})_{(\overline{cl},-k)}\right] \end{cases} \end{cases} \quad (5.4)$$

| 2) il secondo doppietto corrisponde a due forme equivalenti di un'altra coppia di due IQuO ovvero | 2) the second pair corresponds to two equivalent forms of another couple of two IQuO or |

$$II°) \quad \begin{cases} IQuO(cl) \equiv \begin{cases} (\hat{\Psi}_3)_{cl} \equiv \left[(\hat{\bullet}_{1el}^+ \, e^{-i\pi} + \hat{\bullet}_{1in}^+ e^{-i3\pi/2})_{(cl,-k)} + (\hat{o}_{2el} e^{-i\pi/2} + \hat{o}_{2in})_{(cl,k)}\right] \\ (\hat{\Psi}_4)_{cl} \equiv \left[(\hat{o}_{1el}^+ e^{-i\pi} + \hat{o}_{1in}^+ e^{-i3\pi/2})_{(cl,-k)} + (\hat{\bullet}_{2el} e^{-i\pi/2} + \hat{\bullet}_{2in})_{(cl,k)}\right] \end{cases} \\ IQuO(\overline{cl}) \equiv \begin{cases} (\hat{\Psi}_3)_{\overline{cl}} \equiv \left[(\hat{\bullet}_{1el} e^{i\pi/2} + \hat{\bullet}_{1in} e^{i\pi})_{(\overline{cl},k)} + (\hat{o}_{2el}^+ e^{i\pi} + \hat{o}_{2in}^+ e^{i\pi/2})_{(\overline{cl},-k)}\right] \\ (\hat{\Psi}_4)_{\overline{cl}} \equiv \left[(\hat{o}_{2el} e^{i\pi/2} + \hat{o}_{1in} e^{i\pi})_{(\overline{cl},k)} + (\hat{\bullet}_{2el}^+ \, e^{i\pi} + \hat{\bullet}_{2in}^+ e^{i\pi/2})_{(\overline{cl},-k)}\right] \end{cases} \end{cases} \quad (6.4)$$

| Come esempio diamo alcune configurazioni possibili di $\Psi_3$-IQuO ( verso orario ): | As an example we give some possible configurations of the $\Psi_3$-IQuO (cl.): |

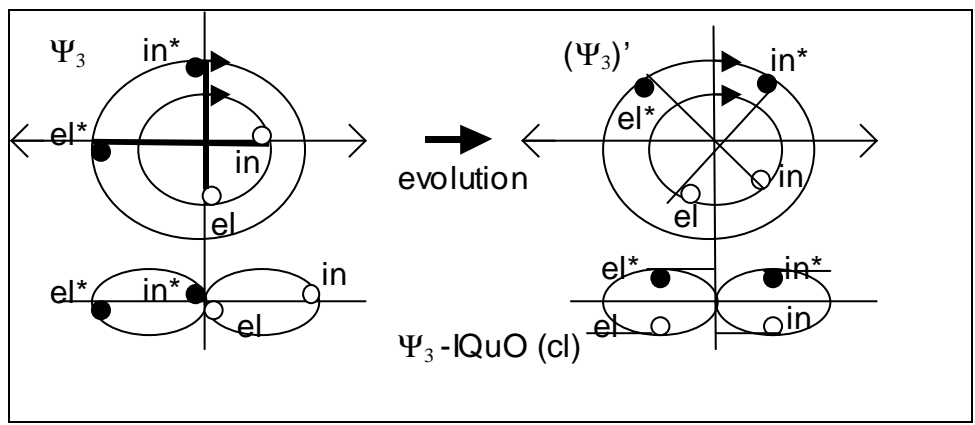

Fig.11

| L'IQuO antiorario ha invece una forma "irregolare"(eq. 3.6b). se confrontata con le configurazioni dell'IQuO orario. Auspichiamo per questo IQuO una forma regolare come: | The anticl. IQuO has a **"no regular"** form if compared with the configurations of the cl-IQuO. But we want that this IQuO has a **"regular"** form as: |
|---|---|

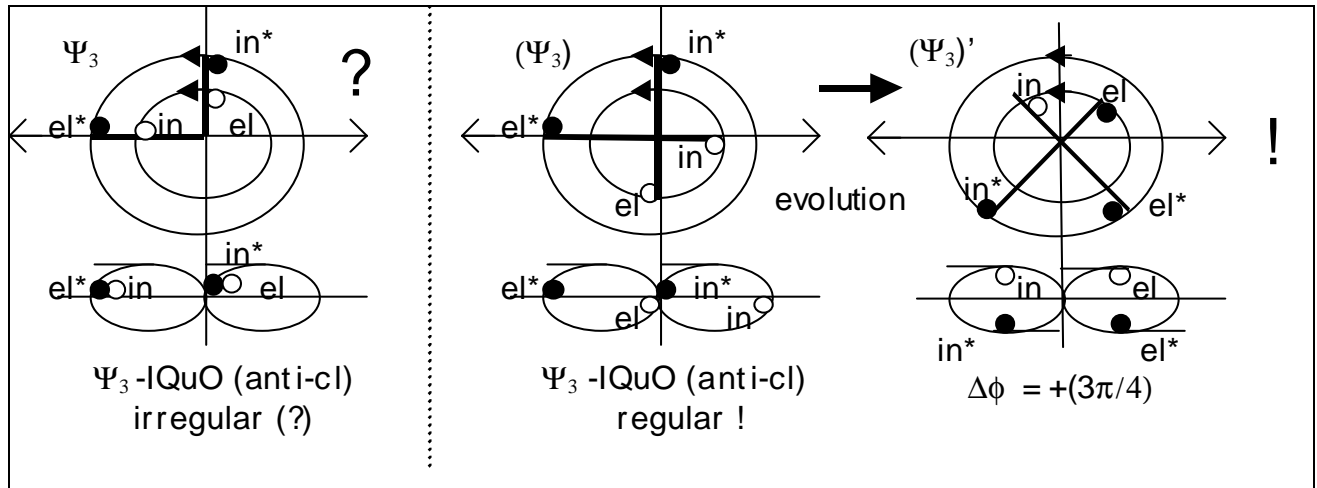

Fig.12

| Si rileva così la possibilità di avere combinazioni di IQuO che possono costituire dei buoni candidati per esprimere un oscillatore quantistico di campo con carica elettrica; vedi i campi K-G: $\Psi \equiv (\Psi_+, \Psi_-)$ Esprimendo in forma matriciale i campi-IQuO $\Psi$ potremmo avere in forma regolare le seguenti matrici: | Note that these possible combinations of IQuO could constitute some acceptable candidates for express a quantum oscillator of field with "electric charge"; you see the K-G fields: $\Psi \equiv (\Psi_+, \Psi_-)$ Expressing in matrix form these $\Psi$ IQuO-fields , we could conjecture, in regular form, the following matrixes : |
|---|---|

$$\begin{cases} \left(\left(\hat{\Psi}\right)\right)_{cl} = \begin{pmatrix} \left[(\hat{\mathbf{o}}_{1el}e^{-i\pi/2} + \hat{\mathbf{o}}_{1in})_{(cl,-k)}\exp(ikx)\right]\exp(-i\omega_{-k}t) \\ \left[(\hat{o}^+_{1el}e^{-i\pi} + \hat{o}^+_{1in}e^{-i3\pi/2})_{(cl,k)}\exp(ikx)\right]\exp(-i\omega_k t) \end{pmatrix} \\ \left(\left(\hat{\Psi}\right)\right)_{\overline{cl}} = \begin{pmatrix} \left[(\hat{\mathbf{o}}^+_{2el}e^{i\pi} + \hat{\mathbf{o}}^+_{2in}e^{i\pi/2})_{(\overline{cl},-k)}\exp(ikx)\right]\exp(i\omega_{-k}t) \\ \left[(\hat{o}_{2el}e^{-i\pi/2} + \hat{o}_{2in})_{(\overline{cl},k)}\exp(ikx)\right]\exp(i\omega_k t) \end{pmatrix} \end{cases} \quad (7.4)$$

| [ dove abbiamo adattato gli indici k secondo le configurazioni di $\Psi$-IQuO ( vedi figure 11 e 12 ): con ($\pm$ k) intendiamo che il semiquanto si sposta verso l'asse $\pm$ x. ]. Inoltre abbiamo cambiato la fase della coppia di semiquanti vuoti ($\mathbf{o_{el}}$, $\mathbf{o_{in}}$) dell'IQuO antiorario per renderlo simile in configurazione a quello orario. | [where the k index are adapted to the $\Psi$-IQuO configurations (you see figures 11 and 12): with ($\pm$ k) we mean that the semi-quanta go in the direction of the $\pm$ x axis.] Besides we have changed the phase of the ($\mathbf{o_{el}}$, $\mathbf{o_{in}}$) couple of the anti-clockwise IQuO, to make it **"regular"** in configuration like the clockwise one. |
|---|---|

Osserviamo che se avessimo effettuato nell'accoppiamento $\Phi_1 \oplus \Phi_2$ uno sfasamento opposto ( $\varphi_1 = +45°$ , $\varphi_2 = -45°$ ) a quello operato nella prima procedura d'accoppiamento ( $\varphi_1 = -45°$ , $\varphi_2 = +45°$ ), avremmo ottenuto l'IQuO orario $\Psi_1$ in forma irregolare mentre quello anti-orario $\Psi_2$ in forma regolare:

$(\Phi_1 \oplus \Phi_2)$ ➜ $(\Psi_{1(\text{irreg})} + \Psi_{2\,(\text{reg})})$.

Sospettiamo allora che deve accadere qualcosa durante l'accoppiamento $\Phi_1 \oplus \Phi_2$ ( oppure subito dopo ) che, operando uno sfasamento su $\Psi_{1(\text{irreg})}$ IQuO lo trasforma in forma regolare!

Una possibilità sarebbe quella di un accoppiamento con un terzo $\Phi_3$-IQuO tale che uno dei due ($\Psi_1$) acquisti una forma regolare a discapito di $\Phi_3$.

Possiamo allora affermare per quanto riguarda gli IQuO $\Psi$ a forma irregolare, che per i relativi campi carichi ci debba essere un $\Phi$ campo esterno che agisca localmente sulla fase di $\Psi$, regolarizzandone la forma dell'IQuO.

Può accadere anche il contrario e cioè un $\Phi_3$ campo esterno potrebbe alterare la forma regolare del $\Psi$-IQuO.

Diremo allora che nel $\Psi$ campo vi è sempre un campo agente $\Phi$ che può indurre variazioni della fase e di configurazioni.

Questo aspetto, come sappiamo dalla letteratura, è analogo a quello descritto dalla Meccanica Quantistica quando tratta di una sorgente carica "vestita" da un campo di fotoni virtuali !

Esaminando l'accoppiamento tra un $\Psi$-IQuO a forma non regolare ed un $\Phi$-IQuO virtuale ( nessun semiquanto pieno), si prova facilmente che $\Phi'$-IQuO è non regolare mentre $\Psi'$-IQuO è regolare in forma.

Adesso proviamo che un $\Psi$ IQuO- campo possiede una carica elettrica.

---

Observe that if we had executed a phase shift ( $\varphi_1 = +45°$ , $\varphi_2 = -45°$) in the ($\Phi_1 \oplus \Phi_2$) coupling opposed to that executed in the first coupling procedure ( $\varphi_1 = -45°$ , $\varphi_2 = +45°$ ) , we would have got the $\Psi_1$ clockwise IQuO as irregular while the $\Psi_2$ anticlockwise one as regular:

$(\Phi_1 \oplus \Phi_2)$ ➜ $(\Psi_{1(\text{irreg})} + \Psi_{2\,(\text{reg})})$

We suspect then that it must happen something, during the $\Phi_1 \oplus \Phi_2$ coupling (or immediately after), that, operating a phase shift on $\Psi_1$ irregular IQuO, transforms this into regular form.

A possibility would be that of a coupling with a third $\Phi_3$-IQuO, such that one of the two ($\Psi_1$) acquires a regular form to the detriment of the third ($\Phi_3$).

One can affirm that for any $\Psi$- IQuO with irregular form, it must exist a $\Phi$ external field which, acting locally on the phase of $\Psi$, adjusts of it the IQuO form.

Can happen also the contrary namely an $\Phi_3$ external field could alter the regular form of the $\Psi$-IQuO.

Thus we will say that around the field $\Psi$ there is always a $\Phi$-agent field which induces in it phase variations and configuration.

This appearance, as we know from literature, is analogous to one described by the quantum mechanics where every electric charge is "dressed" by fields of virtual photons!

Examining the coupling between a $\Psi$-IQuO, no regular form, and a $\Phi$-IQuO, we demonstrate easily, at end, that the $\Phi'$-IQuO outgoing is no regular, while the $\Psi'$-IQuO outgoing is in regular form.

Now we prove that a $\Psi$ IQuO-field has electric charge.

| Calcoliamo la carica elettrica del campo ($\Psi_{cl}$) nella rappresentazione a semi-quanti | We calculate the electric charge of the ($\Psi_{cl}$) field in the semi-quanta representation: |

$$\hat{\Psi}_{cl} \equiv \left[\left(\hat{\bullet}_{1el}^{+}\,e^{-i\pi} + \hat{\bullet}_{1in}^{+}\,e^{-i3\pi/2}\right)_{(cl,-k)} + \left(\hat{o}_{2el}\,e^{-i\pi/2} + \hat{o}_{2in}\right)_{(cl,k)}\right] \qquad (8a.4)$$

| In forma matrice ( vedi eq. 7.4) | in matrix form ( you see the eq. 7.4) |

$$\left(\left(\hat{\Psi}_{cl}\right)\right) \equiv \begin{pmatrix} \left(\hat{o}_{2el}\,e^{-i\pi/2} + \hat{o}_{2in}\right) \\ \left(\hat{\bullet}_{1el}^{+}\,e^{-i\pi} + \hat{\bullet}_{1in}^{+}\,e^{-i3\pi/2}\right) \end{pmatrix} \qquad (8b.4)$$

| Dove gli indici [(cl,k),(cl,-k)] sono stati omessi. Rileviamo che la distribuzione delle coppie di operatori di semi-quanto non è arbitraria, perché, come mostreremo più avanti, la disposizione degli elementi di matrice è correlata al verso di rotazione della fase. Per calcolare Q occorre considerare anche il campo aggiunto $\Psi^{+}$ | where the indices [(cl, k), (cl,-k)] have been omitted. The distribution of the operator pairs of semi-quanta is not arbitrary, because, as will be shown later, the dispositions of the matrix elements is related to the rotation direction of the phase. For calculate Q it's necessary to consider also the $\Psi^{+}$ hermitian field |

$$\left(\left(\hat{\Psi}_{cl}^{+}\right)\right) \equiv \begin{pmatrix} \left(\hat{o}_{el}^{+}\,e^{i\pi/2} + \hat{o}_{in}^{+}\right) \\ \left(\hat{\bullet}_{el}\,e^{i\pi} + \hat{\bullet}_{in}^{i3\pi/2}\right) \end{pmatrix} \qquad (9.4)$$

| Tuttavia dobbiamo congetturare che l'operazione di coniugazione contiene in sé lo scambio ($\bullet$ <=> o ); Avremo pertanto | but we ask that the conjugation operation of the $\Psi$-field contains the exchange ($\bullet \longleftrightarrow$ o ).Therefore we'll have: |

$$\left(\left(\hat{\Psi}_{cl}^{+}\right)\right) \equiv \begin{pmatrix} \left(\hat{\bullet}_{el}^{+}\,e^{i\pi/2} + \hat{\bullet}_{in}^{+}\right) \\ \left(\hat{o}_{el}\,e^{i\pi} + \hat{o}_{in}^{i3\pi/2}\right) \end{pmatrix} \qquad (10.4)$$

| Utilizzando la ben nota relazione che definisce la carica elettrica [1] e le relazioni di commutazioni dei semiquanti ( vedi appendice) segue allora che : | Using the well known definition of the electric charge [1] and the commutation relations with semi-quanta (you see appendix ) it follows then that |

$$Q_{cl} = \int \left[ \left( (\hat{\Psi}^+_{(-)}) \right) \hat{\sigma}_3 \left( (\hat{\Psi}_{(-)}) \right) \right] dV = \int \left( \begin{matrix} \hat{\bullet}^+_{el} \, e^{i\pi/2} + \hat{\bullet}^+_{in} \\ \hat{o}^+_{el} \, e^{i\pi} + \hat{o}^+_{in} \, e^{i3\pi/2} \end{matrix} \right) \left( \begin{matrix} \left( \hat{o}_{el} \, e^{-i\pi/2} + \hat{o}_{in} \right) \\ -\left( \hat{\bullet}^+_{el} \, e^{-i\pi} + \hat{\bullet}^+_{in} \, e^{-i3\pi/2} \right) \end{matrix} \right) dV \equiv$$

$$Q = \left[ \left( (\hat{\bullet}^+_{el} \, e^{i\pi/2}) + \hat{\bullet}^+_{in} \right) \hat{o}_{el} e^{-i\pi/2} + \hat{o}_{in} \right) - \left( \hat{o}_{el} e^{i\pi} + \hat{o}^+_{in} e^{i3\pi/2} \right) \left( \hat{\bullet}^+_{el} \, e^{-i\pi} + \hat{\bullet}^+_{in} \, e^{-i3\pi/2} \right) \right] =$$

$$\left( i \, \hat{\bullet}^+_{el} \, \hat{o}_{in} - i \, \hat{o}_{el} \, \hat{\bullet}^+_{in} \, e^{-i\pi} \hat{o}^+_{in} \right) + \left( \hat{\bullet}^+_{in} \, \hat{o}_{el} e^{-i\pi/2} - \hat{\bullet}^+_{in} \, e^{-i3\pi/2} \hat{o}_{el} e^{i\pi} \right)$$

$$= \left[ \hat{\bullet}^+_{el}, \hat{o}_{el} \right] + \left[ \hat{\bullet}^+_{in}, \hat{o}_{in} \right] + i \left[ \hat{\bullet}^+_{el}, \hat{o}_{in} \right] + i \left[ \hat{o}_{el}, \hat{\bullet}^+_{in} \right] = -1$$

(11.4)

| | |
|---|---|
| Dove $(\Psi_-) = (\Psi_{cl})$ e $\boldsymbol{\sigma}_3$ è una delle matrici di Pauli. | where $(\Psi_-) = (\Psi_{cl})$ and $\boldsymbol{\sigma}_3$ is a Pauli matrix. |
| Calcoliamo anche la carica elettrica dell'IQuO regolare $(\Psi_{anti-cl})$. | We calculate also the electric charge of the $(\Psi_{anti-cl})$ regular IQuO. |
| In forma matrice ( vedi eq. 7.4) si ha | in matrix form ( you see the eq. 7.4) it is |

$$\left( (\hat{\Psi}) \right)_{cl} = \left( \begin{matrix} \left[ (\hat{\bullet}^+_{el} e^{i\pi} + \hat{\bullet}^+_{in} e^{i\pi/2}) \right] \\ \left[ (\hat{o}_{el} e^{-i\pi/2} + \hat{o}_{in}) \right] \end{matrix} \right)$$

(12.4)

| | |
|---|---|
| E non | But it is not |

$$\left( (\hat{\Psi}) \right)_{cl} = \left( \begin{matrix} (\hat{o}_{el} e^{-i\pi/2} + \hat{o}_{in}) \\ (\hat{\bullet}^+_{el} e^{i\pi} + \hat{\bullet}^+_{in} e^{i\pi/2}) \end{matrix} \right)$$

(13.4)

| | |
|---|---|
| Perché, come vedremo più avanti, l'ordine dei coefficienti in $(\Psi_{cl})$ e $(\Psi_{anti-cl})$ è connesso al verso della rotazione di fase. Procedendo come prima segue | Because, as we shall see later, the order of the coefficients in $(\Psi_{cl})$ e $(\Psi_{anti-cl})$ is connected to the direction of phase rotation. Proceeding, it follows |

$$Q_{\overline{cl}} = \int \left[ \left( (\hat{\Psi}^+_{(+)}) \right) \hat{\sigma}_3 \left( (\hat{\Psi}_{(+)}) \right) \right] dV = \int \left( \begin{matrix} (\hat{o}_{el} e^{-i\pi} + \hat{o}_{in} e^{-i\pi/2}) \\ (\hat{\bullet}^+_{el} e^{i\pi/2} + \hat{\bullet}^+_{in}) \end{matrix} \right) \left( \begin{matrix} (\hat{\bullet}^+_{el} e^{i\pi} + \hat{\bullet}^+_{in} e^{i\pi/2}) \\ -(\hat{o}_{el} e^{-i\pi/2} + \hat{o}_{in}) \end{matrix} \right) dV =$$

$$= \left[ \left( \hat{o}_{el} e^{-i\pi} + \hat{o}_{in} e^{-i\pi/2} \right) \hat{\bullet}^+_{el} \, e^{i\pi} + \hat{\bullet}^+_{in} e^{i\pi/2} \right) - \left( \hat{\bullet}^+_{el} \, e^{i\pi/2} + \hat{\bullet}^+_{in} \right) \left( \hat{o}_{el} e^{-i\pi/2} + \hat{o}_{in} \right) \right] =$$

$$= \hat{o}_{el} \, \hat{\bullet}^+_{el} + \hat{o}_{el} \, \hat{\bullet}^+_{in} \, e^{-i\pi/2} + \hat{o}_{in} \, \hat{\bullet}^+_{el} \, e^{i\pi/2} + \hat{o}_{in} \, \hat{\bullet}^+_{in} - \hat{\bullet}^+_{el} \, \hat{o}_{el} - \hat{\bullet}^+_{el} \, \hat{o}_{in} e^{i\pi/2} - \hat{\bullet}^+_{in} \, \hat{o}_{el} e^{-i\pi/2} - \hat{\bullet}^+_{in} \, \hat{o}_{in} =$$

$$= \left[ \hat{o}_{el}, \hat{\bullet}^+_{el} \right] + \left[ \hat{o}_{in}, \hat{\bullet}^+_{in} \right] + i \left[ \hat{o}_{in}, \hat{\bullet}^+_{el} \right] + i \left[ \hat{o}_{in}, \hat{\bullet}^+_{el} \right] = +1$$

(14.4)

| | |
|---|---|
| Rileviamo anche che utilizzando la forma irregolare si ottiene lo stesso risultato! Infatti: | Note that using the irregular form the same result is got. In fact it is: |

$$Q_{\overline{cl}} = \int \left[ \left( (\hat{\Psi}^+_{(+)}) \right) \hat{\sigma}_3 \left( (\hat{\Psi}_{(+)}) \right) \right] dV =$$

$$= \left[ \left( \hat{o}_{el} e^{-i\pi} + \hat{o}_{in} e^{-i\pi/2} \right) \hat{\bullet}^+_{el} \, e^{i\pi} + \hat{\bullet}^+_{in} e^{i\pi/2} \right) - \left( \hat{\bullet}^+_{el} \, e^{-i\pi/2} + \hat{\bullet}^+_{in} \, e^{-i\pi} \right) \left( \hat{o}_{el} e^{i\pi/2} + \hat{o}_{in} e^{i\pi} \right) \right] =$$

$$= \left[ \hat{o}_{el}, \hat{\bullet}^+_{el} \right] + \left[ \hat{o}_{in}, \hat{\bullet}^+_{in} \right] + i \left[ \hat{\bullet}^+_{in}, \hat{o}_{el} \right] + i \left[ \hat{o}_{in}, \hat{\bullet}^+_{el} \right] = +1 \ !$$

15.4)

| | |
|---|---|
| Questo risultato ci dice che lo sfasamento ( operato da un agente esterno per regolarizzare l'IQuO ) non cambia il segno della carica elettrica e ciò corrisponde all'invarianza di gauge delle interazioni elettromagnetiche!<br><br>In breve rileviamo che l'accoppiamento iniziale di due IQuO ($\Phi_1 \oplus \Phi_2$), ha prodotto due particelle – IQuO con carica elettrica opposta: ($\Psi_{anti-cl}$) , ($\Psi_{cl}$).<br>**Si noti come il segno della carica elettrica è connesso al verso di rotazione della fase!** | This result says us that the  phase shift ( operated by an outside agent to regulate the IQuO),  don't changes the sign of the electric charge and this corresponds to the gauge invariance of the electromagnetic interactions!<br><br>In short we notice that: the ($\Phi_1 \oplus \Phi_2$) coupling of two IQuO,  has generated two particles - IQuO with opposite electric charge: ($\Psi_{anti-cl}$)  , ($\Psi_{cl}$).<br>**Note that the sign of electric charge is connected to direction of the phase rotation!** |
| Abbiamo sintetizzato il processo di accoppiamento tra due Φ-IQuO con | the process of coupling between two Φ-IQuO is synthesized with |

$$\hat{\Phi}_1 \oplus \hat{\Phi}_2 = \hat{\Psi}_+ + \hat{\Psi}_-$$ (16.4)

| | |
|---|---|
| Possiamo semplificare il meccanismo di accoppiamento ponendo i due IQuO nella stessa configurazione iniziale ( vedi Fig 10 e Fig. 14): | The coupling mechanism can be simplify by placing the two IQuO in the same initial configuration (see Fig 10 and Fig 14): |

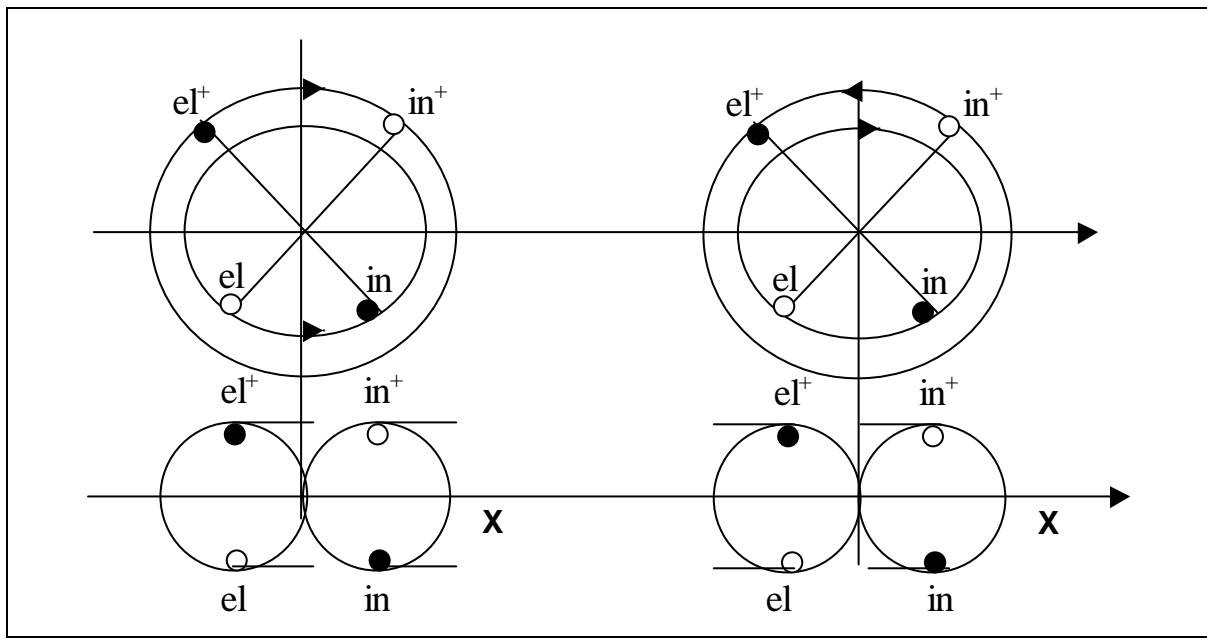

Fig. 13

| Avremo le seguenti matrici | We have the following matrices |
|---|---|

$$\left\{ \left( (\hat{\Phi}_1) \right) \equiv \begin{pmatrix} \hat{\bullet}_{1el}^+ e^{-i(5/4)\pi} + \hat{o}_{1in}^+ e^{-i(7/4)\pi} \\ \hat{o}_{1el} e^{i(5/4)\pi} + \hat{\bullet}_{1in} e^{i(7/4)\pi} \end{pmatrix} \quad \left( (\hat{\Phi}_2) \right) \equiv \begin{pmatrix} \hat{\bullet}_{2el}^+ e^{i(3/4)\pi} + \hat{o}_{2in}^+ e^{i(1/4)\pi} \\ \hat{o}_{2el} e^{-i(3/4)\pi} + \hat{\bullet}_{2in} e^{-i(1/4)\pi} \end{pmatrix} \right\}$$

(17.4)

| Guardando le eq. 7.4 possiamo scrivere in termini generici | Looking at eq. 7.4 we can write in general terms |
|---|---|

$$\left\{ \left( (\hat{\Phi}_1) \right) \equiv \begin{pmatrix} \hat{\bullet}_{1el}^+ e^{-i\alpha} + \hat{o}_{1in}^+ e^{-i\alpha^\circ} \\ \hat{o}_{1el} e^{i\beta} + \hat{\bullet}_{1in} e^{i\beta^\circ} \end{pmatrix} \quad \left( (\hat{\Phi}_2) \right) \equiv \begin{pmatrix} \hat{\bullet}_{2el}^+ e^{i\gamma} + \hat{o}_{2in}^+ e^{i\gamma^\circ} \\ \hat{o}_{2el} e^{-i\rho} + \hat{\bullet}_{2in} e^{-i\rho^\circ} \end{pmatrix} \right\}$$

(18.4)

| Dove [$(\alpha^\circ = \alpha + \pi/2)$, $(\beta^\circ = \beta + \pi/2)$, $(\gamma^\circ = \gamma - \pi/2)$, $(\rho^\circ = \rho - \pi/2)$ ] <br><br> In questo stato è possibile porre uno sfasamento reciproco $\Delta\varphi_1 = -\Delta\varphi_2 = 0$. <br> Si ottiene (vedi le eq. 10.3, eq. 11.3 e 13.3) utilizzando le matrici | Where [$(\alpha^\circ = \alpha + \pi/2)$, $(\beta^\circ = \beta + \pi/2)$, $(\gamma^\circ = \gamma - \pi/2)$, $(\rho^\circ = \rho - \pi/2)$ ] <br><br> It's possible that is $\Delta\varphi_1 = -\Delta\varphi_2 = 0$. <br><br> Using the matrices we obtain (see eq. 10.3, eq. 11.3 and 13.3) |
|---|---|

$$\hat{\Phi}_1 \oplus \hat{\Phi}_2 = \begin{pmatrix} \hat{\Phi}_1 \\ 1 \end{pmatrix} \cdot \left[ \begin{pmatrix} e^{-i\Delta\varphi} & 0 \\ 0 & e^{i\Delta\varphi} \end{pmatrix} \cdot \begin{pmatrix} 1 \\ \hat{\Phi}_2 \end{pmatrix} \right] = \begin{pmatrix} \hat{\Phi}_1 \\ 1 \end{pmatrix} \begin{pmatrix} 1 \\ \hat{\Phi}_2 \end{pmatrix} = \hat{\Psi}_\alpha + \hat{\Psi}_\beta$$

(19.4)

| Una possibile combinazione di semi-quanti (s-q) dell'IQuO 1 con quelli dell'IQuO 2 è riportata nella eq. 5.4. <br> Se usiamo la forma a matrice ( vedi eq. (17.4) ) per rappresentare questa combinazione, possiamo rappresentare il meccanismo di accoppiamento, una volta ammesso uno scambio di energia ($\bullet \leftarrow \rightarrow 0$) tra i due IQuO iniziali. <br> Segue ( vedi eq. 17.4) che: | A possible combination of semi-quanta (s-q) of the IQuO 1 with those of IQuO 2 is shown in Eq. 5.4. <br> If we use the matrix form (see eq. (17.4)) in this combination, we can represent the coupling mechanism between the two initial IQuO (after admitted an exchange of energy ($\bullet \leftarrow \rightarrow 0$) in the following way. <br><br> It follows ( from eq. 17.4): |
|---|---|

$$\left( \hat{\Phi}_1 \oplus \hat{\Phi}_2 \right) = \left\{ \hat{\Phi}_1 \left[ \hat{\Omega} \right] \hat{\Phi}_2 \right\} =$$

$$= \left\{ \begin{pmatrix} \hat{\bullet}_{1el}^+ e^{-i(5/4)\pi} + \hat{o}_{1in}^+ e^{-i(7/4)\pi} \\ \hat{o}_{1el} e^{i(5/4)\pi} + \hat{\bullet}_{1in} e^{i(7/4)\pi} \end{pmatrix} \left( \hat{\Omega} \right) \begin{pmatrix} \hat{\bullet}_{2el}^+ e^{i(3/4)\pi} + \hat{o}_{2in}^+ e^{i(1/4)\pi} \\ \hat{o}_{2el} e^{-i(3/4)\pi} + \hat{\bullet}_{2in} e^{-i(1/4)\pi} \end{pmatrix} \right\} =$$

$$= \begin{pmatrix} \hat{\bullet}_{1el}^+ e^{-i(5/4)\pi} + \hat{\bullet}_{1in}^+ e^{-i(7/4)\pi} \\ \hat{o}_{2el} e^{-i(3/4)\pi} + \hat{o}_{2in} e^{-i(1/4)\pi} \end{pmatrix} + \begin{pmatrix} \hat{\bullet}_{2el}^+ e^{i(3/4)\pi} + \hat{\bullet}_{2in}^+ e^{i(1/4)\pi} \\ \hat{o}_{1el} e^{i(5/4)\pi} + \hat{o}_{1in} e^{i(7/4)\pi} \end{pmatrix}$$

(20.4)

| Dove all'operazione ⊕ si è fatto corrispondere l'operazione Ω definita come: | Where the ⊕-operation is equivalent to the Ω-operation defined as: |

$$\hat{\Omega} = \left(\overline{\chi} \cdot \hat{\Sigma}_{(\bullet \Leftrightarrow o)} \cdot \hat{\phi}(0)\right) = \cdot \left[\left(\overline{\chi}\right)\left(\overline{\Sigma}(\bullet \Leftrightarrow o)\right)\begin{pmatrix} e^{-i\Delta\varphi} & 0 \\ 0 & e^{i\Delta\varphi} \end{pmatrix}_{\varphi=0}\right]$$

$$\text{with} \begin{cases} \hat{\phi}(0) = \begin{pmatrix} e^{-i\Delta\varphi} & 0 \\ 0 & e^{i\Delta\varphi} \end{pmatrix}_{\varphi=0} = \begin{pmatrix} 1 & 0 \\ 0 & 1 \end{pmatrix} \equiv (\text{matrix operator of phase shift}) \\ \left(\overline{\Sigma}(\bullet \Leftrightarrow o)\right) \equiv (\text{operation of energy exchange between s - q }) \\ \left(\overline{\chi}\right) \equiv \quad (\text{operation coupling diagonally of matricial elements}) \end{cases}$$

(21.4)

| Graficamente avremo: | Graphically it is: |

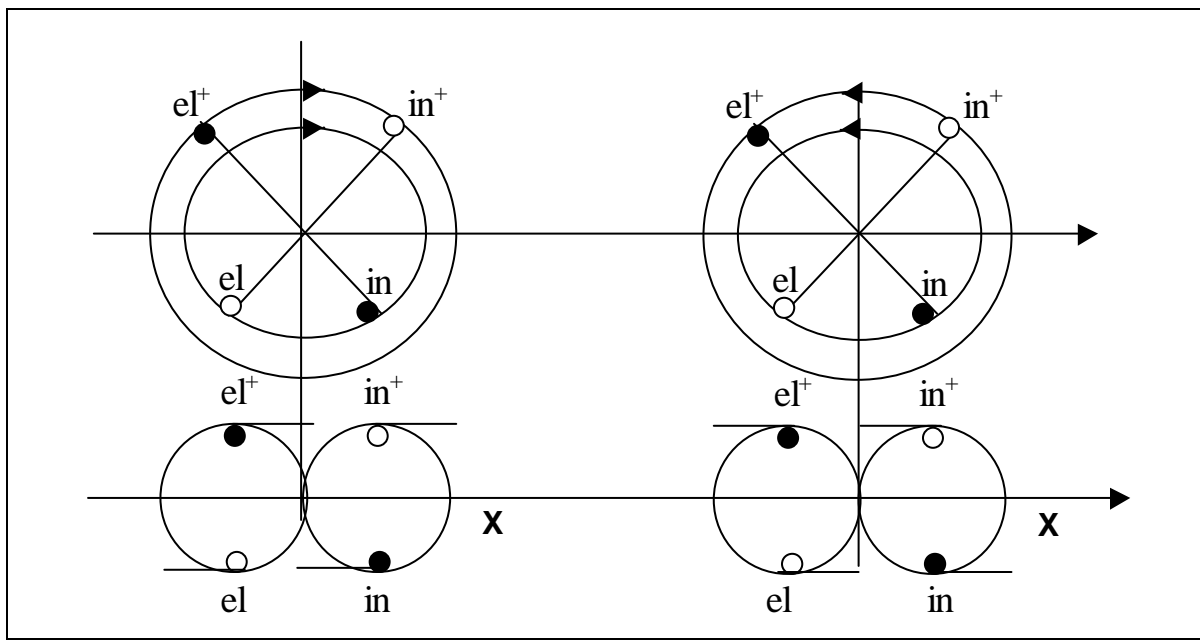

Fig. 14

| **Par.5) Antiparticelle come reinterpretazione delle soluzioni ad energia negativa dell'eq. di K-G** | **Sect. 5) Antiparticles as reinterpretation of negative energy solutions of the Eq. K-G** |

| Ricordiamo la forma del campo scalare complesso $\Psi_c$ di K-G | We recall the $\Psi_c$ complex scalar field of K-G: |

$$\hat{\Psi}_c = \cos t \cdot \sum_{-\infty}^{+\infty}(\hat{a}_p e^{-ip\cdot x} + \hat{b}_p^+ e^{ip\cdot x}) \quad , \quad \hat{\Psi}_c^+ = \cos t \cdot \sum_{-\infty}^{+\infty}(\hat{b}_p e^{-ip\cdot x} + \hat{a}_p^+ e^{ip\cdot x})$$

(1.5)

| $\Psi_c$ descrive particelle con carica elettrica data dall'eq. : | where $\Psi_c$ complex field describes particles with electric charge calculated by following equation |
|---|---|

$$Q = e\sum_k (\hat{a}_{0k}^+ \hat{a}_{0k} - \hat{b}_{0k}^+ \hat{b}_{0k})$$   (2.5)

| Abbiamo già detto che $\Psi_c$ e $\Psi^+_c$ esprimono in maniera "non separata" i campi componenti ($\Psi_+(\mathbf{a},\mathbf{a}^+)$, $\Psi_-(\mathbf{b},\mathbf{b}^+)$ ) associate rispettivamente a due valori opposti di un parametro "interno" che noi riconosciamo come carica elettrica. | We have already said that the fields $\Psi_c$ and $\Psi^+_c$ in a **"not separate envelope"** express the component fields ($\Psi_+(\mathbf{a},\mathbf{a}^+)$, $\Psi_-(\mathbf{b},\mathbf{b}^+)$ ), respectively associated to two opposite values of an **"inside "** parameter that we recognize as the electric charge. |
|---|---|

$$\hat{\Psi}_c \equiv \begin{cases} \hat{\Psi}_+ \equiv (\hat{a}, \hat{a}^+) \\ \hat{\Psi}_- \equiv (b, \hat{b}^+) \end{cases}$$ (3.5)

| Ricordiamo che la soluzione dell'equazione d'onda di K-G esprime i seguenti problemi:<br><br>1) una densità di probabilità negativa ρ<br>2) una soluzione con energia negativa<br><br>In letteratura, la (e) costante, chiamato carica elettrica, viene introdotto al fine di risolvere il primo problema. Questa costante trasforma la ρ in una densità di carica elettrica, che può anche assumere valori negativi.<br>Il secondo problema, come ben si sa dalla letteratura, è determinata dalla forma quadratica dell'equazione d'onda K-G, che ammette anche una soluzione con energia negativa.<br>Se riduciamo l'ordine dell'equazione d'onda (vedi l'equazione di Schrödinger e di Dirac) si potrebbe risolvere questa difficoltà: usiamo una rappresentazione con matrici della funzione d'onda K-G (vedi bibliografia [1], [3]) che trasforma l'equazione K-G in una equazione differenziale lineare del primo ordine, la cui forma è: | We remember that the solution of the K-G wave equation expresses the following problems:<br>1) a negative probability density ρ<br>2) one solution with negative energy<br><br>In the literature, the (**e**) constant, called electric charge, is introduced in order to solve the first problem. This constant transforms the ρ in a density of electric charge, which can also take negative values.<br>The second problem is determined by the quadratic form of the K-G wave equation, which also admit a solution with negative energy.<br><br>If we reduce the order of the wave equation (you see the Schrodinger equation and that of Dirac) we could to resolve this difficulty: we use a representation with matrixes of the K-G wave function ( you see bibliography [1], [3] ) which transforms the K-G equation into a differential linear equation of the first order, whose form is: |
|---|---|

$$\hat{\tilde{H}}\hat{\Psi} = \hat{H}\hat{\Psi}$$   (4.5)

| Con | with |
|---|---|

$$\hat{\Psi} = \begin{pmatrix} \Phi \\ X \end{pmatrix}$$

(5.5)

| Dove $\Phi$ e X sono funzioni complesse e dove | where $\Phi$ and $\Phi$ are complex functions and where |
|---|---|

$$\begin{cases} \hat{H} = \left\{ [(\hat{\sigma}_3 + i\hat{\sigma}_2)(\tilde{p}^2/2m)] + (mc^2\hat{\sigma}_3) \right\} \\ \tilde{p} = -i\hbar \dfrac{\partial}{\partial x} \\ \hat{\tilde{H}} = \sigma_4 \tilde{H} = \hat{\sigma}_4 \left( i\hbar \dfrac{\partial}{\partial t} \right) \end{cases}$$

(6.5)

| Con $(\sigma_i)$ le matrici di Pauli | and $(\sigma_i)$ are the Pauli's matrixes |
|---|---|

$$\hat{\sigma}_1 = \begin{pmatrix} 0 & 1 \\ 1 & 0 \end{pmatrix}, \ \hat{\sigma}_2 = \begin{pmatrix} 0 & -i \\ i & 0 \end{pmatrix}, \ \hat{\sigma}_3 = \begin{pmatrix} 1 & 0 \\ 0 & -1 \end{pmatrix}, \ \hat{\sigma}_4 = \begin{pmatrix} 1 & 0 \\ 0 & 1 \end{pmatrix}$$

(7.5)

| La soluzione completa è data dalla matrice colonna | The complete solution it's given by the matrix column: |
|---|---|

$$((\Psi_{\pm})) = \left( \frac{1}{\sqrt{V}} \right) \begin{pmatrix} \Phi_{\pm} \\ X_{\pm} \end{pmatrix} = \left( \frac{1}{\sqrt{V}} \right) \begin{pmatrix} \Phi_{0\pm} \\ X_{0\pm} \end{pmatrix} \left\{ \exp\left[ \left( \frac{i}{\hbar} \right)(px \mp \varepsilon t) \right] \right\}$$

(8.5)

| dove il segno (-) nel esponenziale si riferisce a onde che si propagano verso destra ($\Psi^{(D)}$), mentre il segno (+) ad onde che si propagano verso sinistra ($\Psi^{(S)}$); invece, nei coefficienti della $\Psi$, il segno (-) è riferito a $\varepsilon = - E_p$ e il segno (+) a $\varepsilon = +E_p$. Rileviamo che nell'esponenziale il termine ($\varepsilon t$) si riferisce alla variazione temporale della fase dell'oscillazione; in questo modo diventa possibile associare a (+ $E_p$) un verso antiorario di rotazione della fase e ad (- $E_p$) un verso orario. Si ottengono quindi quattro possibili funzioni d'onda | where the (-) sign into the exponential is referred to waves propagating toward right ( $\Psi^{(D)}$ ) while the (+) sign is referred to waves propagating toward left ($\Psi^{(S)}$); instead, into the coefficients of $\Psi$, the (-) sign is referred to $\varepsilon = - E_p$ and the (+) sign to $\varepsilon = +E_p$ . We note in exponential that the term ($\varepsilon t$) refers to the time variation of the phase of the oscillation; in this way it is possible to associate with (+$E_p$) a anticlockwise direction of the phase rotation, while to (-$E_p$ ) is associated a clockwise direction; we get therefore four possible wave functions |
|---|---|

$$(\Psi_+^D, \Psi_-^D) \equiv (\Psi_{anticl}^D, \Psi_{cl}^D) \quad ; \quad (\Psi_+^S, \Psi_-^S) \equiv (\Psi_{anticl}^S, \Psi_{cl}^S)$$

(9.5)

| Grazie alla condizione di normalizzazione abbiamo ottenuto [1] quattro soluzioni concrete sia per le onde "di destra" e quello "di sinistra", i cui termini noti sono costituiti da: | The normalization condition allows to obtain [1] four real solutions both for the waves that go " to right" and both for the one going " to left", whose the known terms are constituted by: |
|---|---|

$$\Phi_{0+}^{\ D} \equiv (\Phi_{0+}^{\ S}) = \frac{Mc^2 + E_p}{2\sqrt{E_p Mc^2}}; \quad X_{0+}^{\ D} \equiv (X_{0+}^{\ S}) = \frac{Mc^2 - E_p}{2\sqrt{E_p Mc^2}}$$

$$\Phi_{0-}^{\ D} \equiv (\Phi_{0-}^{\ S}) = \frac{Mc^2 - E_p}{2\sqrt{E_p Mc^2}}; \quad X_{0-}^{\ D} \equiv (X_{0-}^{\ S}) = \frac{Mc^2 + E_p}{2\sqrt{E_p Mc^2}}$$

(10.5)

| Con<br>+ ➜ rotazione antioraria<br>– ➜ rotazione oraria<br><br>**Si noti che nella (10.5) il verso di rotazione della fase è anche indirettamente "mostrato" attraverso i coefficienti della Ψ!**<br>Infatti (considerando la (10.5)) la matrice dei due (**Φ₀,X₀**) coefficienti (si veda la matrice colonna della eq. (8.5)) è diversa nei due casi Ψ± della funzione d'onda.<br>Inoltre ricordiamo che in particelle massive la direzione del movimento "verso destra" o "verso sinistra" è relativa all'osservatore, così i coefficienti differiscono esclusivamente per i segni ±.<br>Se invece gli elementi di matrici di eq. (8.5) non sono presi in considerazione, avremmo che | with<br>+ ➜ the anti-clockwise direction<br>– ➜ the clockwise direction<br><br>**Note that in the eq. (10.5) the direction of phase rotation is also indirectly "displayed" through the coefficients of the Ψ!**<br>In fact (considering the (10.5) relations) the matrix of the two (**Φ₀,X₀**) coefficients (you see the matrix column of the eq. (8.5)) is different in the two cases of Ψ±.<br>Also we remember that in massive particles the direction of the motion "toward right" or "toward left" is relative to observers, so the coefficients differ solely for the ± signs.<br><br>If instead the elements of matrices of the eq. (8.5) aren't taken into account, we would have that |
|---|---|

$$\boxed{\Psi^D(\varepsilon < 0) \equiv \Psi^S(\varepsilon > 0)}$$

(11.5)

| ma la matrice, attraverso i coefficienti, stabilisce in modo assoluto che le onde in relazione con il valore (ε <0) sono diverse da quelle associate a (ε > 0)!<br>Ne segue che con un cambiamento del sistema di riferimento non è possibile trasformare le energie negative in quelle positive. | but the matrix, through the coefficients, establishes in absolute way that the waves with (ε < 0) are different by those associated with (ε > 0)!<br>Therefore with a change of the reference frame is not possible to transform negative energy into positive one. |
|---|---|

| Infatti le quattro diverse soluzioni dell'equazione d'onda sono: | In fact the four different solutions of the wave equation are : |
|---|---|

$$\begin{cases} \left(\left(\Psi_+^D\right)\right)\left(\varepsilon = +E_p > 0 \equiv \omega > 0 \equiv \overline{cl}\right) = \left(\frac{1}{\sqrt{Vol}}\right)\binom{\Phi_{0(+)}}{X_{0(+)}}\left\{\exp\left[\left(\frac{i}{\hbar}\right)(\vec{p}\cdot\vec{x} - E_p t)\right]\right\} \\ \left(\left(\Psi_+^S\right)\right)\left(\varepsilon = +E_p > 0 \equiv \omega > 0 \equiv \overline{cl}\right) = \left(\frac{1}{\sqrt{Vol}}\right)\binom{\Phi_{0(+)}}{X_{0(+)}}\left\{\exp\left[\left(\frac{i}{\hbar}\right)(\vec{p}\cdot\vec{x} + E_p t)\right]\right\} \end{cases} \quad (12a.5)$$

$$\begin{cases} \left(\left(\Psi_-^D\right)\right)\left(\varepsilon = -E_p < 0 \equiv \omega < 0 \equiv cl\right) = \left(\frac{1}{\sqrt{Vol}}\right)\binom{\Phi_{0(-)}}{X_{0(-)}}\left\{\exp\left[\left(\frac{i}{\hbar}\right)(\vec{p}\cdot\vec{x} + E_p t)\right]\right\} \\ \left(\left(\Psi_-^S\right)\right)\left(\varepsilon = -E_p < 0 \equiv \omega < 0 \equiv cl\right) = \left(\frac{1}{\sqrt{Vol}}\right)\binom{\Phi_{0(-)}}{X_{0(-)}}\left\{\exp\left[\left(\frac{i}{\hbar}\right)(\vec{p}\cdot\vec{x} - E_p t)\right]\right\} \end{cases} \quad (12b.5)$$

where $\overline{cl}$ = anti - clockwise direction and cl = clockwise direction

| E la costante Q sarà | And the **Q** constant will be: |
|---|---|

$$Q_{\left[\left(\overline{cl} \equiv +\right),(cl \equiv -)\right]} = Q_\pm = \int\limits_{Vol}\left[\left(\Psi_\pm^+\right)\left(\hat{\sigma}_3\right)\left(\Psi_\pm\right)\right]dV = \left[\left(\Phi_\pm^*\right)\left(\Phi_\pm\right) - \left(X_\pm\right)\left(X_\pm^*\right)\right] =$$

$$= \left[\left(\Phi_{0(\pm)}^*\right)\left(\Phi_{0(\pm)}\right) - \left(X_{0(\pm)}\right)\left(X_{0(\pm)}^*\right)\right] = \pm 1 \quad (13.5)$$

| dove gli indici (+, -) della funzione d'onda sono in corrispondenza con i due versi di rotazione della fase: | where the (+, -) indices of the wave function are in correspondence con the two directions of the phase rotation: |
|---|---|

$$\begin{cases} Q = +1 \quad <=> \text{anti - clockwise direction ((+) index)} \\ Q = -1 \quad <=> \quad \text{clockwise direction ((-) index)} \end{cases}$$

| **Osserviamo subito che: la direzione di rotazione della fase, attraverso i coefficienti ($\varphi_0, \chi_0$) determina il segno della costante Q denominata "carica elettrica"!** Ma l'evidenza sperimentale ci impone che le onde di energia negativa devono essere interpretate in onde a energia positiva. | **We observe immediately that the direction of phase rotation, through the coefficients ($\varphi_0, \chi_0$) determines the sign of the constant Q denominated "electric charge"!** But the experimental evidence imposes that the waves with negative energy must be interpreted as waves with positive energy. |
|---|---|

| Dalle relazioni (8.5) , (10.5)  e dalle (12.5)  si vede subito che: | With the relations (8.5) , (10.5)    and (12.5) we see immediately that: |

$$\left\{ \left(\left((\Psi)_-^D\right)\right)(\varepsilon<0)=\left(\frac{1}{\sqrt{Vol}}\right)\binom{\Phi_{0(-)}}{X_{0(-)}}\left\{\exp\left[\left(\frac{i}{\hbar}\right)\left(\vec{p}\cdot\vec{x}+E_p t\right)\right]\right\}=\left(\frac{1}{\sqrt{Vol}}\right)\binom{X_{0(+)}}{\Phi_{0(+)}}\left\{\exp\left[\left(\frac{i}{\hbar}\right)\left(\vec{p}\cdot\vec{x}+E_p t\right)\right]\right\}\right.$$

$$\left.\left(\left((\Psi)_-^S\right)\right)(\varepsilon<0)=\left(\frac{1}{\sqrt{Vol}}\right)\binom{\Phi_{0(-)}}{X_{0(-)}}\left\{\exp\left[\left(\frac{i}{\hbar}\right)\left(\vec{p}\cdot\vec{x}-E_p t\right)\right]\right\}=\left(\frac{1}{\sqrt{Vol}}\right)\binom{X_{0(+)}}{\Phi_{0(+)}}\left\{\exp\left[\left(\frac{i}{\hbar}\right)\left(\vec{p}\cdot\vec{x}-E_p t\right)\right]\right\}\right\}$$

(14-5)

| Ricordando che | remembering that |

$$\sigma_1\binom{a}{b}=\binom{b}{a}$$
(15 -5)

| Notiamo in questo modo che | we note in this way that |

$$(\Psi)_-^D(\varepsilon<0)\equiv(\Psi)_{cl}^D\equiv\hat{\sigma}_1(\Psi)_+^S(\varepsilon>0)$$
$$(\Psi)_-^S(\varepsilon<0)\equiv(\Psi)_{cl}^S\equiv\hat{\sigma}_1(\Psi)_+^D(\varepsilon>0)$$
(16 -5)

| Avremo così | We'll have so: |

$$\left(\left((\Psi)_-^D\right)\right)(\varepsilon<0)\equiv\hat{\sigma}_1\left(\left((\Psi)_+^S\right)\right)(\varepsilon>0)=\left\{\binom{X_{0(+)}}{\Phi_{0(+)}}\left[\exp\left[\frac{i}{\hbar}\left(\vec{p}\cdot\vec{x}+E_p t\right)\right]\right]\right\}\equiv\left(\left((\overline{\Psi})_{cl}^S\right)\right)(\varepsilon>0)$$

$$\hat{\sigma}_1\left(\left((\Psi)_+^S\right)\right)(\varepsilon>0)\equiv\left(\left((\Psi)_-^D\right)\right)(\varepsilon<0)=\left\{\binom{\Phi_{0(cl)}}{X_{0(cl)}}\left[\exp\left[\frac{i}{\hbar}\left(\vec{p}\cdot\vec{x}+E_p t\right)\right]\right]\right\}\equiv\left(\left((\overline{\Psi})_{cl}^S\right)\right)(\varepsilon>0)$$
(17.5a)

$$\left(\left((\Psi)_-^S\right)\right)(\varepsilon<0)\equiv\hat{\sigma}_1\left(\left(\Psi_+^D\right)\right)(\varepsilon>0)=\left\{\binom{X_{0(+)}}{\Phi_{0(+)}}\left[\exp\left[\frac{i}{\hbar}\left(\vec{p}\cdot\vec{x}-E_p t\right)\right]\right]\right\}\equiv\left(\left((\overline{\Psi})_{cl}^D\right)\right)(\varepsilon>0)$$

$$\hat{\sigma}_1\left(\left(\Psi_+^D\right)\right)(\varepsilon>0)\equiv\left(\left(\Psi_-^S\right)\right)(\varepsilon<0)=\left\{\binom{\Phi_{0(cl)}}{X_{0(cl)}}\left[\exp\left[\frac{i}{\hbar}\left(\vec{p}\cdot\vec{x}-E_p t\right)\right]\right]\right\}\equiv\left(\left((\overline{\Psi})_{cl}^D\right)\right)(\varepsilon>0)$$
(17.5b)

| Dove abbiamo sostituito i segni (+,-) con la (anti-cl, cl) i versi di rotazione. | Where we have replaced the (+, -) signs  with the ( anti-cl. , cl ) directions of rotation. |

| Ma confrontando le eq. (17-5) con l'eq. (12 -5) notiamo che: | But comparing the eq. (17-5)  with the eq. (12-5)  we note that: |

$$(\overline{\Psi})^S_{cl}(\varepsilon > 0) \neq (\Psi)^S_+(\varepsilon > 0) = (\Psi)^{\frac{S}{cl}}(\varepsilon > 0)$$
$$(\overline{\Psi})^D_{cl}(\varepsilon > 0) \neq (\Psi)^D_+(\varepsilon > 0) = (\Psi)^D_{cl}(\varepsilon > 0)$$

(18-5)

| Abbiamo così ottenuto particelle con energia positiva | Therefore we have got particles with positive energy |

$$(\overline{\Psi})^S_{cl}(\varepsilon > 0)$$   (19-5)

| da quelle con energia negativa, ma diverse nella direzione della rotazione di fase da | from the ones with negative energy but different in the direction of phase rotation from the |

$$(\Psi)^{\frac{S}{cl}}(\varepsilon > 0)$$   (20-5)

| particelle già ad energia positiva!<br>Pertanto una particella ad energia negativa potrebbe essere interpretata come particella ad energia positiva se il campo esploratore rileva e distingue la "disposizione dei coefficienti", il verso di rotazione di fase della funzione d'onda e la direzione di movimento della particella nello spazio.<br>Ne consegue che un campo esploratore vede una particella "da destra" con energia negativa (in senso orario di rotazione di fase) come una particella "da sinistra", ma con energia positiva, con un valore (-1) della variabile indicata con Q (vedi eq. 4.12) diverso però da quello (+1) associata a una particella "a sinistra", originariamente di energia positiva!<br>E 'evidente che la caratteristica fisica (Q), , che mette in relazione il verso  di rotazione di fase con i coefficienti della funzione d'onda, è ciò che siamo soliti chiamare come "carica elettrica (Q = ±e).<br>Si noti che queste affermazioni portano alla invarianza delle leggi fisiche per CPT trasformazioni. | particles already by positive energy!<br>Therefore  a particle with negative energy could be interpreted as particle with positive energy if the explorer field detects and distinguishes both the "disposition of the coefficients", both the direction of phase rotation of the wave function and both the motion direction of the particle in the space.<br>So an explorer field sees a negative energy particle coming " from the right" (clockwise direction of phase rotation) as a positive energy particle coming "from the left", but with and a value (-1) of  Q ( see eq. 4.12) different however by the one         (+1) associated to a particle coming " from left" originally with positive energy!<br><br>It's evident that the   physical characteristic (Q), which correlates the direction of phase rotation with the coefficients of the wave function, is what we usually call as " electric charge (Q = ±e).<br>Note that these affirmations lead to the invariance of  physical laws for **CPT** transformations. |

Infatti, interpretare la soluzione con energia negativa come antiparticella è equivalente ad effettuare le seguenti trasformazioni:

1) lo scambio destra ⬅➡ sinistra coincide con la P-Trasformazione
2) lo scambio rotazione oraria ⬅➡ rotazione antiorario coincide con la T-Trasformazione

3) costruire la trasformazione $\sigma_1 \begin{pmatrix} a \\ b \end{pmatrix} = \begin{pmatrix} b \\ a \end{pmatrix} \equiv$ scambio (+)⬅➡(-) coincide con la C-trasformazione ma ricordando che le antiparticelle sono soluzioni della stessa equazione delle particelle, ne consegue che le leggi della fisica sono invarianti per le trasformazioni CPT.
Riassumendo avremo allora

In fact, interpret the solution with negative energy as antiparticle is equivalent to effect the following transformations:

1) the exchange Right ⬅ ➡ Left coincides with P-Transformation
2) the exchange clockwise direction ⬅ ➡ anticlockwise direction = T- Transformation

3) making $\sigma_1 \begin{pmatrix} a \\ b \end{pmatrix} = \begin{pmatrix} b \\ a \end{pmatrix}$ is equivalent to exchange (+)⬅ ➡ (-) = C-Transformation but remembering that the antiparticles are solutions of the same equation associated with the particles, it follows that the physics laws are invariant for CPT transformations.

Summarizing we will have then

$$\hat{\bar{\Psi}}_{cl}^{S}(\varepsilon > 0) = \hat{\bar{\Psi}}_{-}^{S}(e = -1) \neq \hat{\Psi}_{+}^{S}(e = +1)$$
$$\hat{\bar{\Psi}}_{cl}^{D}(\varepsilon > 0) = \hat{\bar{\Psi}}_{-}^{D}(e = -1) \neq \hat{\Psi}_{+}^{D}(e = +1)$$

$$\text{with} \begin{cases} + \equiv \text{anticlockwise direction} \\ - \equiv \text{clockwise direction} \end{cases}$$

(21-5)

Se (C) è l'operazione definita come coniugazione di carica [1]

(C)[Ψ] = σ₁ [Ψ] *(p => -p)

osserviamo che questo coincide proprio con la nostra procedura di interpretazione delle soluzioni a energia negativa.
Infatti se applichiamo l'operatore C alla funzione $\Psi_{+(-)}$, otteniamo una $\Psi^*_{-(+)}$ soluzione con carica opposta a quella iniziale, ma con impulso opposto:

((C))Ψ₊ = σ₁ [ Ψ₊ ]*(p =>-p) = Ψ*₋

If (C) is the operation definite as charge conjugation [1] or

(C)[Ψ] = σ₁ [Ψ] *(p => -p)

observe that this just coincides with our interpretation procedure of the solutions to negative energy.
In fact if we apply the C operator to a $\Psi_{(-)}$ function, we get a $\Psi^*_{-(+)}$ solution with opposite charge to the initial but with opposite impulse:

((C))Ψ₊ = σ₁ [ Ψ₊ ]*(p =>-p) = Ψ*₋

| Ma nel nostro caso avremmo anche che | But in our case we have that |
|---|---|

$$\hat{C}(\Psi)_+^D = \frac{1}{\sqrt{Vol}}\begin{pmatrix} X_{0(+)} \\ \Phi_{0(+)} \end{pmatrix}\exp\left[\frac{-i}{\hbar}\left(-\vec{p}\cdot\vec{x} - E_p t\right)\right] = \frac{1}{\sqrt{Vol}}\begin{pmatrix} \Phi_{0(-)} \\ X_{0(-)} \end{pmatrix}\exp\left[\frac{i}{\hbar}\left(\vec{p}\cdot\vec{x} + E_p t\right)\right] = (\overline{\Psi})_-^S \, (\varepsilon > 0)$$

(22-5)

| | |
|---|---|
| Dove $(\overline{\Psi})_-^S$ (antiparticella di $(\Psi)_+^D$ ) è anche la reinterpretazione di una particella con energia negativa $[(\Psi)_-^D(\varepsilon < 0)]$. Si ottiene che ogni particella ad apparente energia negativa viene interpretato (da un campo esploratore) come antiparticella. <br><br> Abbiamo quindi dimostrato che il processo di reinterpretare le particelle con energia negativa contiene in sé l'invarianza CPT. <br> Se il campo esploratore non rileva la direzione della rotazione di fase, la particella ci appare elettricamente neutra e sarà descritta da una funzione d'onda reale! <br> Nel caso di un campo scalare reale è facile dimostrare che la $\Psi$ matrice contiene i due versi di rotazione di fase, mentre i coefficienti noti, relativi ai (+,-) segni sono "misti". Infatti avremo | Where $(\overline{\Psi})_-^S$ (antiparticle of $(\Psi)_+^D$ ), is also the reinterpretation of a particle with negative energy $[(\Psi)_-^D(\varepsilon < 0)]$. We obtain so that each particle with apparent negative energy is interpreted ( by an explorer field ) as antiparticle. <br><br> Therefore we have shown that the process of reinterpreting the particles with negative energy contains in itself the CPT invariance. <br> If the explorer field does not detect the direction of the phase rotation, the particle will be neutral electrically and it will described by a real wave function! <br> In the case of a real field scalar is easy show that the $\Psi$ matrix contains the both directions of the phase rotation, while the known coefficients, relative to the signs (+,-) are "**mixed**". In fact we'll have |

$$(\Psi)_R = \begin{pmatrix} \Phi_p \\ \Phi_{-p}^* \end{pmatrix} = \begin{pmatrix} \Phi_\pm(t)\left[\exp\left((i/\hbar)\,px\right)\right] \\ X_\mp(t)\left[\exp\left((i/\hbar)\,px\right)\right] \end{pmatrix} =$$
$$= \begin{pmatrix} (\Phi_{0\pm})\left[\exp\left(\pm(i/\hbar)E_p t\right)\right]\left[\exp\left((i/\hbar)\,px\right)\right] \\ (X_{0\mp})\left[\exp\left(\mp(i/\hbar)E_{-p}t\right)\right]\left[\exp\left((i/\hbar)\,px\right)\right] \end{pmatrix}$$

(23-5)

| | |
|---|---|
| In queste condizioni un campo explorer non riesce a rilevare una direzione di rotazione di fase né l'associazione tra questa ed i coefficienti di $(\Psi)$! <br> Infatti troviamo che il parametro Q è zero, perché succede | In this relation an explorer field cannot detect a direction in the phase rotation neither the association between this rotation and the coefficients of $(\Psi)$! <br> In fact the Q parameter is zero, because it is |

$$Q_{\left[(\overline{cl}\equiv+),(cl\equiv-)\right]} = Q_\pm = \int_{Vol}\left[(\Psi_\pm^+)(\hat{\sigma}_3)(\Psi_\pm)\right]dV = \left[(\Phi_\pm^*)(\Phi_\pm) - (\Phi_\pm)(\Phi_\pm^*)\right] =$$
$$= \left[(\Phi_{0(\pm)}^*)(\Phi_{0(\pm)}) - (\Phi_{0(\pm)})(\Phi_{0(\pm)}^*)\right] = 0$$

(24-5)

Quindi ci deve essere qualcosa nel campo esploratore che gli permette di distinguere quegli aspetti fisici descritti dai coefficienti della funzione d'onda della particella.

So there must be something into explorer field that allows it to distinguish those physical aspects which are described by the coefficients of the wave function of the particle.

Allora deve esistere un meccanismo di accoppiamento tra particella e campo esploratore che permette a quest'ultimo, attraverso alcuni indicatori specifici, "di leggere" le particelle ad energia negativa (rotazione di fase in senso orario e coefficienti negativi (**Φ.**, **X.**) come particelle ad energia positiva (antiparticella).

Then there must be a coupling mechanism between particles and field explorer that allows to it, through some specific indicators, "of reading" the particles with negative energy (phase rotation in a clockwise direction and negative coefficients (**Φ.**, **X.**) ) as particles with positive energy (antiparticle).

Lo stesso meccanismo permetterebbe al campo esploratore di distinguere particelle con carica elettrica positiva da quelle con carica negativa.

It's evident that the same mechanism would allow the field explorer to distinguish particles with a positive electric charge from those with a negative charge.

E 'intuitivo capire che questo indicatori sono messe in relazione con gli (**a, a⁺**) operatori di entrambi i campi, come il campo di carica elettrica e l'esploratore.

It's intuitive to understand that these indicators are put in relation with the (**a, a⁺**) operators of both the fields.

La configurazione dei semi-quanti, esposti nelle sezioni precedenti, potrebbero fornire correttamente il corretto "indicatore" che permette a un campo esploratore di distinguere la carica elettrica di una particella!

The configuration of the semi-quanta could furnish to explorer field the correct      " indicator " that may allow to it of distinguish the electric charge of a particle!

Ora, considerando ciò che è stato affermato nella terza sezione, sorge spontanea di stabilire la seguente corrispondenza tra le relazioni ( eq. (17.5)) scritte con operatori :

Now, it rises spontaneous to establish the following correspondence between this equations:

$$\left\{ \begin{array}{l} \left( \left( \Psi_+^S \right) \right) = \begin{pmatrix} \Phi_{0p(+)} \exp\left[ \dfrac{i}{\hbar}(+E_p t) \right] \\ X_{0p(+)} \exp\left[ \dfrac{i}{\hbar}(+E_p t) \right] \end{pmatrix} <=> \left( \left( \hat{\Psi}_+^S \right) \right) = \begin{pmatrix} \hat{d}_{02p(+)}^+ \exp\left[ \dfrac{i}{\hbar}(+E_p t) \right] \\ \hat{d}_{01p(+)} \exp\left[ \dfrac{i}{\hbar}(+E_p t) \right] \end{pmatrix} \\[40pt] \left( \left( \overline{\Psi}_-^D \right) \right) = \begin{pmatrix} \Phi_{0p(-)} \exp\left[ \dfrac{i}{\hbar}(-E_{-p} t) \right] \\ X_{0p(-)} \exp\left[ \dfrac{i}{\hbar}(-E_{-p} t) \right] \end{pmatrix} <=> \left( \left( \hat{\overline{\Psi}}_-^D \right) \right) = \begin{pmatrix} \hat{c}_{02p(-)} \exp\left[ \dfrac{i}{\hbar}(-E_{-p} t) \right] \\ \hat{c}_{01p(-)}^+ \exp\left[ \dfrac{i}{\hbar}(-E_{-p} t) \right] \end{pmatrix} \end{array} \right\} \qquad (25.5)$$

| E le relazioni (7.4), scritte con operatori di semi-quanto. | And the relations (7.4), written with operator of semi-quanta |
|---|---|

$$\begin{cases} \left(\left(\hat{\Psi}_{cl}\right)\right) = \begin{pmatrix} \left[(\hat{\bullet}_{1el}\,e^{-i\pi/2} + \hat{\bullet}_{1in})_{(cl,-k)}\exp(ikx)\right]\exp(-i\omega_{-k}t) \\ \left[(\hat{o}^+_{1el}\,e^{-i\pi} + \hat{o}^+_{1in}\,e^{-i3\pi/2})_{(cl,k)}\exp(ikx)\right]\exp(-i\omega_k t) \end{pmatrix} = \left(\left(\hat{\overline{\Psi}}^{\,D}_-\right)\right) \\[3ex] \left(\left(\hat{\overline{\Psi}}_{\overline{cl}}\right)\right) = \begin{pmatrix} \left[(\hat{\bullet}^+_{2el}\,e^{i\pi} + \hat{\bullet}^+_{2in}\,e^{i\pi/2})_{(\overline{cl},-k)}\exp(ikx)\right]\exp(i\omega_{-k}t) \\ \left[(\hat{o}_{2el}\,e^{-i\pi/2} + \hat{o}_{2in})_{(\overline{cl},k)}\exp(ikx)\right]\exp(i\omega_k t) \end{pmatrix} = \left(\left(\hat{\Psi}^{\,S}_+\right)\right) \end{cases} \quad (27.5)$$

| Da notare che i coefficienti della funzione d'onda si esprimono attraverso la struttura di semi-quanti rappresentante l'oscillatore campo.\n\nIn questo modo l'idea che associa il segno della carica elettrica al parametro interno alla particella è fondata, perché il parametro "interno" corrisponde ad un particolare sub-struttura quantistica dell'oscillatore campo! | Note here that the coefficients of the wave function are expressed through the structure by semi-quanta representative the field oscillator.\n\nIn this way the associating idea the sign of the electric charge to a the inside parameter of the particle is founded, because the "inside" parameter corresponds to a particular sub-quantum structure of the field oscillator! |
|---|---|

| **Par.6) IQuO con carica elettrica non intera** | **Sect. 6) IQuO having electric charge not integer** |
|---|---|
| La presenza di multipletti di mesoni e di Barioni ha indotto Gell-Mann a descrivere gli adroni mediante tre grandezze fisiche: l'ipercarica (Y), l'isospin ($T_3$) e naturalmente la carica elettrica Q.\nLo spazio astratto che descrive gli adroni, costruito su queste tre grandezze, richiede il gruppo di trasformazioni con simmetria SU(3).\n\nI generatori del gruppo sono le matrici di Gell-Mann, associate proprio alle tre grandezze (Y, T,Q).\nFocalizziamo l'attenzione sulle matrici "di base" $\lambda_3$ e $\lambda_8$, due generatori del gruppo [4]: | The presence of multiplets of mesons and baryons has led Gell-Mann to describe the hadrons through three physical quantities: hypercharge (Y), the isospin (T3) and the electrical charge Q.\nThe abstract space to describe the hadrons, built on these three variables, requires the transformation group denoted by SU (3).\n\nThe generators of the group are the Gell-Mann matrices, which are properly associated with the three variables (Y, T, Q).\nWe focus on matrices $\lambda_3$ and $\lambda_8$, two generators of group [4]: |

$$\left\{ \hat{\lambda}_3 = \begin{pmatrix} 1 & 0 & 0 \\ 0 & -1 & 0 \\ 0 & 0 & 0 \end{pmatrix} \quad \hat{\lambda}_8 = \frac{1}{\sqrt{3}}\begin{pmatrix} 1 & 0 & 0 \\ 0 & 1 & 0 \\ 0 & 0 & -2 \end{pmatrix} \right\} \quad (1.6)$$

| | |
|---|---|
| La matrice $\lambda_3$ è la proiezione in SU(3) della matrice di Pauli $\sigma_3$. Quest'ultima nella rappresentazione degli stati con autovettori di base $(\Psi_+, \Psi_-)$ fornisce i valori di Q (vedi l'eq. 11.4 ). | The $\lambda_3$ matrix is the projection in SU (3) of the Pauli matrix $\sigma_3$, which, in the representation of the eigenvectors basic $(\Psi_+, \Psi_-)$, gives the values of Q (see eq. 11.4). |
| Inoltre la matrice $\sigma_3$ viene anche utilizzata per rappresentare la proiezione $(T_3)$ dell'Isospin T definito nello spazio complesso Q(2) e poi proiettato in U(3). | Furthermore, the $\sigma_3$ matrix is also used to represent the projection (T3) of isospin T in the space complex Q (2). |
| Difatti si ha: | In fact it's: |

$$\begin{cases} \left[ \left. \hat{T}_i \right|_{i=1,2,3} = \left(\frac{1}{2}\right) \left. \hat{\sigma}_i \right|_{i=1,2,3} \right] \in U(2) \\ \left[ \left. \hat{T}_i \right|_{i=1,2,3} = \left(\frac{1}{2}\right) \left. \hat{\lambda}_i \right|_{i=1,2,3} \right] \in U(3) \\ \text{con } \hat{T}_3 = \left(\frac{1}{2}\right) \hat{\lambda}_3 = \left(\frac{1}{2}\right) \begin{pmatrix} 1 & 0 & 0 \\ 0 & -1 & 0 \\ 0 & 0 & 0 \end{pmatrix} \end{cases} \quad (2.6)$$

| | |
|---|---|
| Dove le matrici $(T_i)$ con ( i= 1,2,3) generano il sottogruppo SU(2) di SU(3). | Where the matrices $(T_i)$ with (i = 1,2,3) generate the subgroup SU (2) of SU (3). |
| Con tre grandezze-generatori si costruiscono le matrici di G-M: | With the three (Q,T,Y) generators we build the Gell-Mann matrices: |

$$\begin{cases} T => \left[ \hat{\sigma}_1 => \left(\hat{\lambda}_1, \hat{\lambda}_4, \hat{\lambda}_6\right) \; ; \; \hat{\sigma}_2 => \left(\hat{\lambda}_2, \hat{\lambda}_5, \hat{\lambda}_7\right) ; \hat{\sigma}_3 => \left(\hat{\lambda}_3\right) \right] \\ Q => \left[ \hat{\sigma}_3 => \left(\hat{\lambda}_3\right) \right] \\ Y => \left[ \left(\hat{\lambda}_8\right) \right] \end{cases} \quad (3.6)$$

| | |
|---|---|
| Questa analogia formale tra $\sigma_i$ e $(T_i)$, così come tra $\sigma_3$ and $(T_3)$, non può essere casuale! | This formal analogy between $\sigma_i$ e $(T_i)$ as between $\sigma_3$ and $(T_3)$ cannot be random! |
| Se la carica elettrica è stata da noi connessa alla "rotazione interna" della fase di un IQuO-particella, l'isospin si potrebbe connettere (per la struttura analoga a quella dello spin) ad una rotazione "presente all'interno" di un adrone. | If we have connected the electrical charge to the "internal rotation" of the phase of a IQuO-particle, one could connect the isospin (for the structure similar to that of the spin) to a rotation "inside" an hadron. |

Se la conservazione della Q è conseguenza dell'invarianza del verso di rotazione della fase in seguito a sfasamenti indotti dall'accoppiamento elettromagnetico tra l'IQuO-particella ed l'IQuO-fotone, la conservazione dello isospin sarebbe conseguenza dell'invarianza del verso di rotazione interna in seguito a sfasamenti indotti dall'interazione forte.

Lo stesso vale per le trasformazioni "interne" tra adroni indotte dalle interazioni forti: la conservazione di isospin totale, deriva da una invarianza globale che coinvolge le rotazioni "interne" a ciascun adrone che partecipa all'interazione.

Come ben sappiamo, le trasformazioni interne degli adroni evidenziano l'esistenza di multipletti.

Se $T_3$ differenzia i membri di un sotto-multipletto individuato dal valore di T (ad es. il gruppo di pioni $\pi$), l'ipercarica Y differenzia i gruppi di sotto-multipletti all'interno di un gruppo di adroni.

Risulta evidente che la presenza di un secondo numero quantico oltre $T_3$ (T) estende la simmetria di Isospin SU(2) ad un gruppo più largo di rango 2, ovvero SU(3).

Le "organizzazioni" di adroni in multipletti degeneri in massa e le trasformazioni interne che fanno passare da un sotto multipletto ad un altro, inducono a sospettare l'esistenza di diversi modi di comporre particelle "più elementari" contenuti all'interno di un adrone, che Gell-Mann chiamò quark!

La "rotazione" interna ipotizzata esistere in un adrone riguarderebbe così i quarks; se lo spin di un quark riguarda il suo orientamento spaziale interno, l'isospin non potrebbe riguardare "ulteriori" orientamenti spaziali.

Potremmo allora ipotizzare l'esistenza di una corrente interna di "semi-quanti" che collegherebbe i quarks componenti un adrone (corrente di gluoni?).

If the conservation of Q electric charge is a consequence of the invariance of the direction of the phase rotation for phase shifts induced by the electromagnetic coupling between electric charge and a photon, the conservation of the isospin would be a consequence of the invariance of the direction of internal rotation for phase shifts induced by the interaction strong.

The same goes for the transformations "internal" between hadrons induced by strong interactions: the conservation of total isospin, derived from a global invariance which involves the rotations "inside" to each hadron participating in the interaction.

As you well know, the transformations between hadrons shows the existence of multiplets.

If T3 differentiates the members of a sub-multiplet identified by the value of T (eg. The group of pions $\pi$), hypercharge Y differentiates the sub-groups of multiplets within a group of hadrons.

It is evident that the presence of a second quantum number than T3 (T) extends the symmetry of Isospin SU (2) to a larger group of rank 2, or SU (3).

The "organizations" of hadrons in multiplets degenerate in mass and the hadron transformations from a multiplet to another, they prove grounds for suspecting that the hadrons are composed by " more elementary" particles, which Gell-Mann called quarks!

The "rotation" inside a hadron could relate well to quarks; if the spin of a quark regards its internal spatial orientation, the isospin cannot relate to spatial orientations.

The "rotation" of isospin from us assumed could be correlated to an internal current of "semi-quanta" that connect the quarks constituting a hadron (gluon current?).

Si rileva che l'analogia formale tra ($T_3$) e lo spin di una particella "composta" induce ad assegnare anche ai quarks un valore di ($T_3$), (come anche un valore di ipercarica Y).

Si ricorda difatti il carattere additivo dell'isospin (ereditato dallo spin), per cui l'isospin di un adrone è somma degli isospin assegnati ai singoli quark.

Inoltre, dato il legame tra Q e $T_3$ (evidenziato dalla matrice comune $\sigma_3$) possiamo asserire che il verso di rotazione della fase (segno di Q) ed il verso di rotazione interna di isospin potrebbero in qualche modo essere correlati.

Ciò accade in maniera evidente nei singoli quark, dove si evidenziano le seguenti associazioni:

1) Quark (u) : Q ($> 0$) ←→($T_3$)'= (+1/2)
2) Quark (d) : Q ($< 0$) ←→($T_3$)'= (-1/2)

Così, ad un IQuO-quark vengono, in definitiva, associate tre intrinseche "rotazioni interne":

1) Rotazione della fase ( segno di Q)
2) Rotazione di Isospin (segno di $T_3$)
3) Rotazione "spaziale" di Spin (segno di $\sigma$)

Ciò può sembrare in contrasto con l'aspetto "puntuale" assegnato ad una particella elementare ( sia leptone che quark) dalla realtà sperimentale, in perfetto accordo con la teoria della relatività quantistica.

Tuttavia, prendendo in considerazione la fenomenologia delle interazioni tra particelle elementari, può diventare accettabile assumere la seguente posizione epistemologica:

una particella elementare pur manifestandosi come "puntuale" nello spazio ad un osservatore "esterno", può esibire indirettamente l'esistere di uno spazio "interno" in cui si possono individuare delle rotazioni "intrinseche" i cui generatori sono da ricondurre alle osservabili fisiche identificate nella carica elettrica (Q), nello spin ($\sigma$) e nell'isospin (T).

Note that the formal analogy between ($T_3$) and the spin of a composite particle induces assign a value of ($T_3$) also to the quarks (as well as a value of hypercharge Y).

It is recalled in fact the summative character isospin (inherited from the spin), for which the isospin of a hadron is the sum of the isospin assigned to the individual quark.

Furthermore, the correlation between $T_3$ and Q, leads us to admit the existence of a correlation between the direction of rotation of the phase and the one of the current isospin.

This clearly happens in individual quarks, because there is the following associations:

1) Quark (u) : Q ($> 0$) ←→($T_3$)'= (+1/2)
2) Quark (d) : Q ($< 0$) ←→($T_3$)'= (-1/2)

Thus, ultimately, we can assign three intrinsic "internal rotations" to a IQuO-quark:

1) Phase rotation ( Q sign )
2) Isopin rotation ($T_3$ sign)
3) Spatial rotation of spin ( $\sigma$ sign)

This may seem at odds with the aspect of "point" given to an elementary particle (like quarks and leptons ) from the experimental reality, in perfect agreement with the quantum theory of relativity.

However, taking into account the phenomenology of the interactions between elementary particles, can become acceptable to take the following position epistemological:

an elementary particle, even if it occurs as a "point" in space for an observer "outside", can indirectly show the existence of an "inside" space where one can find some "intrinsic" rotations of whose the generators are attributable to physical observables identified in the electric charge (Q), in spin ($\sigma$) and isospin (T).

Può così risultare accettabile l'idea di assegnare ad un singolo quark una **"struttura interna".**

Se all'isospin (T) associamo una corrente interna (corrente di semi-quanti) occorre individuare di essa un "percorso" tra elementari "punti interni di oscillazione", considerando sempre un quark un sistema fisico "oscillante".

Un quark, pur mantenendo l'aspetto esterno di un quanto di energia in un punto spaziale, potrebbe essere invece costituito da una struttura interna di IQuO "virtuali" che si possono chiamare Sub-IQuO.

Due altri indici di questa "virtuale" struttura elementare interna possono essere la carica di colore e l'Ipercarica Y.

Quest'ultima, oltre ad essere stata introdotta per differenziare i multipletti adronici, viene assegnata anche ad un singolo quark, rafforzando l'idea di una sua sub-struttura correlata a diverse configurazioni di accoppiamenti del sub-IQuO costituenti.

Un altro sostegno a questa idea è individuabile nelle correlazioni tra le rotazioni "interne" (quella di fase e quella di isospin) durante le interazioni adroniche.

Queste correlazioni vengono a manifestarsi sia sia nella matrice $\lambda_8$ sia nella combinazione dei rispettivi generatori durante le interazioni

$Q \longleftrightarrow J_{(Q)} \longleftrightarrow |J_{(Q)}'>$
$T_3 \longleftrightarrow J_{(T)} \longleftrightarrow |J_{(T3)}'>$

Ricordiamo che quando si compongono rotazioni (generatori-momenti angolari **J**) si ricorre ai coefficienti di Clebesch- Gordan.

Pertanto le correlazioni tra rotazioni interne possono essere espresse dai coefficienti associati agli autostati dei generatori corrispondenti ovvero dai coefficienti di Clebesch- Gordan.

Ricordiamo, ad esempio, la reazione **[π⁻ + p ➔ π⁻ + p]** dove si evidenziano le seguenti ampiezze riguardo l'isospin:

---

Then we conjecture that a quark has an "**internal structure**".

In fact, if we associate to the isospin (T) an internal current (a semi-quanta current), it is necessary to identify of this a "path" between elementary "internal points of oscillation", always considering a quark as an "oscillating" physical system.

A quark, while maintaining the external appearance of a quantum of energy in a spatial point, it could instead be constituted by an internal structure of IQuO "virtual" which can be called Sub-IQuO.

Two other indices of this "virtual" structure can be the color charge and hypercharge Y.

The latter, as well as being introduced to differentiate the hadron multiplets, is assigned to an individual quark, reinforcing the idea of its sub-structure related to coupling different configurations of sub-IQuO constituents.

Another support for this idea is identifiable in the correlations between the rotations "internal" (the one of phase and that of isospin) during the hadron interactions.

These correlations are exhibited both by the matrix $\lambda_8$ and by the combination of the respective generators during interactions

$Q \longleftrightarrow J_{(Q)} \longleftrightarrow |J_{(Q)}'>$
$T_3 \longleftrightarrow J_{(T)} \longleftrightarrow |J_{(T3)}'>$

Remember that when you compose rotations (generators - angular moments (**J**)) you use the Clebesch-Gordan coefficients.

Therefore, the correlations between internal rotations, can be expressed by the coefficients associated with the eigenstates of the generators or by the corresponding coefficients Clebesch - Gordan.

Recall, for example, the reaction **[π⁻ + p ➔ π⁻ + p]** where there are evident the following amplitudeabout the isospin:

$$\begin{array}{l} \pi^- + p \quad \Rightarrow \quad \pi^- + p \\ |1,0\rangle \left|\frac{1}{2}, -\frac{1}{2}\right\rangle \Rightarrow \left(\sqrt{\frac{1}{3}}\right)\left|\frac{3}{2}, -\frac{1}{2}\right\rangle + \left(\sqrt{\frac{2}{3}}\right)\left|\frac{1}{2}, -\frac{1}{2}\right\rangle \end{array} \quad (4.6)$$

| Importanti informazioni si ricavano dalle corrispondenti ampiezze di probabilità: | Relevant information is obtained from the corresponding probability amplitude: |
|---|---|

$$A_{(\pi,p)} = \left[\left(\sqrt{\frac{1}{3}}\right)\left(\sqrt{\frac{1}{3}}\right)\right]\left\langle\frac{3}{2}, -\frac{1}{2}\right|S\left|\frac{3}{2}, -\frac{1}{2}\right\rangle + \left[\left(\sqrt{\frac{2}{3}}\right)\left(\sqrt{\frac{2}{3}}\right)\right]\left\langle\frac{1}{2}, -\frac{1}{2}\right|S\left|\frac{1}{2}, -\frac{1}{2}\right\rangle =$$
$$= \left(\frac{1}{3}\right)A_{(1)} + \left(\frac{2}{3}\right)A_{(2)} \qquad (5.6)$$

| | |
|---|---|
| Non riteniamo che sia un caso che i coefficienti delle ampiezze di probabilità di reazione coincidono con i due possibili valori della carica elettrica associate ai quark. | We do not think it's a case that the coefficients of the "probability" amplitude of the reaction coincide with the two possible values of the electric charge associated with the quarks. |
| Avendo evidenziato la correlazione di segni tra i valori della carica elettrica ( rotazione interna della fase) e quelli di isospin (rotazione di correnti interne), deriva che, nelle interazioni adroniche, gli accoppiamenti di isospin devono essere sempre correlati ai valori della carica elettrica delle particelle interagenti. | Being demonstrated the correlation of signs between the values of the electric charge (rotation of the internal phase) and those of isospin (rotation of internal currents), it follows that, in hadron interactions, the couplings of isospin must always be correlated to the values of the electric charge of the interacting particles. |
| È evidente che tale correlazione influenza le varie probabilità di combinazione degli accoppiamenti di isospin, ovvero, queste probabilità sono in correlazione con i valori della carica elettrica interna ad un quark. | It is evident that such a correlation influence the various probability of combination of the coupling of isospin, ie, these probabilities are in correlation with the values of the electric charge internal to a quark. |
| Ciò fa pensare anche che la carica non intera di un quark sia espressione di un aspetto probabilistico associato all'osservazione sperimentale. | This also suggests that the non-integer value of the charge of a quark is expression of a probabilistic aspect associated with the experimental observation. |
| Se il verso di rotazione della fase di un IQuO-particella ci da il segno della carica elettrica (aspetto interno), il suo valore invece può essere messo in relazione con la probabilità che un fotone ha di rilevare "il quanto" associato alla particella carica. | If the direction of phase rotation of a particle-IQuO gives to us the sign of the electric charge, its value can instead be placed in relation with the probability that a photon detects the quanta of the charged particle . |
| Se per un elettrone questa probabilità è uno, in un quark la probabilità sembra non essere più uno. | If this probability has value equal to one for an electron, in the quarks the probability is not integer. |

Quest'ultimo aspetto tuttavia può essere accettato solo se al quark si da una struttura interna di "oscillatori" accoppiati (sub-IQuO) entro cui si muove il quanto di energia (coppia di semi-quanti) assegnato ad ogni quark.

La presenza di più sub-IQuO dentro un quark non permetterebbe al fotone ( accoppiandosi con un sub-IQuO "interno") di trovare sempre il quanto.

Per la definizione statistica della probabilità il quark potrebbe risultare neutro non solo ad un singolo fotone ma ad un numero infinito di essi con cui si accoppia! Come dire che il quark sarebbe invisibile!

Ma se il singolo quark (separatamente) potrebbe apparire a volte  neutro, un adrone composto da quark non può mai apparire neutro se la somma delle cariche dei quark componenti raggiunge il valore di una unità. Ciò perché due o tre quark legati diventano un'unica unità o un unico centro diffusore sia per accoppiamenti elastici che non elastici con i fotoni.

L'informazione del valore non intero della carica elettrica Q di un quark la possiamo trovare anche nella matrice $\lambda_8$ di Gell-Mann.

Come da letteratura, Gell-Mann, dopo avere assegnati i valori di Y ai vari sottomultipletti adronici, trovò la forma dell'ottava matrice del gruppo SU(3) compatibile con la struttura dell'intero gruppo generato dalle tre grandezze $Q, T_3$ e Y.  Pertanto nella matrice $\lambda_8$ devono essere già contenute quelle informazioni relative alla carica elettrica dei quark. Non è un caso infatti che la matrice $\lambda_8$ di G-M , commuti proprio con la $\lambda_3$.

Richiamiamo la correlazione tra Y e la matrice $\lambda_8$ :

This latter aspect, however, may be accepted only if the quark is an structure of "oscillators" coupled (sub-IQuO) within which moves the quantum of energy (pair of semi-quanta) assigned to each quark.

Having more than one sub-IQuO inside quark, the photon  (in coupling with a "internal" sub-IQuO ) could never find the quanta of the quark.

For the definition of statistical probability, the quarks may be neutral not only to a single photon, but to an infinite number of them with which is coupled. How to say that the quarks would be invisible!

However, if the single quark occasionally could appear neutral, a hadron composed of quarks can never appear neutral if the sum of the charges of the component quarks reaches the value of a unit. This is because two or three quarks bound become a single unit (or a single center diffuser) for elastic couplings and not with photons.

The information of the non-integer value of the electric charge of a quark  can also be found in the Gell-Mann matrix $\lambda_8$.

As per literature, Gell-Mann, assigned the values of Y to the  various multiplets of the hadrons, found the Eighth matrix form of the group SU (3) which is compatible with the structure of the entire group generated by the three variables Q, $T_3$, and Y. Therefore the matrix  $\lambda_8$  must already contain the informations related to the electric charge of quarks. It is no coincidence that the matrix $\lambda_8$ precisely commutes with the $\lambda_3$.

We call the correlation between Y and the matrix $\lambda_8$:

$$\hat{Y} = \left(\frac{1}{\sqrt{3}}\right)\hat{\lambda}_8 \quad \Rightarrow \quad \hat{Y} = \left(\frac{1}{3}\right)\begin{pmatrix} 1 & 0 & 0 \\ 0 & 1 & 0 \\ 0 & 0 & -2 \end{pmatrix}$$

(6.6)

| Gli autovalori di Y sono: | The eigenvalues of Y are: |
|---|---|

$$\hat{Y}' \Rightarrow \left\{ \begin{vmatrix} [(1/3) - \lambda'] & 0 & 0 \\ 0 & [(1/3) - \lambda'] & 0 \\ 0 & 0 & [(-2/3) - \lambda'] \end{vmatrix} = 0 \right\} \Rightarrow \left\{ [(1/3 - \lambda')^2][(-2/3) - \lambda']\} = 0 \right.$$

$$Y' = \left\{ \begin{matrix} 1/3 \\ 1/3 \\ -2/3 \end{matrix} \right\}$$

(7.6)

| | |
|---|---|
| Da notare anche qui che gli autovalori di Y sono connessi ai valori di carica elettrica Q. Ovviamente tutto ciò è perfettamente in accordo con la relazione: [Y = 2 (Q -T₃)]). Data la correlazione tra la carica Q e l'isospin T ha senso considerare lo stato di un adrone come stato simultaneo composto da autostati delle rotazioni e di Y. Pertanto possiamo allora prendere in considerazione la matrice prodotto $\hat{\lambda}_3 = (\lambda_3)(Y)$ | Also note here that the eigenvalues of Y are related to the values of electric charge Q. Obviously everything is perfectly in accord with the relation: [Y = 2 (Q-T₃)]). Given the correlation between the charge Q and the isospin T, has the meaning consider the state of a hadron as a state composed of simultaneous eigenstates of "rotations" and Y. Therefore we can consider the matrix product $\hat{\lambda}_3 = (\lambda_3)(Y)$ |

$$\hat{\lambda}_3 = \hat{\lambda}_3 \hat{Y} = \left(\frac{1}{3}\right)\begin{pmatrix} 1 & 0 & 0 \\ 0 & -1 & 0 \\ 0 & 0 & 0 \end{pmatrix}\begin{pmatrix} 1 & 0 & 0 \\ 0 & 1 & 0 \\ 0 & 0 & -2 \end{pmatrix} = \left(\frac{1}{3}\right)\begin{pmatrix} 1 & 0 & 0 \\ 0 & -1 & 0 \\ 0 & 0 & 0 \end{pmatrix} = \begin{pmatrix} (1/3) & 0 & 0 \\ 0 & -(1/3) & 0 \\ 0 & 0 & 0 \end{pmatrix}$$

(8.6)

| | |
|---|---|
| In conclusione, $\lambda_3$ ci può dire quali valori sono permessi alla carica Q di un quark. Per calcolare tali valori ricorriamo alla forma matrice di un IQuO. Nel par. 4 l'accoppiamento tra due IQuO "neutri" nella direzione della rotazione di fase ma con autovalori dell'operatore (r) opposti [(r')₁ = +1, (r')₂ = -1] ha prodotto una coppia di IQuO con cariche elettriche a valore intero [(Q')₁ = +1, (Q')₂ = -1]. L'esistenza dei quark spinge a cercare un possibile accoppiamento tra IQuO producenti coppie di IQuO con carica elettrica non intera.<br><br>Una diversa forma della matrice rappresentativa di un IQuO è dovuta in modo tale che applicando le relazioni eq. 11.4 e 12.4 sia possibile trovare valori non interi degli autovalori di Q. | In conclusion, the matrix $\lambda_3$ can tells what values are allowed to charge Q of a quark. To calculate these values we resort to the form of a matrix IQuO. In Sect. 4, the coupling between two IQuO "neutral" in the direction of phase rotation but with eigenvalues of the operator (r) opposite [(r') 1 = +1, (r') 2 = -1], has produced a pair of IQuO electric charges to integer [(Q') = 1 +1, (Q') 2 = -1]. Quarks leads one to seek a possible coupling between IQuO pairs that produce outgoing IQuO pairs having not integer electric charge.<br><br>A different form of the representative matrix an IQuO is to be found in such a way that it is possible find, by the relations eq. 11.4 and 12.4, non-integer values of the eigenvalues of Q. |
| Non resta adesso che trovare la matrice colonna (Φ) rappresentativa di un "sub-IQuO" di un quark per applicare la relazione definita dall'eq. 11.4. | Now we find the column matrix (Φ) which is representative of a "sub-IQuO" of a quark, to apply the relationship defined by Eq. 11.4. |
| E' evidente che con matrici 3x3 si deve avere una matrice colonna (Φ) con tre elementi (matrice 1x3) mentre noi abbiamo operato, sino a questo momento, con IQuO rappresentati da matrici (1x2) (vedi eq. 8.4), la cui procedura di accoppiamento produce sempre IQuO con matrici (1x2). | It 'obvious that with 3x3 matrices one must have a matrix column (Φ) with three elements (1x3 matrix). |
| Si rileva che la matrice Y è l'unica matrice di G-M che operando su una matrice colonna (1x3) mantiene inalterato il numero 3 dei suoi elementi. | Note that the matrix Y is the only G-M matrix that by operating on a matrix column (1x3) keeps unchanged the number 3 of its elements. |
| Per avere un IQuO a semi-quanti con tre elementi di matrice dobbiamo pensare ad un accoppiamento coinvolgente simultaneamente tre IQuO Φᵢ. Poniamo: | To having an IQuO by semi-quanta and with three matrix elements, it is necessary to resort simultaneously to coupling between three IQuO Φᵢ . We put: |

$$\hat{\Phi}_1 \oplus \hat{\Phi}_2 \oplus \hat{\Phi}_3 = \hat{\Psi}_+ + \hat{\Psi}_- \quad (9.6)$$

| | |
|---|---|
| In forma matriciale ( vedi le eq . 18.4 , 20.4 e 21.4 e 22.4 ) otteniamo | In matrix formi it is |

$$\left\{ \hat{\Psi}_{cl} = \begin{pmatrix} \left( \hat{o}_{el} e^{-i\rho} + \hat{\bullet}_{in} e^{-i\rho°} \right) \\ \left( \hat{\bullet}_{el}^+ e^{-i\alpha} + \hat{o}_{in}^+ e^{-i\alpha°} \right) \\ \left( \hat{o}_{el} e^{-i\sigma} + \hat{\bullet}_{in} e^{-i\sigma°} \right) \end{pmatrix} \quad , \quad \hat{\Psi}_{\overline{cl}} = \begin{pmatrix} \left( \hat{o}_{el}^+ e^{i\gamma} + \hat{\bullet}_{in}^+ e^{i\gamma°} \right) \\ \left( \hat{\bullet}_{el} e^{i\beta} + \hat{o}_{in} e^{i\beta°} \right) \\ \left( \hat{\bullet}_{el}^+ e^{i\varepsilon} + \hat{o}_{in}^+ e^{i\varepsilon°} \right) \end{pmatrix} \right\} \quad (10.6)$$

| | |
|---|---|
| Possiamo calcolare adesso la carica elettrica: | We can now calculate the electric charge: |

$$Q_{cl} = \left( \left( \hat{\Psi}^+ \right) \right)_{cl} \hat{\lambda}_3 \left( \left( \hat{\Psi} \right) \right)_{cl} = \begin{pmatrix} \left( \hat{\bullet}_{el}^+ e^{i\rho} + \hat{o}_{in}^+ e^{i\rho°} \right) \\ \left( \hat{o}_{el} e^{i\alpha} + \hat{\bullet}_{in} e^{i\alpha°} \right) \\ \left( \hat{\bullet}_{el}^+ e^{i\sigma} + \hat{o}_{in}^+ e^{i\sigma°} \right) \end{pmatrix} \begin{pmatrix} 1/3 & 0 & 0 \\ 0 & -1/3 & 0 \\ 0 & 0 & 0 \end{pmatrix} \begin{pmatrix} \left( \hat{o}_{el} e^{-i\rho} + \hat{\bullet}_{in} e^{-i\rho°} \right) \\ \left( \hat{\bullet}_{el}^+ e^{-i\alpha} + \hat{o}_{in}^+ e^{-i\alpha°} \right) \\ \left( \hat{o}_{el} e^{-i\sigma} + \hat{\bullet}_{in} e^{-i\sigma°} \right) \end{pmatrix}$$

(11.6)

$$Q_{cl} = (1/3) \begin{pmatrix} (\hat{\bullet}^+_{el}\, e^{i\rho} + \hat{o}^+_{in}\, e^{i\rho^\circ}) \\ (\hat{o}^+_{el}\, e^{i\alpha} + \hat{\bullet}_{in}\, e^{i\alpha^\circ}) \\ (\hat{o}^+_{el}\, e^{i\sigma} + \hat{\bullet}^+_{in}\, e^{i\sigma^\circ}) \end{pmatrix} \begin{pmatrix} (\hat{o}_{el}\, e^{-i\rho} + \hat{\bullet}_{in}\, e^{-i\rho^\circ}) \\ -(\hat{\bullet}^+_{el}\, e^{-i\alpha} + \hat{o}^+_{in}\, e^{-i\alpha^\circ}) \\ 0 \end{pmatrix} =$$

$$= (1/3)\left\{ (\hat{\bullet}^+_{el}\, e^{i\rho} + \hat{o}^+_{in}\, e^{i\rho^\circ})(\hat{o}_{el}\, e^{-i\rho} + \hat{\bullet}_{in}\, e^{-i\rho^\circ}) - (\hat{o}^+_{el}\, e^{i\alpha} + \hat{\bullet}_{in}\, e^{i\alpha^\circ})(\hat{\bullet}^+_{el}\, e^{-i\alpha} + \hat{o}^+_{in}\, e^{-i\alpha^\circ}) \right\} =$$

$$= (1/3)\left\{ \begin{aligned} & [\hat{\bullet}^+_{el}, \hat{o}_{el}] + [\hat{o}^+_{in}, \hat{\bullet}_{in}] + \\ & + (\hat{\bullet}^+_{el}\,\hat{\bullet}_{in}\, e^{i(\rho - \rho^\circ)} - \hat{\bullet}_{in}\,\hat{\bullet}^+_{el}\, e^{i(\alpha^\circ - \alpha)}) + (\hat{o}^+_{in}\,\hat{o}_{el}\, e^{i(\rho^\circ - \rho)} - \hat{o}_{el}\,\hat{o}^+_{in}\, e^{i(\alpha - \alpha^\circ)}) \end{aligned} \right\}$$

$$= (1/3)\left\{ [\hat{\bullet}^+_{el}, \hat{\bullet}_{in}]\, e^{i(\pi/2)} + [\hat{o}^+_{in}, \hat{o}_{el}]\, e^{-i(\pi/2)} \right\} = (1/3)\left\{ i\left(\frac{i}{2}\right) - i\left(\frac{-i}{2}\right) \right\} = -(1/3) \quad !$$

(12.6)

| | |
|---|---|
| Dove [(α° = α + π/2), (ρ° = ρ - π/2) ] Si calcola facilmente che $Q_{anti\text{-}cl}$ = +1/3. Abbiamo così ottenuto un valore non intero della carica elettrica da assegnare di certo ad un quark, componente un adrone. Consideriamo la matrice differenza $((\Delta\hat{\lambda}_3))$ delle matrici $[((\hat{\lambda}_3)), ((Q))]$ | Where [(α° = α + π/2), (ρ° = ρ - π/2) ] Is easily calculated that $Q_{anti\text{-}cl}$ = +1 / 3. We have thus obtained a non-integer value of the electric charge to be assigned to a quark, which is a component of hadron. Consider the $((\Delta\hat{\lambda}_3))$ matrix difference of the matrices $[((\hat{\lambda}_3)), ((Q))]$ |

$$\Delta\hat{\lambda}_3 = (\hat{\lambda}_3 \hat{Y} - \hat{Q}) = \begin{pmatrix} (1/3) & 0 & 0 \\ 0 & -(1/3) & 0 \\ 0 & 0 & 0 \end{pmatrix} - \begin{pmatrix} 1 & 0 & 0 \\ 0 & -1 & 0 \\ 0 & 0 & 0 \end{pmatrix} = \begin{pmatrix} -(2/3) & 0 & 0 \\ 0 & +(2/3) & 0 \\ 0 & 0 & 0 \end{pmatrix}$$  (13.6)

| | |
|---|---|
| La carica elettrica calcolata con questa matrice fornisce l'altro valore di carica associata ai quark; tuttavia si rileva che le matrici $((\hat{\lambda}_3))$ e $((\Delta\hat{\lambda}_3))$ sono "opposte" in numeri relativi ovvero forniranno valori opposti di carica elettrica Q➜Q'=-Q. Avremo allora: | The electric charge calculated with this matrix gives the other value of charge associated with the quark, but it is noted that the matrices $((\hat{\lambda}_3))$ and $((\Delta\hat{\lambda}_3))$ are "opposite" in relative way and therefore they give opposite values of electric charge Q➜Q'=-Q. We will then: |

$$Q'_{cl} = \left((\hat{\Psi}^+)\right)_{cl} \hat{\lambda}_3 \left((\hat{\Psi})\right)_{cl} = \begin{pmatrix} (\hat{\bullet}^+_{el}\, e^{i\rho} + \hat{o}^+_{in}\, e^{i\rho^\circ}) \\ (\hat{o}_{el}\, e^{i\alpha} + \hat{\bullet}_{in}\, e^{i\alpha^\circ}) \\ (\hat{o}^+_{el}\, e^{i\sigma} + \hat{\bullet}_{in}\, e^{i\sigma^\circ}) \end{pmatrix} \begin{pmatrix} -2/3 & 0 & 0 \\ 0 & 2/3 & 0 \\ 0 & 0 & 0 \end{pmatrix} \begin{pmatrix} (\hat{o}_{el}\, e^{-i\rho} + \hat{\bullet}_{in}\, e^{-i\rho^\circ}) \\ (\hat{\bullet}^+_{el}\, e^{-i\alpha} + \hat{o}^+_{in}\, e^{-i\alpha^\circ}) \\ (\hat{o}_{el}\, e^{-i\sigma} + \hat{\bullet}_{in}\, e^{-i\sigma^\circ}) \end{pmatrix}$$

(14.6)

$$Q'_{cl} = (2/3) \begin{pmatrix} \left(\hat{\bullet}_{el}^+ e^{i\rho} + \hat{o}_{in}^+ e^{i\rho^\circ}\right) \\ \left(\hat{o}_{el} e^{i\alpha} + \hat{\bullet}_{in} e^{i\alpha^\circ}\right) \\ \left(\hat{o}_{el}^+ e^{i\sigma} + \hat{\bullet}_{in}^+ e^{i\sigma^\circ}\right) \end{pmatrix} \begin{pmatrix} -\left(\hat{o}_{el} e^{-i\rho} + \hat{\bullet}_{in} e^{-i\rho^\circ}\right) \\ \left(\hat{\bullet}_{el}^+ e^{-i\alpha} + \hat{o}_{in}^+ e^{-i\alpha^\circ}\right) \\ 0 \end{pmatrix} =$$

$$= (2/3)\left\{ -\left(\hat{\bullet}_{el}^+ e^{i\rho} + \hat{o}_{in}^+ e^{i\rho^\circ}\right)\left(\hat{o}_{el} e^{-i\rho} + \hat{\bullet}_{in} e^{-i\rho^\circ}\right) + \left(\hat{o}_{el} e^{i\alpha} + \hat{\bullet}_{in} e^{i\alpha^\circ}\right)\left(\hat{\bullet}_{el}^+ e^{-i\alpha} + \hat{o}_{in}^+ e^{-i\alpha^\circ}\right) \right\} =$$

$$= (2/3)\left\{ -\left(\hat{\bullet}_{el}^+ \hat{\bullet}_{in} e^{i(\rho-\rho^\circ)} - \hat{\bullet}_{in} \hat{\bullet}_{el}^+ e^{i(\alpha^\circ-\alpha)}\right) - \left(\hat{o}_{in}^+ \hat{o}_{el} e^{i(\rho^\circ-\rho)} - \hat{o}_{el} \hat{o}_{in}^+ e^{i(\alpha-\alpha^\circ)}\right) \right\}$$

$$= (2/3)\left\{ -\left[\hat{\bullet}_{el}^+, \hat{\bullet}_{in}\right] e^{i(\pi/2)} - \left[\hat{o}_{in}^+, \hat{o}_{el}\right] e^{-i(\pi/2)} \right\} = (2/3)\left\{ -i\left(\frac{i}{2}\right) + i\left(\frac{-i}{2}\right) \right\} = (2/3)$$

$$Q_{cl} = -Q'_{cl} = -(2/3)$$

(15.6)

| | |
|---|---|
| Abbiamo così ottenuto IQuO con carica elettrica non intera ma con valori possibili dati solo dalla coppia ($\pm 1/3, \pm 2/3$). | We have thus obtained IQuO electrically charged but with non-integer values: ($\pm 1/3, \pm 2/3$). |

| **Considerazioni finali** | **Final considerations** |
|---|---|
| Come abbiamo sottolineato in questo studio, la struttura a semi-quanti di un oscillatore e il meccanismo di accoppiamento di due IQuO possono determinare un insieme di situazioni che risultano fortemente analoghe a una fenomenologia tipica delle particelle elettricamente cariche e del processo di creazione coppie. | As it has been underlined in this study, the semi-quanta structure of an oscillator and the coupling mechanism of two IQuO determine a set of situations that is strongly analogous to a typical phenomenology of the electrically charged particles and of the process of pair creation. |
| Ciò non è affatto casuale e ci porta a pensare, con estrema umiltà, che un tale approccio teorico originale rappresenta qualcosa di fondamentale che sta alla base della creazione delle particelle elementari e del loro comportamento in relazione alla carica elettrica. | We think that this original approach represents something of fundamental which stands on the base of the elementary particles creation and of their behavior in relation to the electric charge. |
| Così potremmo essere in presenza di un più profondo modello di struttura dei campi quantistici in grado di spiegare in modo più chiaro la complessa natura delle particelle e le loro interazioni. | So we could be in presence of a model of structure of the quantum fields able of explaining in a clearer way the complex nature of the particles and their interactions. |

A sostegno della efficacia descrittiva di questo modello si evidenziano i seguenti aspetti :

1) interpretando le particelle di energia negativa come antiparticelle è equivalente ad affermare l'invarianza delle leggi fisiche per CPT trasformazioni;

2) collegando il segno della carica elettrica al senso di rotazione di fase si evidenzia la correlazione tra i campi con carica elettrica e la invarianza per trasformazioni di gauge delle loro interazioni;

3) associando alla carica elettrica e all'isospin T delle rotazioni "interne" si evidenzia la possibilità di descrivere i quark come strutture di accoppiamenti di più IQuO.

4) La possibilità di una struttura interna ai quark ci permette di interpretare in termini probabilistici il valore non intero della carica elettrica

Possiamo anche dire che alcuni aspetti di questo studio possono preludere ad una nuova chiave di lettura di alcune problematiche ancora non risolte dalla fisica moderna:

• l'"interazione" di fase tra due IQuO potrebbe aiutare a comprendere meglio il processo fisico di riduzione della funzione d'onda

• Poiché le particelle senza massa ma con carica elettrica non esistono, il processo di accoppiamento di due $\Phi$-IQuO che conduce alla creazione di due $\Psi$-IQuO, con carica elettrica, che potrebbe contribuire a spiegare meglio la creazione di particelle massive.

To support this model, we underline the following aspects:

1) interpreting the particles to negative energy like antiparticles is equivalent to assert the invariance of the physical laws for CPT transformations;

2) connecting the sign of the electric charge to the direction of phase rotation it highlights the correlation between the fields with electric charge and the invariance for gauge transformations of their interactions.

3) if we associate to the electric charge and to the isospin T the internal rotations, then we can give to quarks a internal structure of quantum coupled oscillators (IQuO)

4) give a internal structure to the quark allows us to interpret in terms of probability of the non-integer electric charge

Some aspects of this study can prelude to a reading new key of some problematic still no resolved by the modern physics :

• the phase "interaction" between two IQuO could help to understand better of the physical process of reduction of the wave function.

• Since the massless particles, but with charge electric, don't exist, the coupling process of two $\Phi$-IQuO that conducts to the creation of two $\Psi$-IQuO, with electric charge, it could contribute to explain better the creation of the massive particles.

- Il meccanismo in cui un campo di gauge (**Φ–IQuO**) regolarizza la forma irregolare di un campo **Ψ-IQuO,** richiama a noi, le cariche elettriche (come un elettrone) circondate da nuvole di fotoni virtuali.

- La struttura a semi-quanti degli oscillatori di campo Ψ-IQuO e - **Φ**-IQuO potrebbe dare a noi alcune informazioni sulla struttura delle particelle e sulla differenza fondamentale tra fermioni e bosoni, come anche quella tra leptoni e quark.

In conclusione, la struttura a semi-quanti di un oscillatore di campo potrebbe costituire la chiave fondamentale per comprendere meglio la natura profonda delle particelle, la loro formazione e le loro interazioni.

- The mechanism in which a gauge field (**Φ–IQuO**) regularizes the form of an "irregular" field **Ψ-IQuO**, reminds us the surrounded electric charges (like an electron) in clouds of virtual photons.

- The structure by semi-quanta of the field oscillators Ψ-IQuO and **Φ**–IQuO could give to us some information about structure of particle e about fundamental difference between Fermions and Bosons, as also the one between Leptons and Quarks.

In conclusion, the structure by semi-quanta of the field oscillator could be the fundamental key to understand the deep nature of the particles, their formation and their interactions.

**Appendice**

**Parentesi di commutazione degli operatori di semi-quanto**

Consideriamo le seguenti relazioni

**Appendix**

**Commutation relations of the semi-quanta operators**

We consider the following relations

$$\left[\hat{a},\hat{a}^+\right]=1 \quad ; \quad \hat{H}=\hbar\omega(\hat{a}^+\hat{a}+1/2)=\frac{1}{2}\hbar\omega(\hat{a}^+\hat{a}+\hat{a}\hat{a}^+) \quad ; \quad \hat{a}^+\hat{a}=\hat{n}=\frac{\hat{H}}{\hbar\omega}-\frac{1}{2}$$

$$(\hat{a}^+\hat{a}+\hat{a}\hat{a}^+)=(\hat{k}_{el}^2+\hat{k}_{in}^2) \quad ; \quad \hat{H}=\frac{1}{2}\hbar\omega(\hat{k}_{el}^2+\hat{k}_{in}^2)$$

(A.1)

Esaminiamo gli (a, a⁺) operatori espressi dall' eq. (32.2)

We examine the (a, a⁺) operators which are expressed by the eq. 32.2

$$\hat{a}_1^+(t)=(\hat{\bullet}^+)_{el}\exp(ir'\omega t)+(\hat{o}^+)_{cin}\exp(i(r'\omega t-\pi/2))=[(\hat{\bullet}_{el}^+-i\hat{o}_{in}^+)]\exp(i(\omega t))$$

$$\hat{a}_1(t)=(\hat{o})_{el}\exp(-ir'\omega t)+(\hat{\bullet})_{cin}\exp(-i(r'\omega t-\pi/2))=[(\hat{o}_{el}+i\hat{\bullet}_{in})]\exp(-i(\omega t))$$

(A.2)

| dove   r' = +1.<br>Calcoliamo le relazioni di Commutazione.<br>Partiamo dalle seguenti equazioni | Where  r' = +1.<br>We calculate the Commutation  Parenthesis.<br>We depart from the followings relations |
|---|---|

$$\hat{a}^+\hat{a} = [(\hat{\bullet}_{el}^+ - i\hat{o}_{in}^+)]\,[(\hat{o}_{el} + i\,\hat{\bullet}_{in})] = (\hat{\bullet}_{el}^+ \hat{o}_{el} + \hat{o}_{in}^+\,\hat{\bullet}_{in}) + i(\hat{\bullet}_{el}^+\,\hat{\bullet}_{in} - \hat{o}_{in}^+\hat{o}_{el})$$
$$\hat{a}\hat{a}^+ = [(\hat{o}_{el} + i\,\hat{\bullet}_{in})][(\hat{\bullet}_{el}^+ - i\hat{o}_{in}^+)] = (\hat{o}_{el}\,\hat{\bullet}_{el}^+ + \hat{\bullet}_{in}\,\hat{o}_{in}^+) + i(\hat{\bullet}_{in}\,\hat{\bullet}_{el}^+ - o_{el}\hat{o}_{in}^+)$$

(A.3)

| Ricordando l'eq. 14.2, abbiamo: | Remembering the eq. 14.2 , we have |
|---|---|

$$a^+\hat{a} = (\hat{k}_{el}e^{i\hat{r}\alpha t} - i\hat{k}_{in}e^{i\hat{r}\alpha t})(e^{-i\hat{r}\alpha t}\hat{k}_{el} + ie^{-i\hat{r}\alpha t}\hat{k}_{in}) = (\hat{k}_{el}^2 + \hat{k}_{in}^2) + i(\hat{k}_{el}\hat{k}_{in} - \hat{k}_{in}\hat{k}_{el})$$
$$\hat{a}\hat{a}^+ = (e^{-i\hat{r}\alpha t}\hat{k}_{el} + ie^{-i\hat{r}\alpha t}\hat{k}_{in})(\hat{k}_{el}e^{i\hat{r}\alpha t} - i\hat{k}_{in}e^{i\hat{r}\alpha t}) = (\hat{k}_{el}^2 + \hat{k}_{in}^2) + i(\hat{k}_{in}\hat{k}_{el} - \hat{k}_{el}\hat{k}_{in})$$

(A.4)

| Confrontando l'eq. (A.3) con la (A.4) otteniamo | Comparing the eq. (3) with the (4) we obtain |
|---|---|

$$\hat{a}^+\hat{a} = (\hat{\bullet}_{el}^+\hat{o}_{el} + \hat{o}_{in}^+\,\hat{\bullet}_{in}) + i(\hat{\bullet}_{el}^+\,\hat{\bullet}_{in} - \hat{o}_{in}^+\hat{o}_{el}) = (\hat{k}_{el}^2 + \hat{k}_{in}^2) + i(\hat{k}_{el}\hat{k}_{in} - \hat{k}_{in}\hat{k}_{el})$$

(A.5)

| Eguagliando i termini dello stesso tipo nelle due parti dell'eq., segue | Equalizing the terms of the same type in the two parts of the equation, it follows |
|---|---|

$$\hat{k}_{el}^2 = \hat{\bullet}_{el}^+\hat{o}_{el} \quad ; \quad \hat{k}_{in}^2 = \hat{o}_{in}^+\,\hat{\bullet}_{in} \quad ; \quad (\hat{k}_{el}\hat{k}_{in} - \hat{k}_{in}\hat{k}_{el}) = (\hat{\bullet}_{el}^+\,\hat{\bullet}_{in} - \hat{o}_{in}^+\hat{o}_{el})$$

(A.6)

| Nello stesso modo calcoliamo | In this same way we calculate |
|---|---|

$$\hat{a}\hat{a}^+ = (\hat{k}_{el}^2 + \hat{k}_{in}^2) + i(\hat{k}_{in}\hat{k}_{el} - \hat{k}_{el}\hat{k}_{in})$$
$$\hat{a}\hat{a}^+ = (\hat{o}_{el}\,\hat{\bullet}_{el}^+ + \hat{\bullet}_{in}\,\hat{o}_{in}^+) + i(\hat{\bullet}_{in}\,\hat{\bullet}_{el}^+ - o_{el}\hat{o}_{in}^+)$$

(A.7)

| Uguagliando, segue | Equalizing, it follows |
|---|---|

$$\hat{k}_{el}^2 = \hat{o}_{el}\,\hat{\bullet}_{el}^+ \quad ; \quad \hat{k}_{in}^2 = \hat{\bullet}_{in}\hat{o}_{in}^+ \quad ; \quad (\hat{k}_{in}\hat{k}_{el} - \hat{k}_{el}\hat{k}_{in}) = (\hat{\bullet}_{in}\,\hat{\bullet}_{el}^+ - o_{el}\hat{o}_{in}^+)$$

(A.8)

| Da $[\mathbf{a}, \mathbf{a}^+] = 1$ troviamo | From the $[\mathbf{a}, \mathbf{a}^+] = 1$ we'll find |
|---|---|

$$\left(\hat{a}\hat{a}^+\right) - \left(\hat{a}^+\hat{a}\right) = \left[(\hat{o}_{el}\,\hat{\bullet}_{el}^+ + \hat{\bullet}_{in}\,\hat{o}_{in}^+) + i(\hat{\bullet}_{in}\,\hat{\bullet}_{el}^+ - o_{el}\hat{o}_{in}^+)\right] - \left[(\hat{\bullet}_{el}^+\hat{o}_{el} + \hat{o}_{in}^+\,\hat{\bullet}_{in}) + i(\hat{\bullet}_{el}^+\,\hat{\bullet}_{in} - \hat{o}_{in}^+\hat{o}_{el})\right] = 1$$

(A.9)

| o | or |
|---|---|

$$\left[\left(\hat{o}_{el}\,\hat{\bullet}_{el}^+ - \hat{\bullet}_{el}^+\hat{o}_{el}\right) + \left(\hat{\bullet}_{in}\,\hat{o}_{in}^+ - \hat{o}_{in}^+\,\hat{\bullet}_{in}\right)\right] + \left[i\left(\hat{\bullet}_{in}\,\hat{\bullet}_{el}^+ - \hat{\bullet}_{el}^+\,\hat{\bullet}_{in}\right) + i\left(\hat{o}_{in}^+\hat{o}_{el} - o_{el}\hat{o}_{in}^+\right)\right] = 1$$

(A.10)

| e | and |
|---|---|

$$\left\{ \left[\hat{o}_{el}, \hat{\bullet}_{el}^+\right] + \left[\hat{\bullet}_{in}\ \hat{o}_{in}^+\right]\right\} + \left\{ i\left[\hat{\bullet}_{in}, \hat{\bullet}_{el}^+\right] + i\left[\hat{o}_{in}^+, \hat{o}_{el}\right]\right\} = 1 \tag{A.11}$$

| da eq. (A.8a) e  eq. (A.8b), troviamo | by the eq. (A.8a) and eq. (A.8b), we find |
|---|---|

$$\hat{k}_{el}^2 = \hat{o}_{el}\ \hat{\bullet}_{el}^+ = \hat{\bullet}_{el}^+ \hat{o}_{el} \quad ; \quad \hat{k}_{in}^2 = \hat{\bullet}_{in}\hat{o}_{in}^+ = \hat{o}_{in}^+ \hat{\bullet}_{in} \quad \Longrightarrow \quad \left\{\left[\hat{o}_{el}, \hat{\bullet}_{el}^+\right] = 0 ; \left[\hat{\bullet}_{in}, \hat{o}_{in}^+\right] = 0\right\} \tag{A.12}$$

| allora è evidente ( vedi l'eq. (A.11)) che segue | then it's evident (from the eq. (A.11)) that it follows |
|---|---|

$$\left\{\left[\hat{\bullet}_{in}, \hat{\bullet}_{el}^+\right] = \frac{-i}{2} ; \left[o_{el} \hat{o}_{in}^+\right] = \frac{i}{2}\right\} \tag{A.13}$$

| Mentre da eq. (A.1), (A.5), (A.6),  (A.7),  (A.8) ricaviamo | While from the eq. (A.1), (A.5), (A.6),  (A.7), (A.8) we find |
|---|---|

$$\left(\hat{\bullet}_{in}\ \hat{\bullet}_{el}^+ - o_{el}\hat{o}_{in}^+\right) = -\frac{1}{2}$$

$$\left(\hat{\bullet}_{el}^+\ \hat{\bullet}_{in} - \hat{o}_{in}^+ \hat{o}_{el}\right) = -\frac{1}{2} \tag{A.14}$$

| Se invece esaminiamo le alter equivalenti forme dell' IQuO date da | If we instead examine the other equivalent form of the IQuO given by |
|---|---|

$$\hat{a}_1^+(t) = [(\hat{o}_{el}^+ - i\ \hat{\bullet}_{in}^+)]\exp(i(\omega t))$$

$$\hat{a}_1(t) = [(\hat{\bullet}_{el} + i\hat{o}_{in})]\exp(-i(\omega t)) \tag{A.15}$$

| Dove abbiamo scambiato un semi-quanto pieno con uno vuoto e viceversa; le seguenti parentesi di commutazione sono date: | Where we have changed a full semi-quanta with a empty semi-quanta and vice versa; the following commutation parenthesis is gotten: |
|---|---|

$$\left\{\left[\hat{\bullet}_{el}, \hat{o}_{el}^+\right] = 0 ; \left[\hat{o}_{in}, \hat{\bullet}_{in}^+\right] = 0\right\} \tag{A.16}$$

$$\left\{\left[o_{in}, o_{el}^+\right] = \frac{-i}{2} ; \left[\hat{\bullet}_{el}\ \hat{\bullet}_{in}^+\right] = \frac{i}{2}\right\} \tag{A.17}$$

| Se invece esaminiamo un IQuO del tipo: | If we instead examine an IQuO of the type: |
|---|---|

$$\hat{a}^+(t) = [(\hat{\bullet}_{el}^+ - i\ \hat{\bullet}_{in}^+)]\exp(i(\omega t))$$

$$\hat{a}_1(t) = [(\hat{o}_{el} + i\hat{o}_{in})]\exp(-i(\omega t)) \tag{A.18}$$

| da eq. (22.2) , eq. 24.2 e relazione (A.1) | by eq. eq. (22.2) , eq. 24.2 and the relations (A.1) |
|---|---|

$$E = K + U$$
$$\hat{H} = \frac{1}{2}\hbar\omega(\hat{k}_{el}^2 + \hat{k}_{in}^2) \equiv \frac{1}{2}\hbar\omega\big[(\bullet + o) + (\bullet + o)\big]$$

(A.19)

| otteniamo | we obtain |
|---|---|

$$\hat{a}^+\hat{a} = (\hat{\bullet}_{el}^+\hat{o}_{el} + \hat{\bullet}_{in}^+\hat{o}_{in}) + i\big(\hat{\bullet}_{el}^+\hat{o}_{in} - \hat{\bullet}_{in}^+\hat{o}_{el}\big)$$
$$\hat{k}_{el}^2 = \hat{\bullet}_{el}^+\hat{o}_{el} \quad ; \quad \hat{k}_{in}^2 = \hat{\bullet}_{in}^+\hat{o}_{in} \quad ; \quad \big(\hat{\bullet}_{el}^+\hat{o}_{in} - \hat{\bullet}_{in}^+\hat{o}_{el}\big) = -\frac{1}{2}$$

(A.20)

| e | and |
|---|---|

$$\hat{a}\hat{a}^+ = (\hat{o}_{el}\,\hat{\bullet}_{el}^+ + \hat{o}_{in}\,\hat{\bullet}_{in}^+) + i(\hat{o}_{in}\,\hat{\bullet}_{el}^+ - o_{el}\,\hat{\bullet}_{in}^+)$$
$$\hat{k}_{el}^2 = \hat{o}_{el}\,\hat{\bullet}_{el}^+ \quad ; \quad \hat{k}_{in}^2 = \hat{\bullet}_{in}\hat{o}_{in}^+ \quad ; \quad (\hat{o}_{in}\,\hat{\bullet}_{el}^+ - o_{el}\,\hat{\bullet}_{in}^+) = -\frac{1}{2}$$

(A.21)

| Usando sempre la stessa procedura troviamo le seguenti parentesi di commutazione: | Always using the same procedure we find the following commutation relations: |
|---|---|

$$\left\{ \big[\hat{o}_{in}, \hat{\bullet}_{el}^+\big] = \frac{-i}{2} \; ; \big[\hat{o}_{el}\,\hat{\bullet}_{in}^+\big] = \frac{i}{2} \right\}$$

(A.22)

| Se invece abbiamo un IQuO della forma | If instead we have an IQuO with form |
|---|---|

$$\hat{a}^+(t) = [(\hat{o}_{el}^+ - i\hat{o}_{in}^+)(\exp(i(\omega)t))]$$
$$\hat{a}_1(t) = [(\hat{\bullet}_{el} + i\,\bullet_{in})]\exp(-i(\omega)t))$$

(A.23)

| Per simmetria  avremo | For symmetry |
|---|---|

$$\left\{ \big[\hat{\bullet}_{in}, \hat{o}_{el}^+\big] = \frac{-i}{2} \; ; \big[\hat{\bullet}_{el}\,\hat{o}_{in}^+\big] = \frac{i}{2} \right\}$$

(A.24)

| Riassumendo avremo | Summarizing we'll have |
|---|---|

$$\left\{[\hat{o}_{el},\hat{\bullet}^+_{el}]=0\;;\;[\hat{\bullet}_{in},\hat{o}^+_{in}]=0\right\}\quad;\quad\left\{[\hat{\bullet}_{in},\hat{\bullet}^+_{el}]=\frac{-i}{2}\;;\;[o_{el}\,\hat{o}^+_{in}]=\frac{i}{2}\right\}$$

$$\left\{[\hat{\bullet}_{el},\hat{o}^+_{el}]=0\;;\;[\hat{o}_{in},\hat{\bullet}^+_{in}]=0\right\}\quad;\quad\left\{[\hat{o}_{in},\hat{\bullet}^+_{el}]=\frac{-i}{2}\;;\;[\hat{o}_{el}\;\hat{\bullet}^+_{in}]=\frac{i}{2}\right\}$$

$$\left\{[\hat{o}_{el},\hat{\bullet}^+_{el}]=0\;;\;[\hat{o}_{in},\hat{\bullet}^+_{in}]=0\right\}\quad;\quad\left\{[\hat{o}_{in},\hat{\bullet}^+_{el}]=\frac{-i}{2}\;;\;[\hat{o}_{el}\;\hat{\bullet}^+_{in}]=\frac{i}{2}\right\}$$

$$\left\{[\hat{\bullet}_{el},\hat{o}^+_{el}]=0\;;\;[\hat{\bullet}_{in},\hat{o}^+_{in}]=0\right\}\quad;\quad\left\{[\hat{\bullet}_{in},\hat{o}^+_{el}]=\frac{-i}{2}\;;\;[\hat{\bullet}_{el}\;\hat{o}^+_{in}]=\frac{i}{2}\right\}$$

(A.25)

**Author : Guido Giovanni**
**Department of Mathematics and Physics**
**Senior High School " C.Cavalleri " Parabiago ( Milano )**
**e-mail: liceocavalleri@libero.it ; gioguido54@libero.it**